\documentclass[twocolumn,appendixfloats]{aastex6}

\usepackage{amsmath}
\usepackage{amssymb}
\usepackage{amsfonts}
\usepackage{graphicx}
\usepackage{wasysym}
\usepackage{multirow}
\usepackage{mdframed}

\newcommand{\methanol}{\mbox{CH$_3$OH}}
\newcommand{\twelveco}{\mbox{$^{12}$CO}}

\newcommand{\thirteenco}{\mbox{$^{13}$CO}}
\newcommand{\ceighteeno}{\mbox{C$^{18}$O}}
\newcommand{\fmh}{\mbox{H$_2$CO}}
\newcommand{\mthc}{\mbox{CH$_3$CN}}
\newcommand{\cyacet}{\mbox{HC$_3$N}}
\newcommand{\water}{\mbox{H$_2$O}}
\newcommand{\amm}{\mbox{NH$_3$}}

\newcommand{\kms}{\mbox{km\,s$^{-1}$}}
\newcommand{\sqc}{\mbox{cm$^{-2}$}}
\newcommand{\cc}{\mbox{cm$^{-3}$}}
\newcommand{\lsol}{\mbox{$L_\odot$}}
\newcommand{\msol}{\mbox{$M_\odot$}}
\newcommand{\msolpyr}{\mbox{$M_\odot$\,yr$^{-1}$}}
\newcommand{\ctt}{20~\kms{} cloud}

\newcommand{\mjypbm}{\mbox{mJy\,beam$^{-1}$}}
\newcommand{\jypbm}{\mbox{Jy\,beam$^{-1}$}}
\newcommand{\hii}{\mbox{H\,{\sc ii}}}

\shorttitle{Molecular Gas in the 20 \kms{} Cloud}
\shortauthors{Lu et al.}

\begin{document}

\title{The Molecular Gas Environment in the 20 km\,s$^{-1}$ Cloud in the Central Molecular Zone}

\author{Xing Lu\altaffilmark{1,2,3}, 
Qizhou Zhang\altaffilmark{3}, Jens Kauffmann\altaffilmark{4}, Thushara Pillai\altaffilmark{4}, Steven N.~Longmore\altaffilmark{5}, J.~M.~Diederik Kruijssen\altaffilmark{6}, Cara Battersby\altaffilmark{3}, Hauyu Baobab Liu\altaffilmark{7}, Adam Ginsburg\altaffilmark{8}, Elisabeth A.~C.~Mills\altaffilmark{9}, Zhi-Yu Zhang\altaffilmark{10,7}, and Qiusheng Gu\altaffilmark{1}}
\altaffiltext{1}{School of Astronomy and Space Science, Nanjing University, Nanjing, Jiangsu 210093, China; \href{mailto:xinglv.nju@gmail.com}{xinglv.nju@gmail.com}}
\altaffiltext{2}{National Astronomical Observatory of Japan, 2-21-1 Osawa, Mitaka,Tokyo, 181-8588, Japan}
\altaffiltext{3}{Harvard-Smithsonian Center for Astrophysics, 60 Garden Street, Cambridge, MA 02138, USA}
\altaffiltext{4}{Max Planck Institut f\"ur Radioastronomie, Auf dem H\"ugel 69, D-53121 Bonn, Germany}
\altaffiltext{5}{Astrophysics Research Institute, Liverpool John Moores University, 146 Brownlow Hill, Liverpool L3 5RF, UK}
\altaffiltext{6}{Astronomisches Rechen-Institut, Zentrum f\"{u}r Astronomie der Universit\"{a}t Heidelberg, M\"{o}nchhofstra\ss e 12-14, 69120 Heidelberg, Germany}
\altaffiltext{7}{European Southern Observatory, Karl-Schwarzschild-Str.~2, D-85748 Garching, Germany}
\altaffiltext{8}{National Radio Astronomy Observatory, Socorro, NM 87801, USA}
\altaffiltext{9}{Department of Physics and Astronomy, San Jose State University, One Washington Square, San Jose, CA 95192, USA}
\altaffiltext{10}{Institute for Astronomy, University of Edinburgh, Royal Observatory, Blackford Hill, Edinburgh EH9 3HJ}

\begin{abstract}
We recently reported a population of protostellar candidates in the \ctt{} in the Central Molecular Zone of the Milky Way, traced by \water{} masers in gravitationally bound dense cores. In this paper, we report high-angular-resolution ($\sim$3\arcsec{}) molecular line studies of the environment of star formation in this cloud. Maps of various molecular line transitions as well as the continuum at 1.3~mm are obtained using the Submillimeter Array. Five \amm{} inversion lines and the 1.3~cm continuum are observed with the Karl~G.~Jansky Very Large Array. The interferometric observations are complemented with single-dish data. We find that the \methanol{}, SO, and HNCO lines, which are usually shock tracers, are better correlated spatially with the compact dust emission from dense cores among the detected lines. These lines also show enhancement in intensities with respect to SiO intensities toward the compact dust emission, suggesting the presence of slow shocks or hot cores in these regions. We find gas temperatures of $\gtrsim$100~K at 0.1-pc scales based on RADEX modelling of the \fmh{} and \amm{} lines. Although no strong correlations between temperatures and linewidths/\water{} maser luminosities are found, in high-angular-resolution maps we notice several candidate shock heated regions offset from any dense cores, as well as signatures of localized heating by protostars in several dense cores. Our findings suggest that at 0.1-pc scales in this cloud star formation and strong turbulence may together affect the chemistry and temperature of the molecular gas.
\end{abstract}


\section{INTRODUCTION}\label{sec:intro}

The Central Molecular Zone (CMZ) is the inner $\sim$500~pc of the Galaxy \citep{morris1996} and contains more than $10^7$~\msol{} of dense molecular gas \citep[mean density in clouds $\sim$a few $10^3$~\cc{};][]{ferriere2007}. Within the CMZ are a series of massive molecular clouds with typical projected scales of 10~pc and masses of $10^5$~\msol{}. These clouds are characterized by large turbulent linewidths \citep[FHWM$\sim$10--10$^2$~\kms{};][]{shetty2012,kruijssen2013}. Spectral line studies using single-dish observations have found interesting gas properties related to such strong turbulence. Mappings of shock tracers, such as SiO, revealed that their emission is widespread but non-uniform, suggesting large-scale shocks at $\gtrsim$1-pc scales \citep{martinpintado1997,riquelme2010,jones2012}. Widespread emission of organic molecules has been detected and these molecules are suggested to be released from grain mantles by shocks \citep{martinpintado2001,requenatorres2006,requenatorres2008}. In addition, efforts have been made to map gas temperatures in the CMZ using multiple transitions of \amm{} or \fmh{}, which have revealed ubiquitously high temperatures (50--100~K or higher) at $\gtrsim$1-pc scales and suggested that turbulent shocks could be the heating source \citep{mills2013,ao2013,ginsburg2016,immer2016}. These studies indicate that strong turbulence plays a vital role in shaping the molecular gas environment in the CMZ clouds at $>$1-pc scales, though the origin of the turbulence is under debate \citep[e.g.,][]{rodriguez2006,kruijssen2015}.

However, an ambiguity exists in single-dish observations that use linewidths to indicate turbulent strength: at angular resolutions of $\sim$30\arcsec{} (1~pc at the distance of the CMZ) and above, linewidths may have contributions from unresolved systematic motions (e.g., rotation, infall) thus may be questionable to be a good indicator of turbulence \citep[e.g., discussions in][]{henshaw2016}. Higher angular resolutions of 3\arcsec{} (0.1~pc) would help to resolve systematic motions of dense cores within clouds in order to evaluate the impact of turbulence on gas.

Star formation is one of the key factors in shaping the gas environment in galaxies \citep{kennicutt2012}. Active star formation can reveal itself by heating the ambient gas \citep[e.g., hot molecular cores,][]{araya2005} as well as changing the chemical composition of gas \citep[e.g., enhancement of SiO emission by outflows,][]{garay2000}. However, it remains unclear whether star formation has a major impact on the molecular gas in addition to strong turbulence in the CMZ clouds. The overall star formation rate (SFR) in the CMZ measured with infrared luminosities is $\sim$0.05--0.15~\msolpyr{} \citep{barnes2017}, which is significantly lower than expected from the well-established correlation between amount of dense gas and star formation \citep{longmore2013a}, e.g., the Kennicutt-Schmidt relations \citep{kennicutt1998,kennicutt2012}. \citet{kruijssen2014} discussed mechanisms that may explain the currently low SFR in the CMZ, including strong turbulence, high virial ratios, acoustic instabilities, and episodic cycling. They present a self-consistent scenario in which star formation in the CMZ may proceed episodically, regulated by turbulence driving and feedback. This idea was quantified by \citet{KK2015} and \citet{krumholz2017}, who showed that the combination of these mechanisms implies that a wide range of star-forming environments should be present within the CMZ, with widely varying degrees of star formation activity.

Except for a few active star forming regions such as Sgr~B2, most of the clouds in the CMZ have been found to be relatively quiescent in star formation \citep{gusten1983,lis1994,immer2012b,kauffmann2016a}. An example is G0.253+0.016, which shows quite inactive star formation (e.g., a weak \water{} maser, \citealt{lis1994}; a likely gravitationally bound dense core, \citealt{kauffmann2013a,johnston2014,rathborne2014b}; and no evidence of free-free emission from \hii{} regions, \citealt{rodriguez2013,mills2015}). Observations of spectral lines show that the gas in G0.253+0.016 is dominantly influenced by strong turbulence \citep{rathborne2014a,rathborne2015}, but little evidence of being affected by star formation has been found. In other CMZ clouds, to what extent star formation can affect the environment remains to be answered, and observations with angular resolutions of $\lesssim$3\arcsec{} (0.1~pc) are required to match with the scale of dense cores where star formation takes place \citep[$\lesssim$0.1~pc;][]{kauffmann2008,lu2014}.

Therefore, to examine the impact of turbulence and star formation on the molecular gas environment in CMZ clouds, interferometric observations that can provide $\sim$3\arcsec{} or better angular resolutions are necessary. We also need an optimized sample for our study, e.g., a CMZ cloud with more active star formation than G0.253+0.016.

In our recent work \citep{lu2015b}, we studied star formation in the \ctt{}, a massive ($\gtrsim$10$^5$~\msol{}) molecular cloud in the Sgr~A complex \citep{ferriere2012}. Single-dish observations have found this cloud to be extremely turbulent with linewidth of $\gtrsim$10~\kms{} at 1-pc scales and large velocity gradient of 2--5~\kms{}\,pc$^{-1}$ along its major axis \citep{coil1999,tsuboi2011}. In \citet{lu2015b}, we detected 18 \water{} masers and associated dense cores in this cloud, thus revealing a population of deeply embedded protostellar candidates. Therefore, with both strong turbulence and relatively active star formation activity, the \ctt{} is an appropriate sample for understanding the interplay between star formation and the environmental molecular gas.

Here we continue to use high-angular-resolution interferometric observations of molecular lines to study the molecular gas environment in the \ctt{}. We use a number of molecular lines at 1.3~mm wavelengths including both dense gas tracers (e.g., \fmh{}) and shock tracers (e.g., SiO), and five \amm{} lines at 1.3~cm wavelengths (K band) that are conventional dense gas tracers. Moreover, multiple transitions of \fmh{} and \amm{} can be used as thermometers to reveal impact of star formation and turbulence on gas temperatures, in which the two lines may trace different gas components, in terms of chemistry and/or density, thus complement each other. In \autoref{sec:obs}, we outline our interferometric and single-dish observations as well as archival data used in this paper. In \autoref{sec:results}, we present the results, including the 1.3~mm continuum and spectral lines, the 1.3~cm continuum, the \amm{} lines, and gas temperatures derived from \fmh{} and \amm{}. In \autoref{sec:discussion}, we focus on two points: the impact of star formation and turbulence on the chemical composition of gas (spatial correlations between dust and spectral line emission, enhancement of emission of shock tracers), and the heating of gas at 0.1-pc scales by star formation and turbulence. We draw our conclusions and summarize our results in \autoref{sec:conclusions}. In this paper we adopt the distance to the Galactic Center of 8.34~kpc \citep{reid2014}.

\addtocounter{footnote}{10}

\section{OBSERVATIONS AND DATA REDUCTION}\label{sec:obs}

\subsection{Submillimeter Array (SMA) Observations}\label{subsec:obs_sma}

We observed eight positions as a mosaic in the \ctt{} in 2013 with the SMA\footnote{The Submillimeter Array is a joint project between the Smithsonian Astrophysical Observatory and the Academia Sinica Institute of Astronomy and Astrophysics, and is funded by the Smithsonian Institution and the Academia Sinica.} \citep{ho2004} at 1.3~mm wavelengths in the compact and sub-compact configurations. Spectral lines between 216.9--220.9~GHz and 228.9--232.9~GHz as well as the continuum were obtained at the same time. Details of the SMA observations and data reduction, using the IDL superset MIR\footnote{\url{https://www.cfa.harvard.edu/~cqi/mircook.html}}, MIRIAD \citep{sault1995}, and CASA \citep{mcmullin2007}, have been reported in \citet{lu2015b}. In the end we obtained image cubes of molecular lines, with typical clean beams of $5\farcs{0}\times2\farcs{8}$ at position angles of 5\arcdeg, and rms of 0.11--0.14~\jypbm{} per 1.1~\kms{} velocity bin. In \autoref{subsec:obs_combine}, we will discuss the combination of the SMA and single-dish data.

\subsection{Karl G.~Jansky Very Large Array (VLA) Observations}\label{subsec:obs_vla}

We observed three positions as a mosaic in the \ctt{} in 2013 with the National Radio Astronomy Observatory (NRAO\footnote{The National Radio Astronomy Observatory is a facility of the National Science Foundation operated under cooperative agreement by Associated Universities, Inc.}) VLA at K band in the DnC configuration. The \water{} maser transition at 22.2~GHz, five \amm{} inversion transitions from ($J$,~$K$)=(1,~1) to (5,~5), two CCS lines and one NH$_2$D line, as well as the continuum using a total bandwidth of 1~GHz were observed. Details of the VLA observations and reduction of the \water{} maser and continuum data, using CASA, have been reported in \citet{lu2015b}. The CCS and NH$_2$D lines were not detected at a sensitivity of 2.0~mJy per $3\arcsec\times2\arcsec$ beam per 1~\kms{}.

For the \amm{} (1,~1) and (2,~2) transitions, the 16-MHz-wide subbands were split into 1024 channels, leading to a velocity coverage of 200~\kms{} and a channel width of 0.2~\kms{}, which can cover all the hyperfine splittings of these two transitions. For \amm{} (3,~3) to (5,~5), the satellite components of the hyperfine splittings are usually much weaker, therefore the 8-MHz-wide subbands, split into 512 channels, were used to cover the main components only, with a velocity coverage of 100~\kms{} and a channel width of 0.2~\kms{}. To increase the signal-to-noise ratios, we regridded the visibility data to a velocity bin of 1~\kms{} before CLEANing.

Then we imaged the \amm{} data with CASA~4.5.0. From the early versions of the CLEANed images, we noticed that the emission of \amm{}, especially that of \amm{} (3,~3), is extended in nature, thus the canonical CLEAN with pixel-by-pixel clean component finding algorithm did not recover such emission well. This is evidenced by the extended `stripes'  and strong `negative bowls' seen in the images. Therefore, we used the multi-scale CLEAN in CASA, which implements multi-scale clean component finding algorithm, to improve the imaging of extended structures. We used multi-scale parameters of 0, 5, 20, and 80 pixels, corresponding to angular scales of 0\arcsec{}, 3\arcsec{}, 12\arcsec{}, and 48\arcsec{}. The stripes and negative bowls are significantly suppressed in the obtained images. Typical rms of the images is 1.8--2.0~\mjypbm{} per 1~\kms{} velocity bin, with beams of $3\farcs{0}\times2\farcs{1}$ at position angles of 10\arcdeg. In \autoref{subsec:obs_combine}, we will discuss the combination of the VLA and single-dish \amm{} data. Similarly, we reprocessed the VLA 1.3~cm continuum data that also present extended structures with multi-scale CLEAN.

\subsection{Caltech Submillimeter Observatory (CSO) Heterodyne Observations}\label{subsec:obs_cso}

In April 2014, we observed a $2\farcm{5}\times4\farcm{0}$ area in the \ctt{} in the on-the-fly position switching mode with the CSO. The heterodyne receiver was tuned to cover the same frequency range of the SMA observations in the double-sideband mode. The wideband Fast Fourier Transform Spectrometer (FFTS2) and the Fast Fourier Transform Spectrometer (FFTS1) backends were connected to the receiver in parallel. The FFTS2 was used to cover 228.9--232.9 GHz, which in the double-sideband mode also sampled 216.9--220.9~GHz. The channel width is $\sim$0.27~MHz, corresponding to $\sim$0.35~\kms{} at 230~GHz. The FFTS1 was used to cover 231.9--232.9~GHz as well as 216.9--217.9~GHz in the double-sideband mode, in order to observe the SiO 5--4 line at $\sim$217.1~GHz with a finer channel width of $\sim$0.12~MHz than the FFTS2, corresponding to $\sim$0.16~\kms{} at 230~GHz. However the FFTS2 data are adequate for our purpose hence the FFTS1 data are not used in this paper. The on-the-fly mapping was done in two orthogonal directions, between which we shifted the IF frequency by 50~MHz to separate spectral lines from the two sidebands. The optical depth at 225~GHz relative to zenith during the observation was 0.07--0.10, corresponding to precipitable water vapor of 1.3--2.0~mm. The system temperature of FFTS2 was 250--300~K during the observation. We derived a beam efficiency of 0.66 by observing Mars.

There is an atmospheric feature, probably a \water{} absorption line, at $\sim$231.3~GHz, corresponding to $\sim$218.6~GHz in the lower-sideband with the double-sideband mode. The full width at zero intensity of the feature is 0.3--0.4~GHz, depending on the choice of the continuum baseline. Therefore, two \fmh{} lines (3$_{2,2}$--2$_{2,1}$ and 3$_{2,1}$--2$_{2,0}$), the \methanol{} 4$_{2,2}$--3$_{1,0}$ line, and possibly the \cyacet{} 24--23 line (cf.~\autoref{fig:sma_spec}) were affected. All other lines detected by the CSO were free from atmospheric contamination or double-sideband confusion. We processed the data with the CLASS package\footnote{\url{http://www.iram.fr/IRAMFR/GILDAS}} and obtained the final images after a baseline subtraction and a correction for the beam efficiency.

\subsection{Archival Atacama Pathfinder EXperiment (APEX) Heterodyne and Bolocam Galactic Plane Survey (BGPS) Data}\label{subsec:obs_archival}
We used spectral line data from APEX at 218--219~GHz that cover the three \fmh{} lines as well as the \cyacet{} and \methanol{} lines, which were released by \citet{ginsburg2016}. These data readily complement our CSO heterodyne data that missed this frequency range due to the atmospheric contamination.

The 1.1~mm continuum data were obtained from the BGPS v2.1 \citep{ginsburg2013}. The intensities were multiplied by a factor of 0.5 to convert to those at 1.33~mm (equivalent to the IF frequency of the SMA observations, 225~GHz) for the use of data combination, assuming a dust emissivity index
of 2 \citetext{Walker et al., in prep.}.

\subsection{Green Bank Telescope (GBT) \amm{} Data}\label{subsec:obs_gbt}
We used the \amm{} inversion lines from ($J$,~$K$)=(1,~1) to (5,~5) from the NRAO GBT observations (PI: H.~B.~Liu). The observations were done on November 7, 2011. The ($J$,~$K$)=(3,~3) to (5,~5) data have been published in \citet{minh2013} and details of the observations can be found therein. The GBT observations fully cover the VLA mosaic field hence can be combined with the latter to recover diffuse \amm{} emission.

\subsection{Combination of the Interferometer and Single-Dish Data}\label{subsec:obs_combine}
In order to recover the missing fluxes in the interferometric observations, we combined the SMA and VLA data with single-dish data from CSO, APEX or GBT, using the \textit{feather} task in CASA.

\begin{figure*}[!t]
\centering
\includegraphics[width=0.9\textwidth]{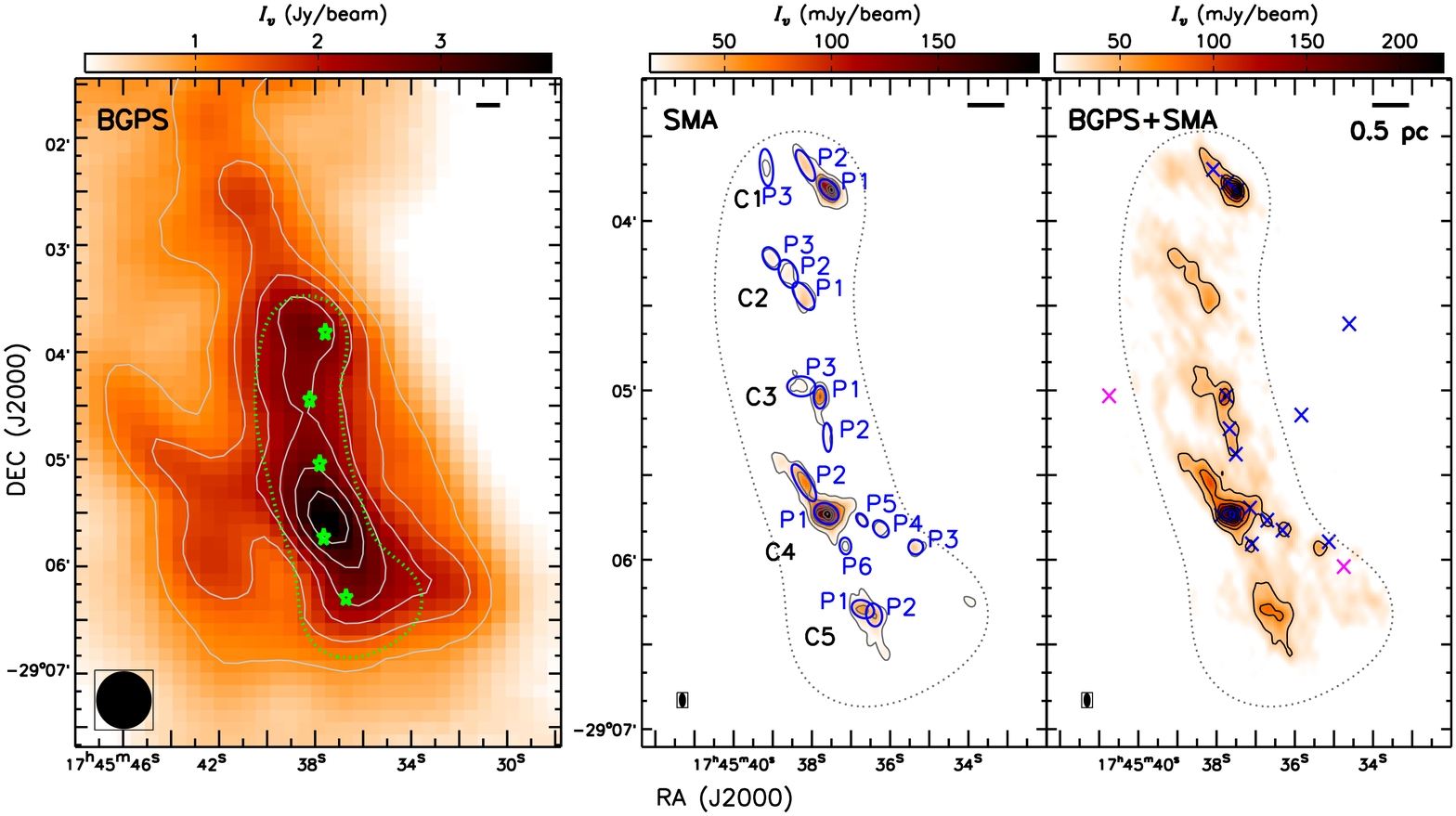}
\caption{1.3~mm continuum maps of \ctt{}. Left: BGPS continuum map, which has been multiplied by factor 0.5 to convert to intensities at the frequency of the SMA observations, is shown in the background image, with contours at 1.0--3.5~\jypbm{} in steps of 0.5~\jypbm{}. The blue dotted loop shows the FWHM of the SMA primary-beams. The 33\arcsec{} effective resolution of the BGPS is shown in the lower left corner. Stars mark the most massive dense cores in each of the five clumps (all named as `P1' in the middle panel). Middle: both contours and image show the SMA 1.3~mm continuum emission (not corrected for primary beam responses in order to have a uniform rms). The contours are between 5$\sigma$ and 65$\sigma$ levels in steps of 10$\sigma$, where 1$\sigma$=3~\mjypbm{}. The dotted loop shows the FWHM of the SMA primary beams. The synthesized beam of the SMA, 5\farcs{0}$\times$2\farcs{8} with a position angle of 5\arcdeg{}, is shown in the lower left corner. The five clumps, from C1 to C5, are labeled. The blue ellipses are the results of Gaussian fittings to dense cores. Right: both contours and image show the SMA+BGPS combined 1.3~mm continuum emission. The contours levels are selected at 24~\mjypbm{}+3~\mjypbm{}$\times$[5, 15, 25, 35, 45, 55, 65], in order to highlight the similarity with the middle panel at 0.1-pc scales. The \water{} masers detected by the VLA are marked by crosses, with magenta ones showing cataloged AGB stars \citep{lu2015b}.}
\label{fig:dust}
\end{figure*}

First, we combined the SMA 1.3~mm continuum data and the BGPS data. We weighted the BGPS image with the primary beams of the SMA, to suppress the signal outside of the SMA primary beams. Then we used the \textit{feather} task to combine the two datasets, with a lowpass filtering at a spatial scale of 10~m (corresponding to an angular scale of 36\arcsec{} at the wavelength of 1.33~mm) for the BGPS data to filter out the high spatial frequency data\footnote{In principle we should use a lowpass filtering at 12~m (31\arcsec{} at the wavelength of 1.33~mm), which provides the same spatial frequency as the original BGPS data with a 10-m dish at the wavelength of 1.1~mm. However due to the limitation of the \textit{feather} task we are unable to set such a large parameter. Using 10~m is more aggressive for filtering high spatial frequency data. We have confirmed that varying the lowpass filtering parameters does not affect the resulting image (variation in total fluxes $<$ 1\%, variation in intensities $<$ 1/3 of the rms).}. At last we compared the fluxes of the combined image and the BGPS image to check the consistency. The difference between total fluxes within the FWHM of the SMA primary beams (see the dotted loop in \autoref{fig:dust}) for the combined image and the BGPS image (weighted by the SMA primary beams) is 1.6\%, which we considered to be acceptable. The results are presented in \autoref{subsec:results_dust}.

Similarly, we combined the SMA and single-dish spectral line data. For several spectral lines around 218.6~GHz, including \fmh{} 3$_{2,2}$--2$_{2,1}$ and 3$_{2,1}$--2$_{2,0}$, \methanol{} 4$_{2,2}$--3$_{1,0}$, and \cyacet{} 24--23, the CSO data were affected by an atmospheric feature, therefore we used the APEX data which are free from contamination. The \fmh{} 3$_{0,3}$--2$_{0,2}$ line in the CSO data was not affected by the atmospheric feature. However, later in \autoref{subsec:results_tkin} we will use line ratios of \fmh{} to derive gas temperatures. To avoid any offset between \fmh{} images, we made combined images with the APEX data also for \fmh{} 3$_{0,3}$--2$_{0,2}$. For the other lines we used the CSO data. We first weighted the single-dish images with the primary beams of the SMA, then combined them with the SMA images using the \textit{feather} task with a lowpass filtering of spatial scales at 10~m (36\arcsec{} at the wavelength of 1.33~mm) for the CSO data or 12~m (31\arcsec{} at the wavelength of 1.33~mm) for the APEX data.

The VLA and GBT \amm{} data were combined following the same procedures. The \textit{feather} task was run with a lowpass filtering of spatial scales at 100~m (36\arcsec{} at the wavelength of 1.3~cm) for the GBT data.

All the combined images processed above have been convolved with primary beams of the interferometers. We also applied primary-beam corrections to them to obtain correct fluxes. In the following sections, we identified structures using the combined images and measured fluxes using the primary-beam corrected images.

\section{RESULTS}\label{sec:results}

\subsection{SMA+BGPS Dust Emission}\label{subsec:results_dust}
In \citet{lu2015b}, we reported the discovery of 17 dense cores traced by the SMA 1.3~mm continuum emission. Here we use the combined SMA and BGPS continuum data to trace the diffuse dust emission in the cloud. Later in \autoref{subsec:disc_2dxcorr} we will use the SMA-only data and the combined SMA+BGPS data to represent compact and diffuse dust emission, respectively.

The BGPS map, the continuum image made with the SMA data, and that made with the combined SMA+BGPS data are shown in \autoref{fig:dust}. The total 1.3~mm continuum flux, within the FWHM of the SMA primary beams, estimated from the primary-beam-corrected SMA+BGPS image, is 19.2 Jy. The 1.3~cm continuum flux in the same area observed by the VLA is $\sim$0.4~Jy \citep{lu2015b}. There is no evidence of any optically thick free-free emission in the area at the considered frequencies. Assuming a completely flat spectral index from 1.3~cm to shorter wavelengths, the flux of 0.4~Jy does not make significant difference to the total flux at 1.3~mm. Therefore, we conclude that the 1.3~mm continuum is dominated by dust emission.

The dust emission maps in \autoref{fig:dust} demonstrate that the SMA observations are able to resolve dense cores at 0.1-pc scales. Therefore, in the following we will use them as well as VLA observations that have an even higher angular resolution to study the gas environment of star formation in the \ctt{}.

\begin{figure*}[!t]
\centering
\includegraphics[width=0.9\textwidth]{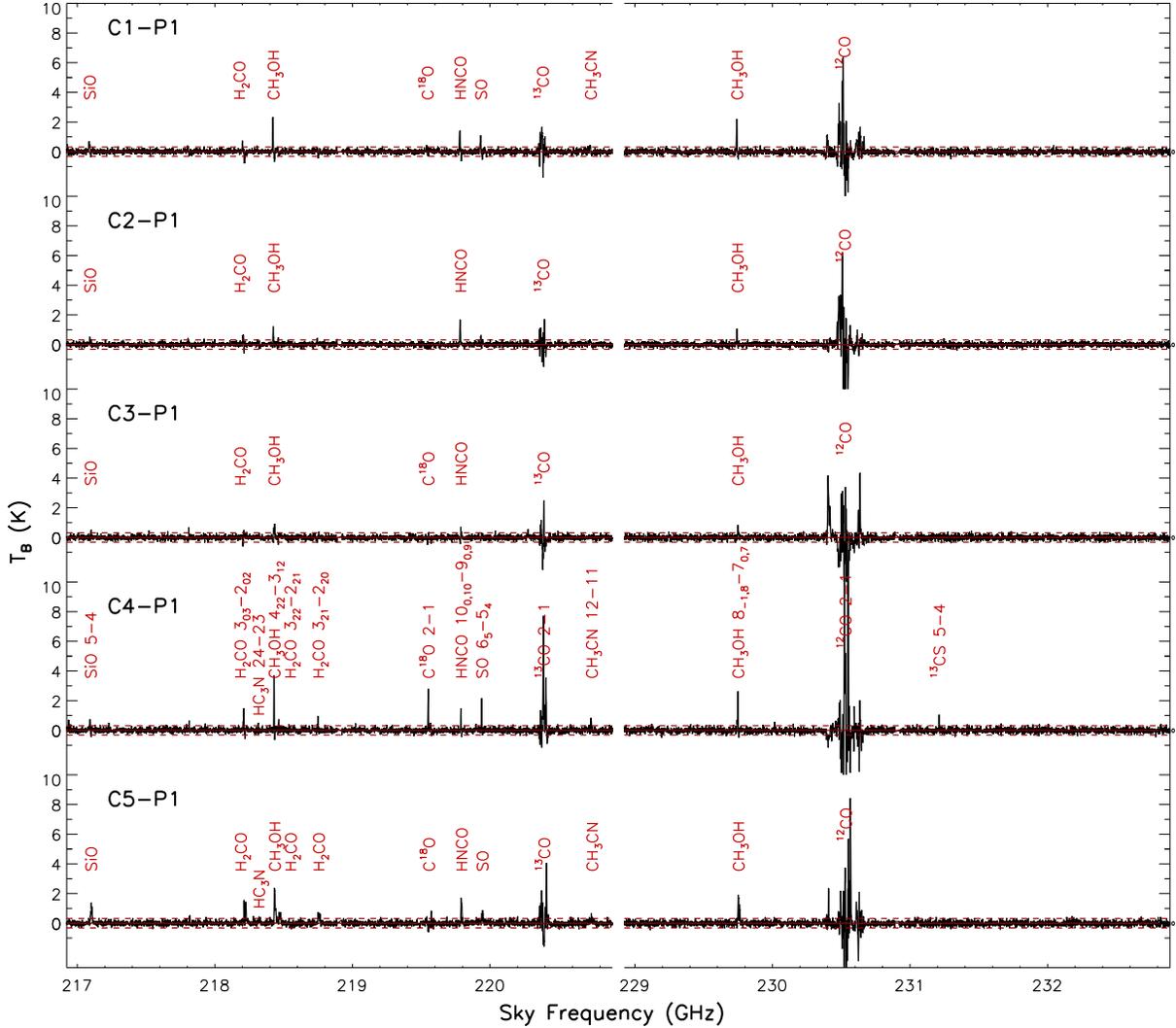}
\caption{Spectral lines detected by the SMA toward the most massive dense cores in each of the five clumps. Names of molecular species are marked; for C4-P1 where all the species are detected, transitions are also labeled. Horizontal dashed lines in each spectrum mark the $\pm$3$\sigma$ levels. For \twelveco{} and \thirteenco{} lines, absorption features can be clearly seen.}
\label{fig:sma_spec}
\end{figure*}

\begin{figure*}[!t]
\begin{tabular}{p{3.2cm}p{3.2cm}p{3.2cm}p{3.2cm}p{3.2cm}}
\multicolumn2c{\raisebox{-0.5em}{\includegraphics[height=0.32\textwidth]{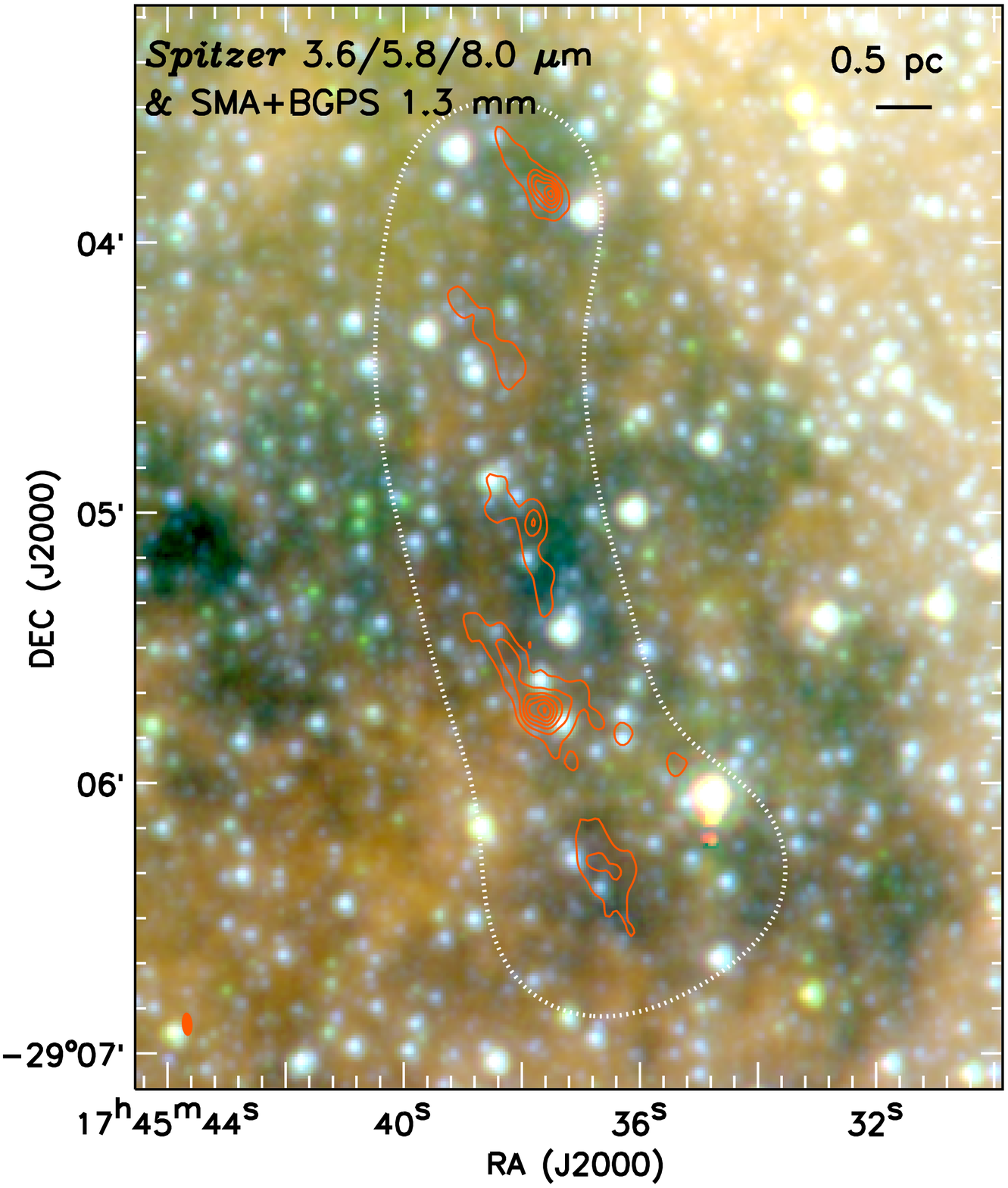}}} &  \includegraphics[height=0.31\textwidth]{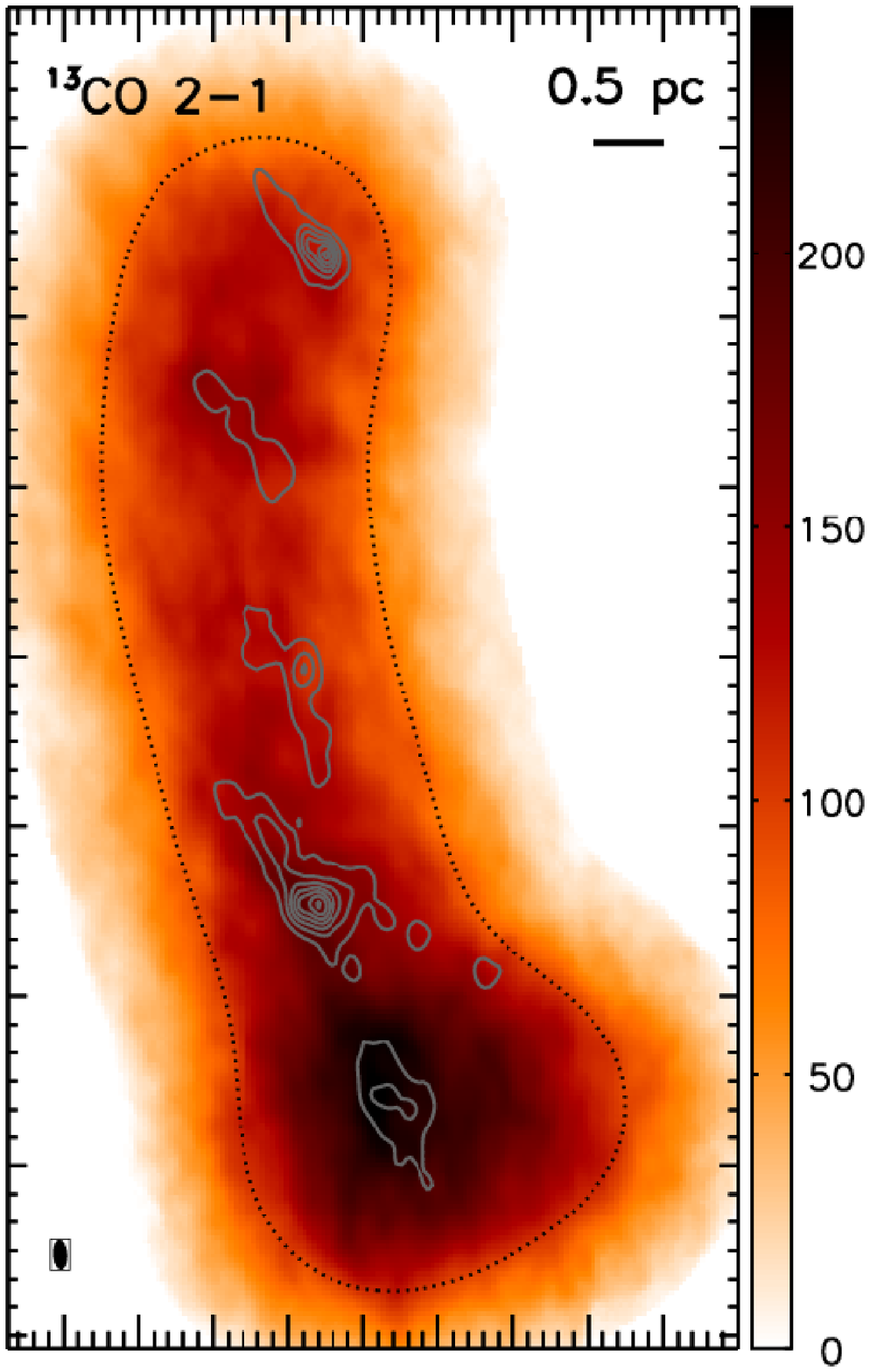} & \includegraphics[height=0.31\textwidth]{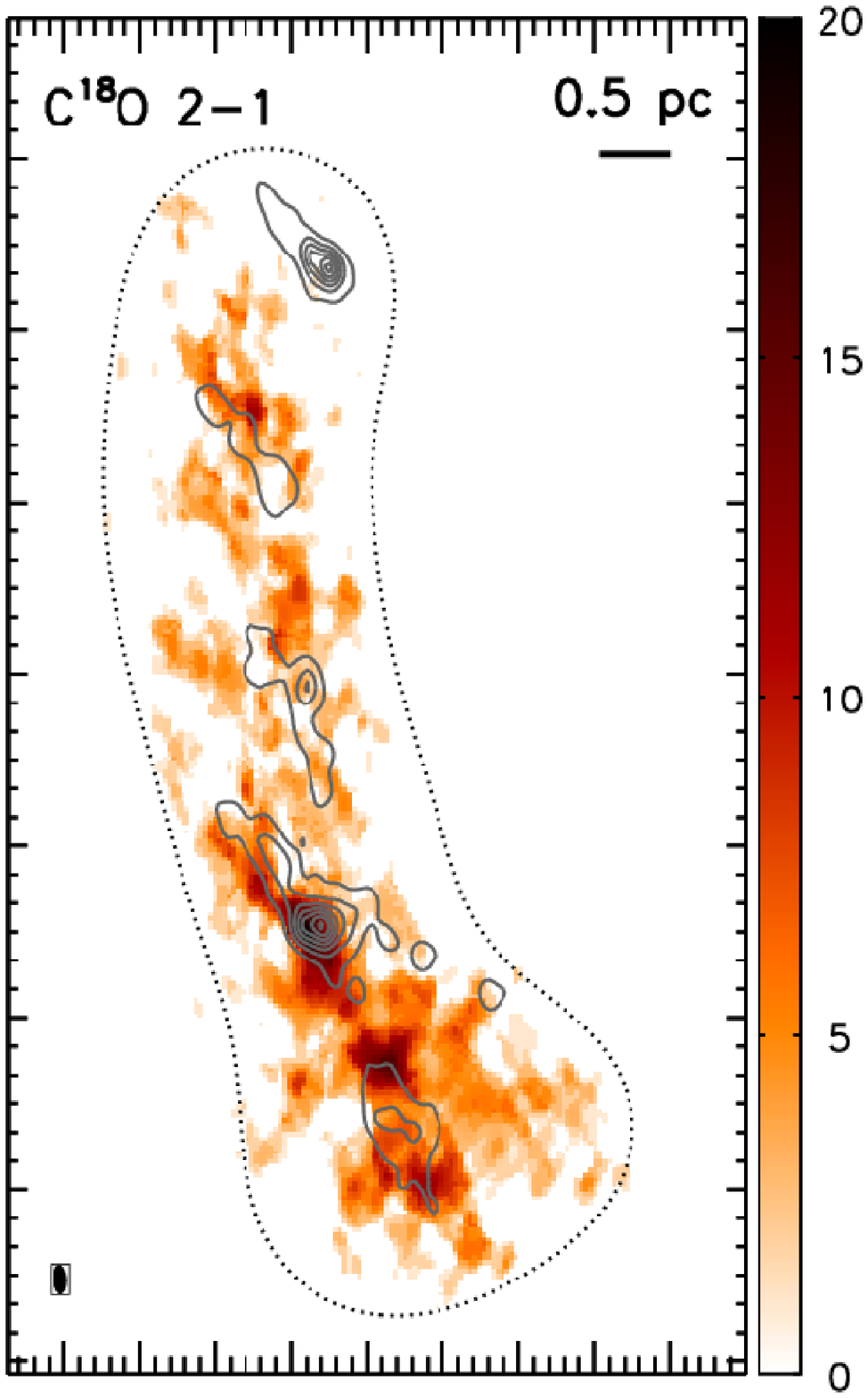} & \includegraphics[height=0.31\textwidth]{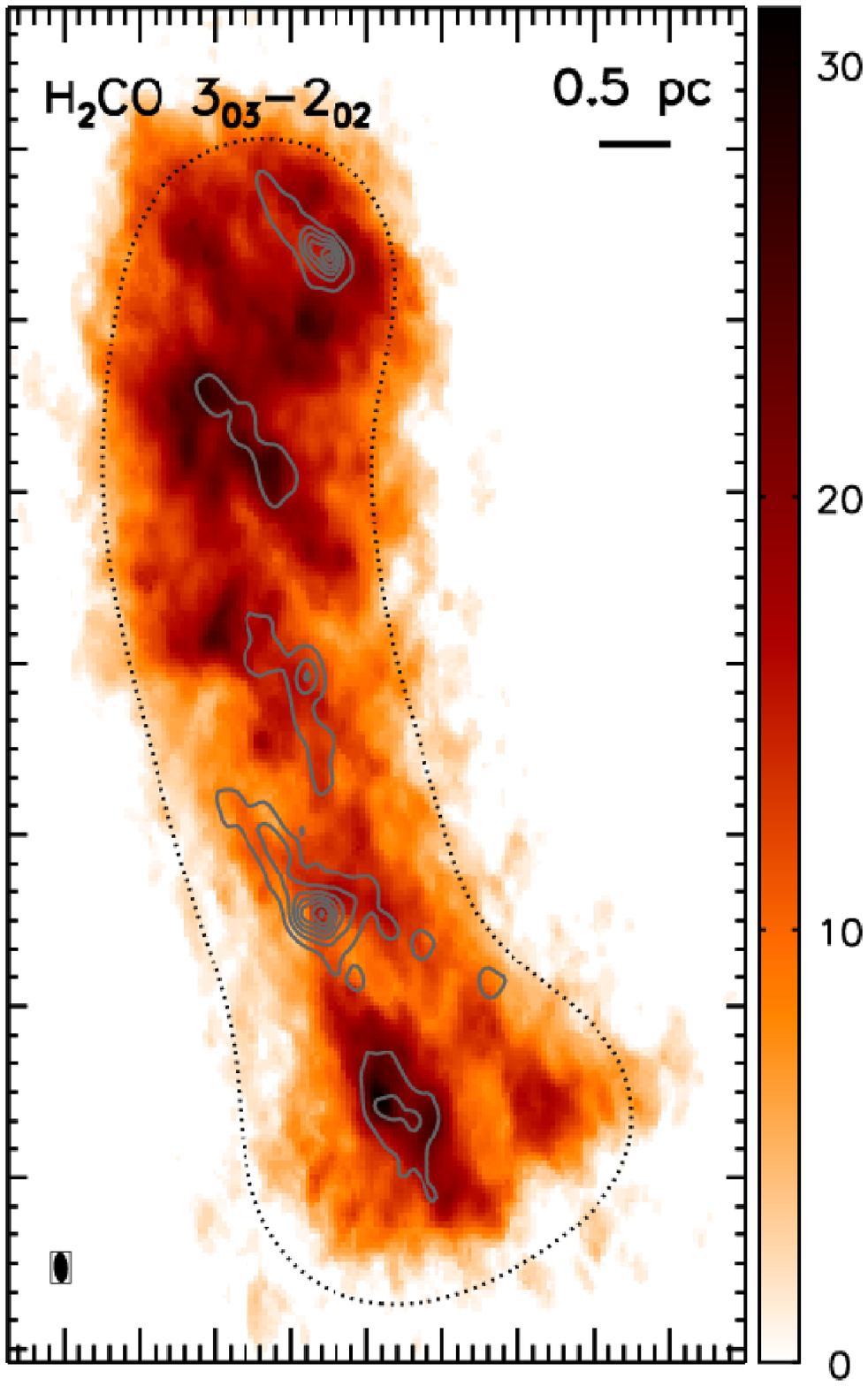} \\
\includegraphics[height=0.31\textwidth]{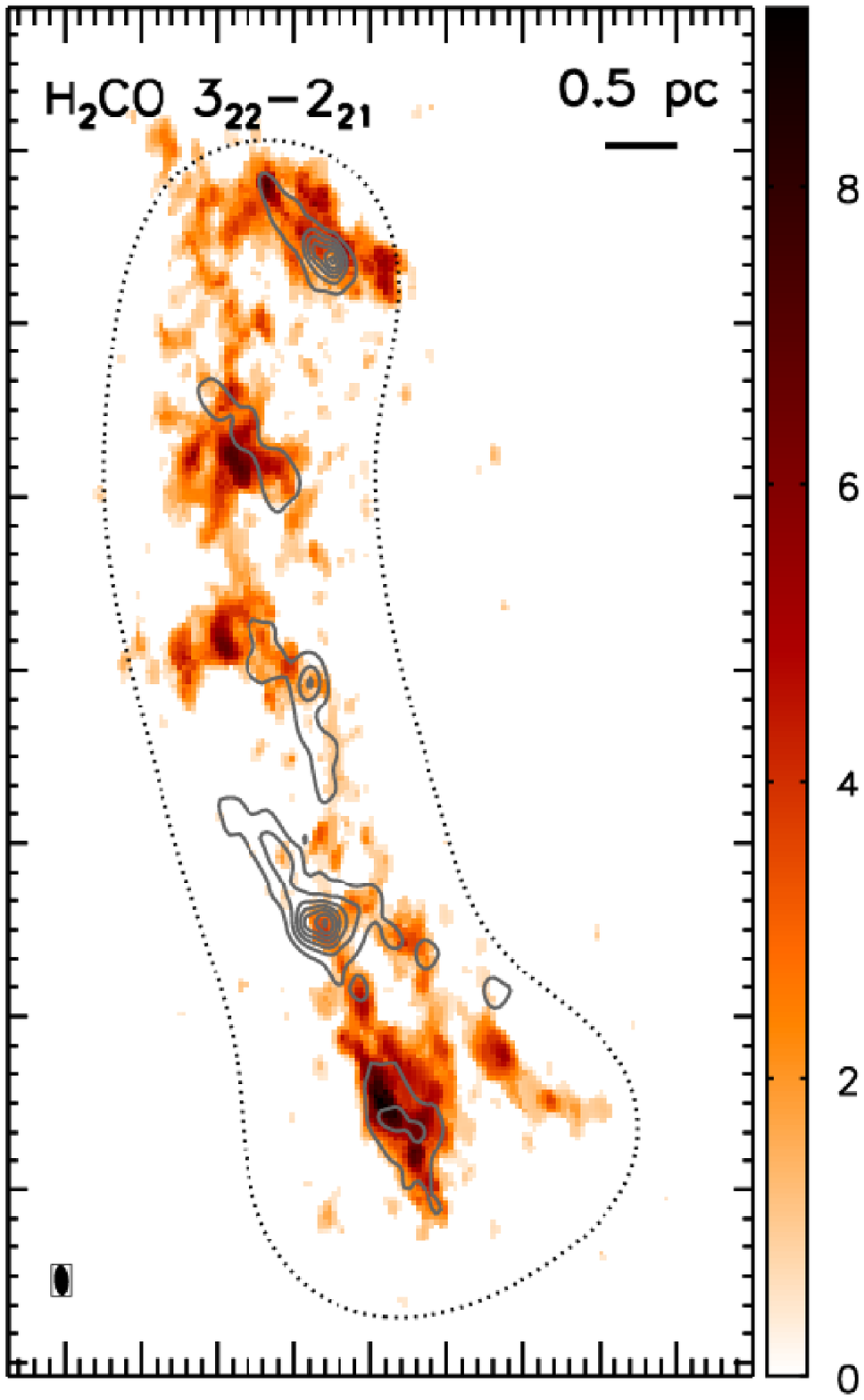} & \includegraphics[height=0.31\textwidth]{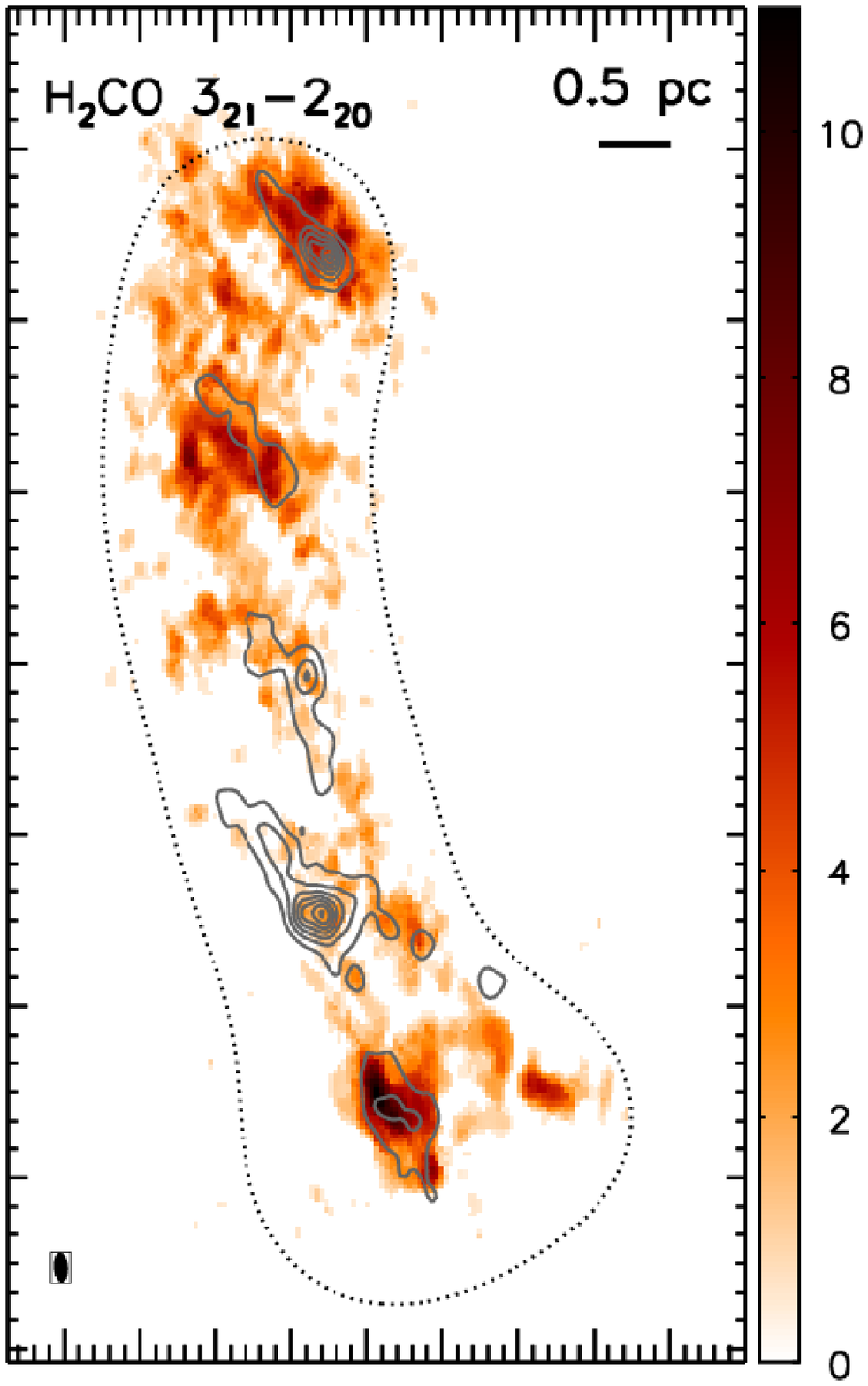} & \includegraphics[height=0.31\textwidth]{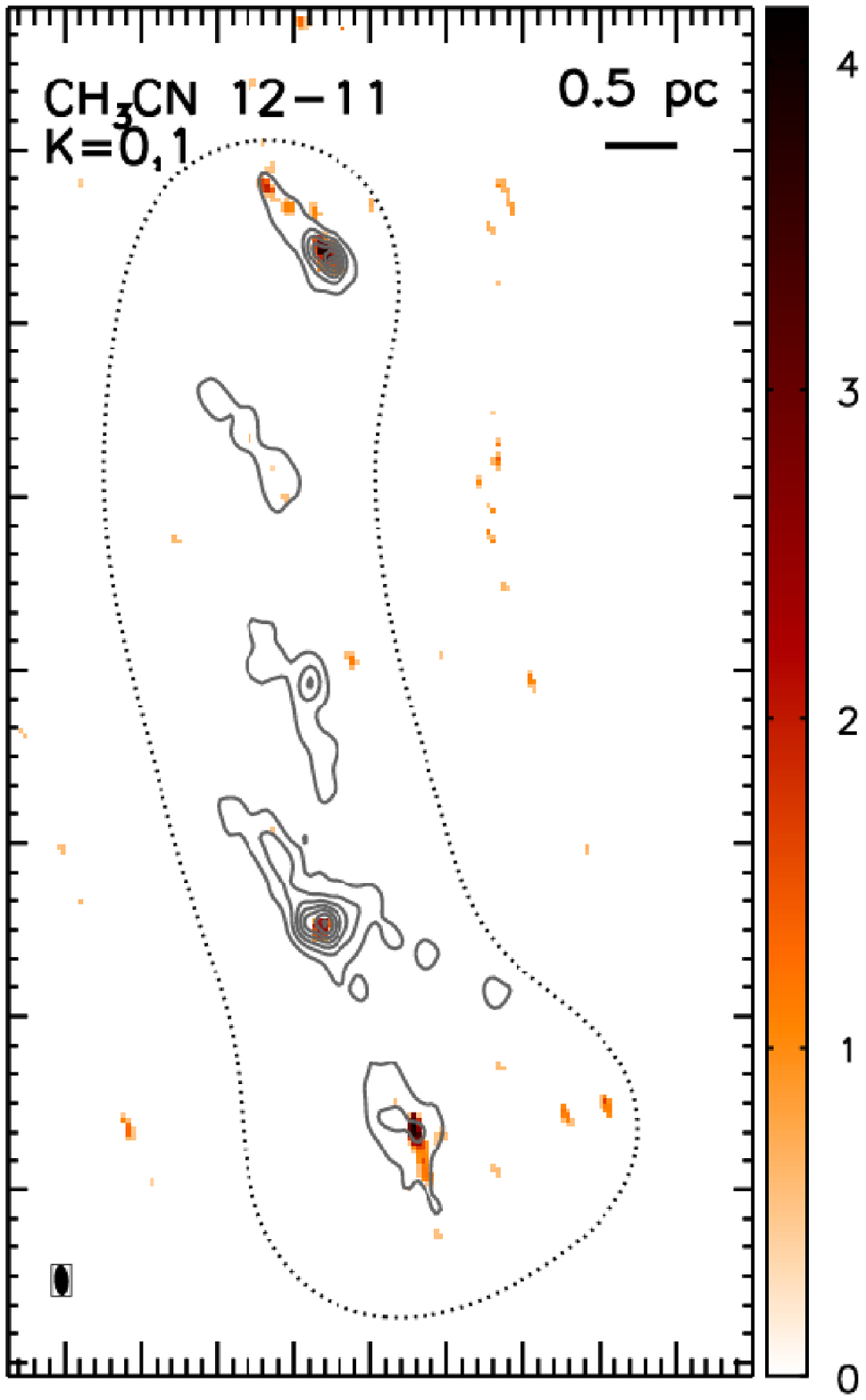} & \includegraphics[height=0.31\textwidth]{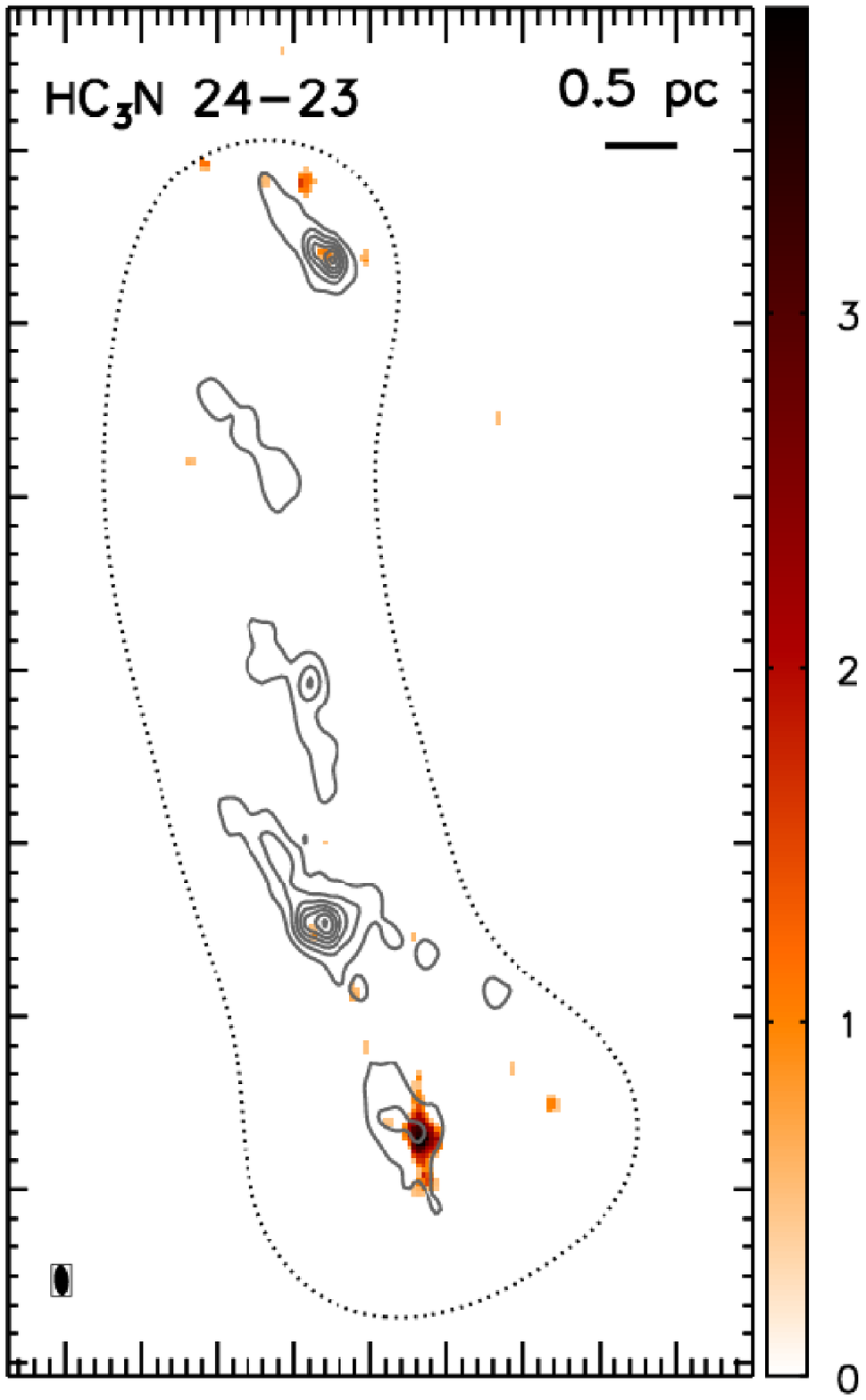} & \includegraphics[height=0.31\textwidth]{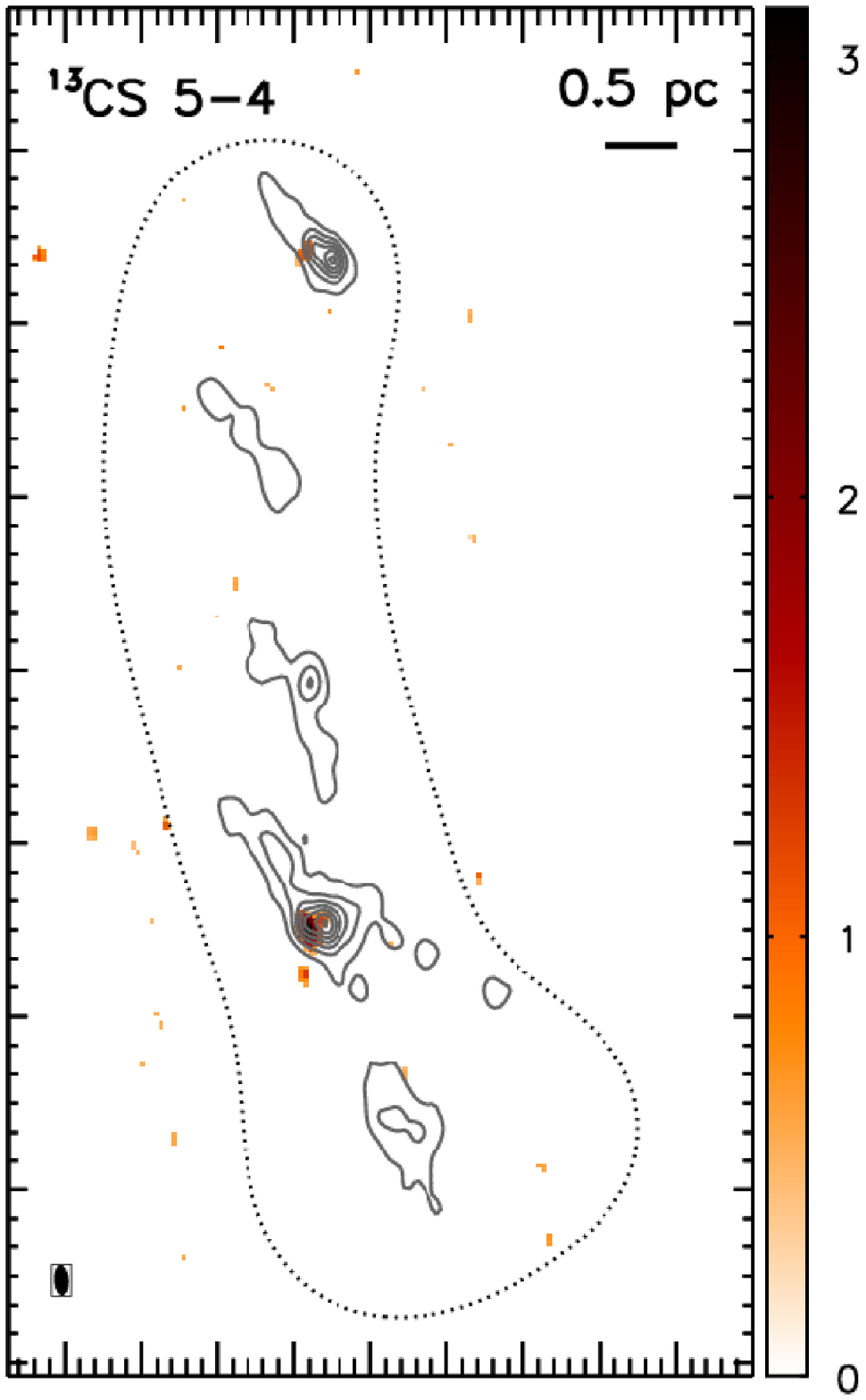} \\
\raisebox{-1.2em}{\hspace*{-1.9em}\includegraphics[height=0.333\textwidth]{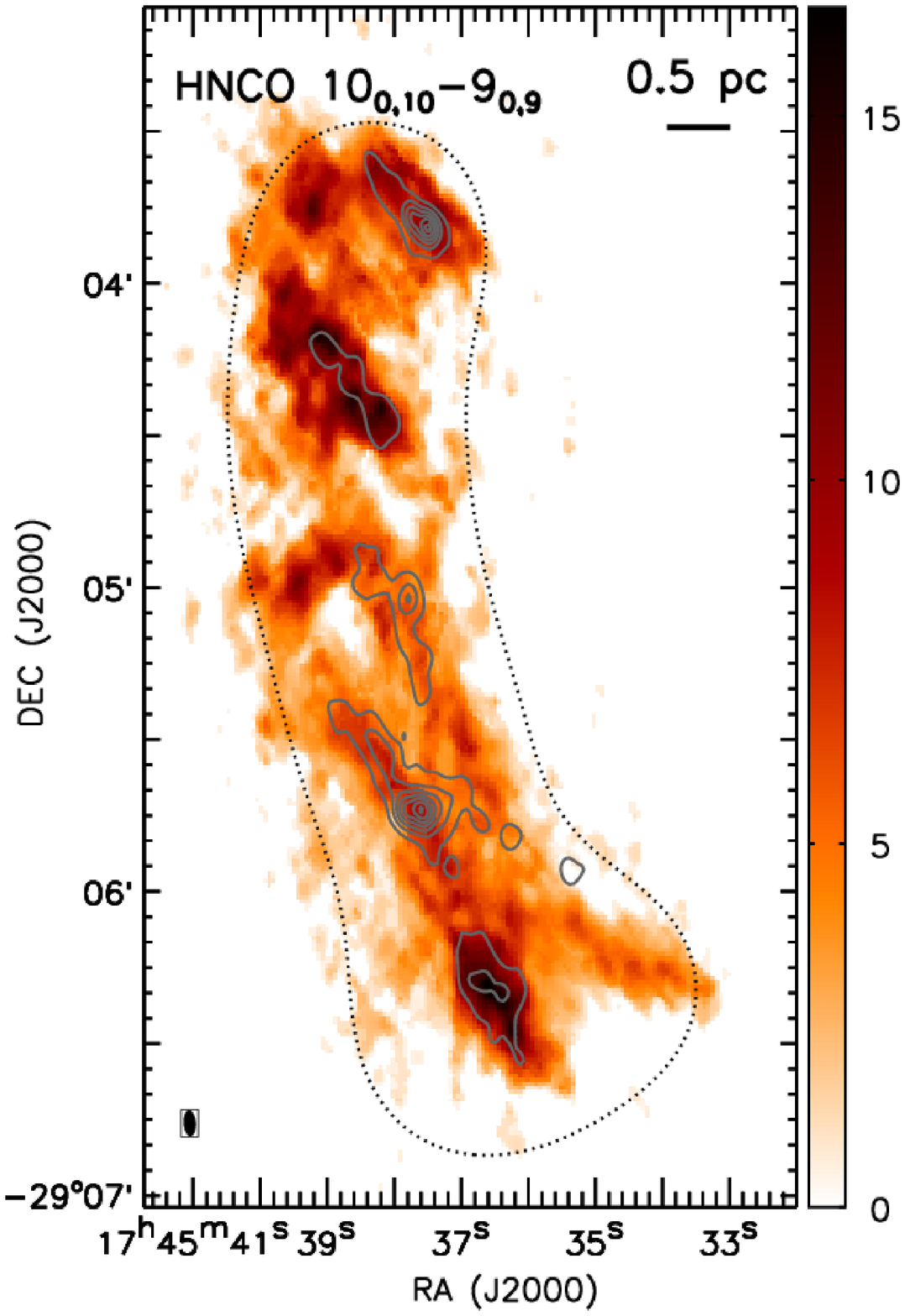}} & \includegraphics[height=0.31\textwidth]{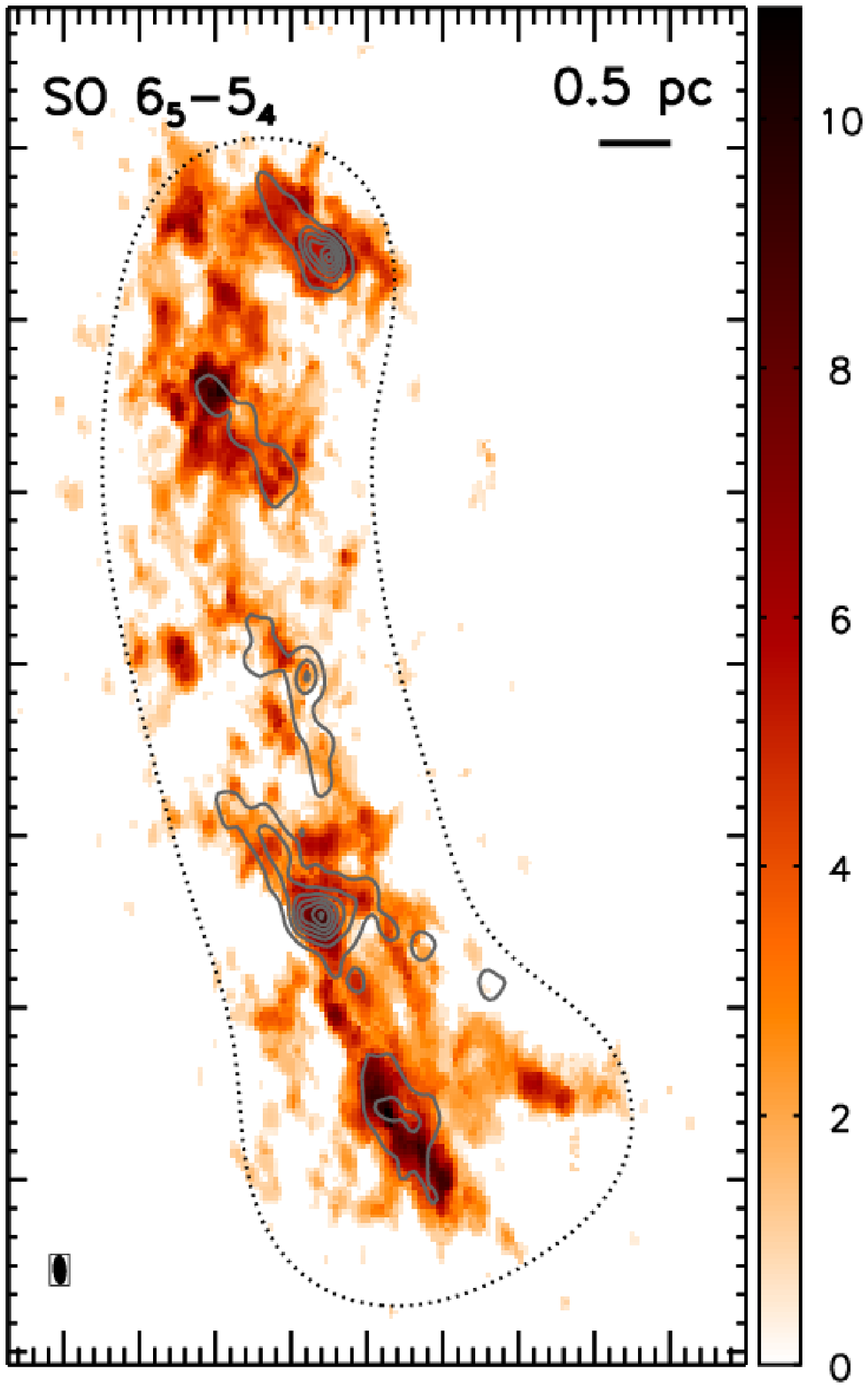} & \includegraphics[height=0.31\textwidth]{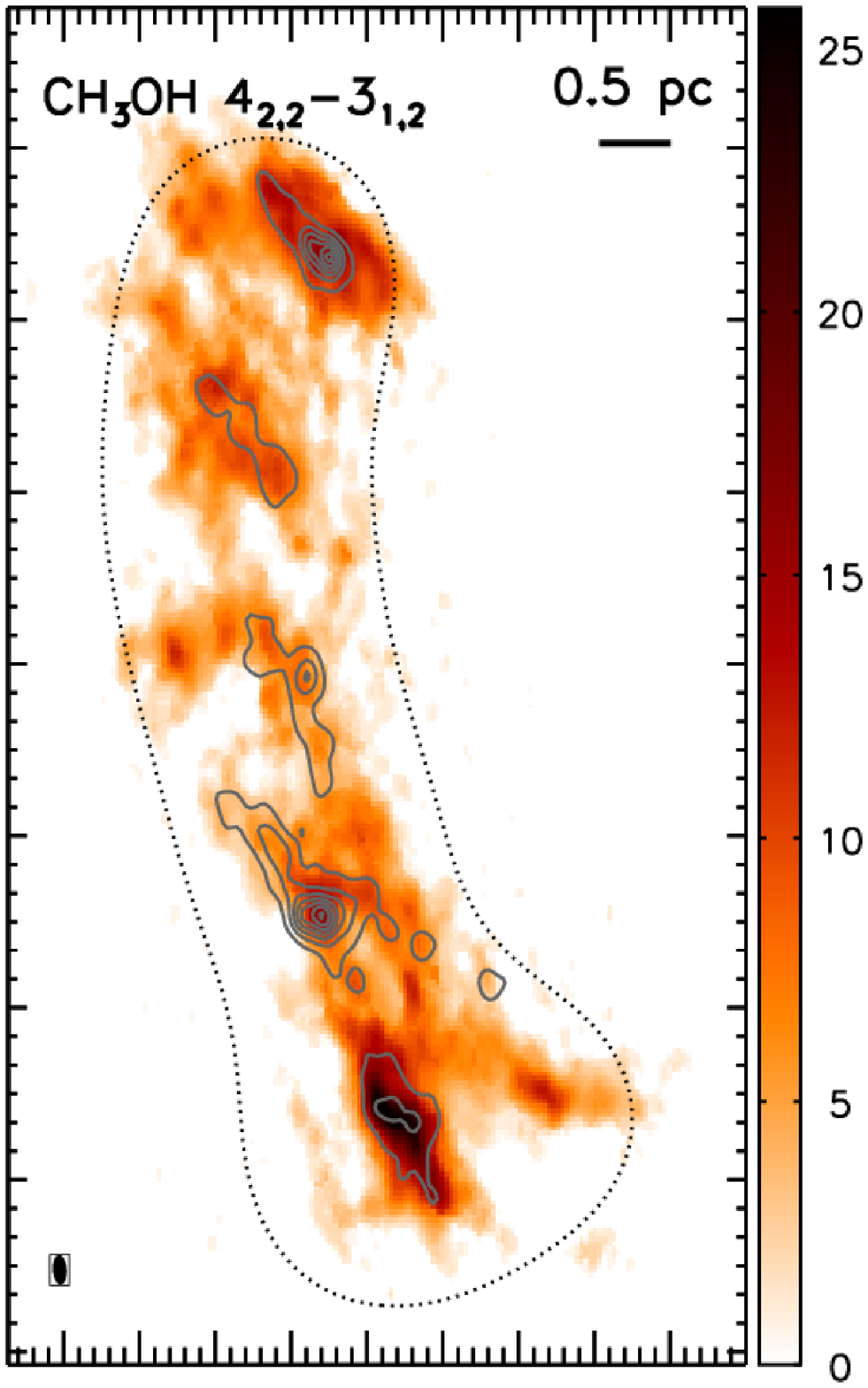} & \includegraphics[height=0.31\textwidth]{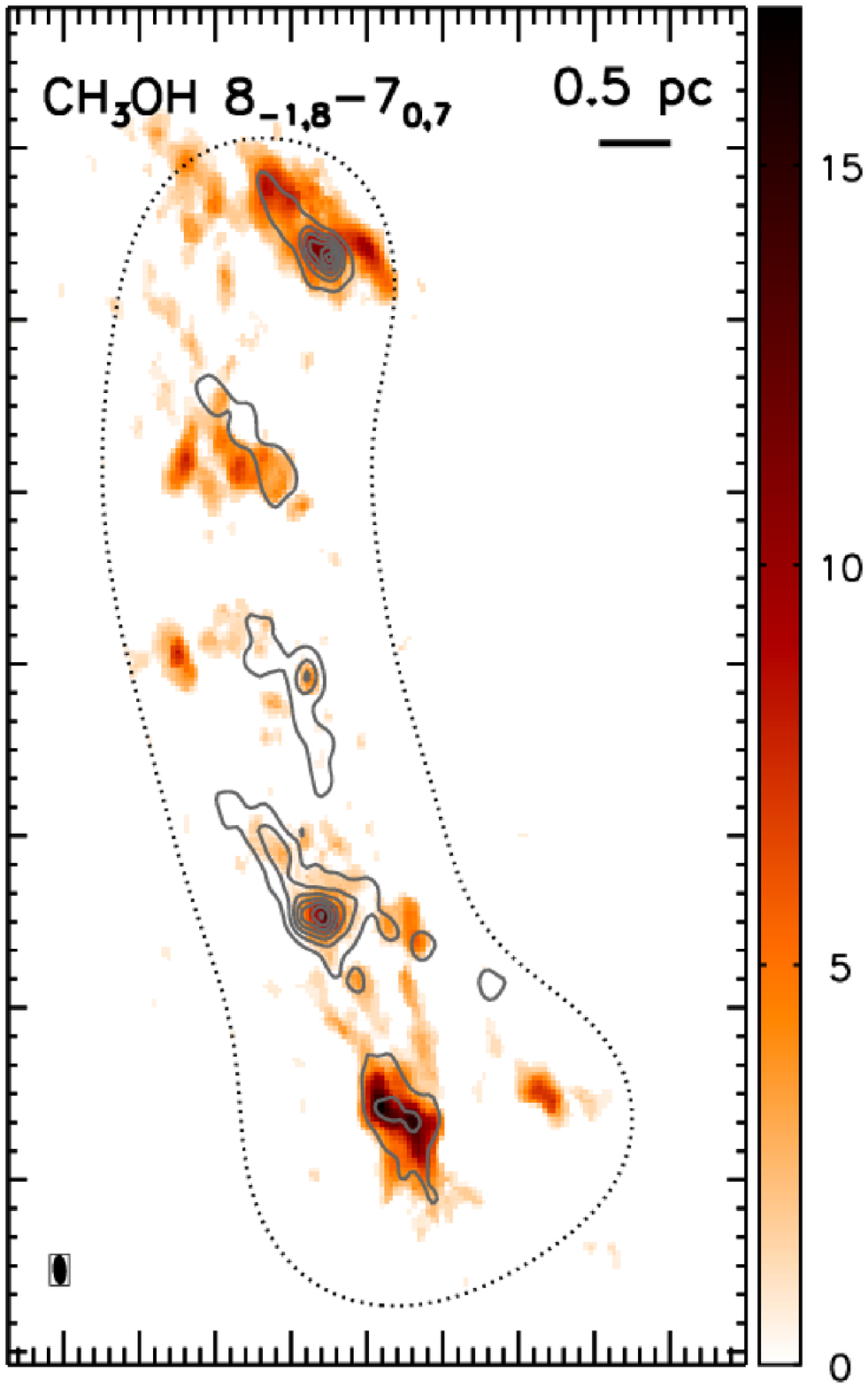} & \includegraphics[height=0.31\textwidth]{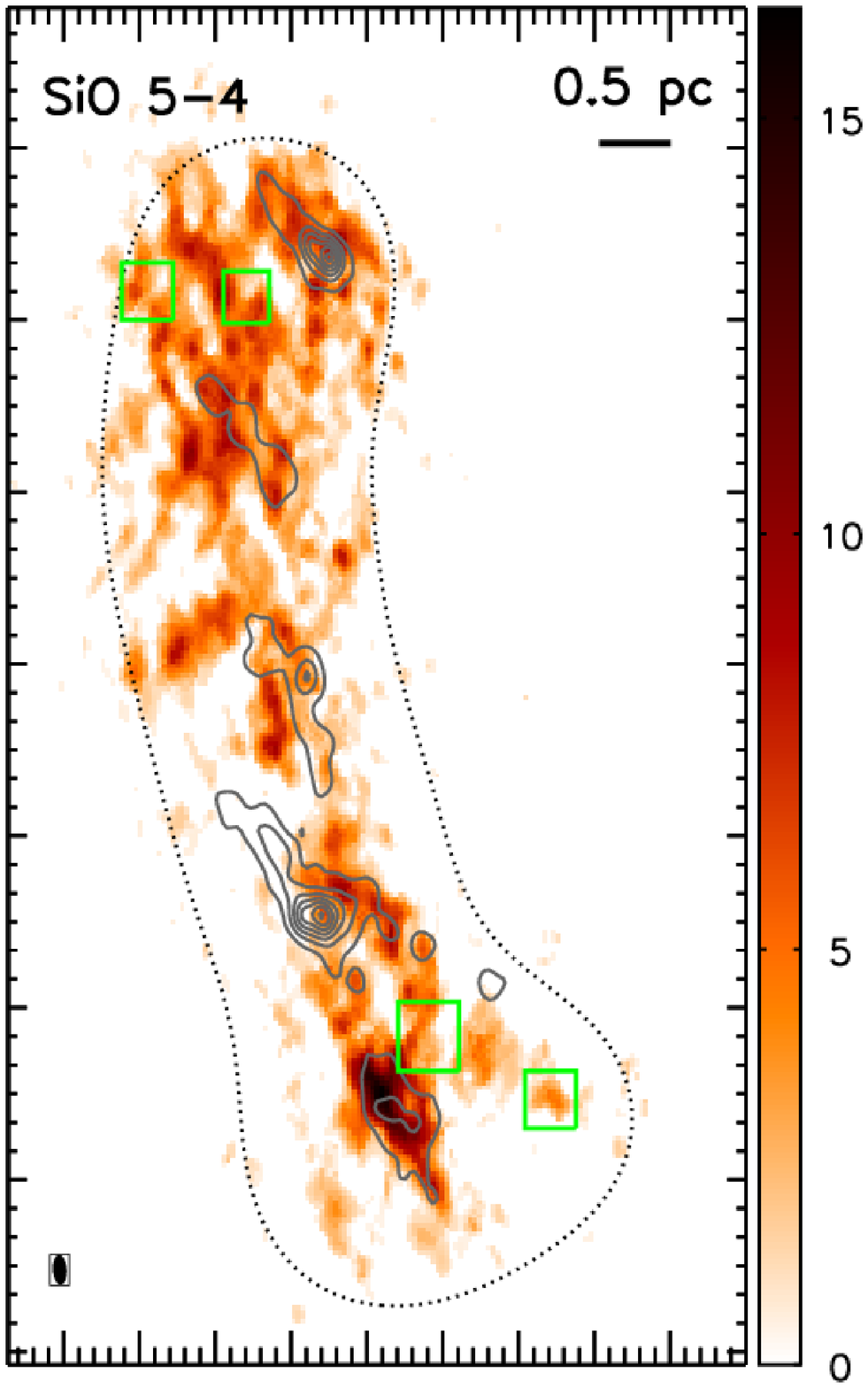}  \\
\end{tabular}
\caption{The first panel shows a three-color image at 3.6~\micron{}, 5.8~\micron{}, and 8.0~\micron{} from the \textit{Spitzer} Space Telescope/IRAC \citep{stolovy2006}. All the other panels show the integrated intensity maps (not corrected for primary beam responses in order to have a uniform rms) of the lines detected in the SMA+CSO/APEX observations (except $^{13}$CS which includes only the SMA observation), in unit of \jypbm{}\,\kms{}. In all panels, contours show the SMA+BGPS combined 1.3~mm continuum emission with identical levels as in the right panel of \autoref{fig:dust}. Dashed loops show the FWHM of the SMA primary beams. The synthesized beam of the SMA is shown in the lower left corner of each panel. In the SiO map, four candidates of shock heated regions are marked by boxes, which are also marked in Figures~\ref{fig:h2co_tkin} \& \ref{fig:nh3_tkin} and discussed in \autoref{subsubsec:disc_heating}.}
\label{fig:sma_maps}
\end{figure*}

\subsection{SMA+CSO/APEX 1.3~mm Spectral Lines}\label{subsec:results_smalines}
The SMA, CSO, and APEX observations detected a number of molecular lines in the \ctt{}. In \autoref{fig:sma_spec}, spectra of the full SMA 8~GHz band are presented for the most massive dense cores in each of the five clumps. The detected spectral lines in general can be classified into several groups based on their excitation conditions (e.g., critical densities\footnote{The critical density, assuming optically thin and without a background, is the density for which the net radiative decay rate from the upper level to the lower level equals the rate of collisional depopulation out of the upper level for a multilevel system \citep{shirley2015}. A temperature of 100~K is assumed when calculating critical densities in this section.}): \twelveco{} and \thirteenco{} 2--1 lines have low critical densities ($10^2$--$10^3$~\cc{}) and become optically thick quickly, therefore are usually diffuse gas tracers; \fmh{}, \cyacet{}, HNCO, \methanol{}, \mthc{}, and $^{13}$CS transitions have high critical densities ($10^5$--$10^6$~\cc{}) and are usually dense gas tracers; SiO, HNCO, SO, and \methanol{} molecules are usually released to gas phase from dust by shocks hence their lines are usually shock tracers, though they may not uniquely trace shocks in the CMZ.

The integrated intensities over velocities, represented by the zeroth moment maps, of all the lines detected in the SMA observations are shown in Figures~\ref{fig:sma_maps}. Except otherwise noted, we integrated over $V_\text{lsr}$ between $-$20 and 40~\kms{}, and discarded pixels below 5$\sigma$ in the datacubes when making the moment maps. For the \thirteenco{} line that is broad in velocity, we integrated between $-$40 and 60~\kms{}. For the \fmh{} 3$_{2,2}$--2$_{2,1}$ and \methanol{} 4$_{2,2}$--3$_{1,2}$ lines that are blended with each other along several lines of sight, we integrated them between $-$20 and 28~\kms{}, and between $-$15 and 40~\kms{}, respectively, to avoid blending. In the following we discuss these lines by groups.

\subsubsection{Diffuse Gas Tracers}\label{subsubsec:results_smadiffuse}
\twelveco{} and \thirteenco{} 2--1 in \ctt{} spread diffusely, in both emission and absorption. They are both extended beyond the mapped area according to single-dish observations \citep[e.g.,][]{oka1998,oka2012}, hence image fidelity may suffer from strong side-lobes from emission outside of the region, even after combining with single-dish data. We attempted to search for signatures of bipolar outflows traced by \twelveco{} or \thirteenco{} emission, but was unable to confirm any due to the complex kinematics. Therefore, we only show an integrated intensity map of \thirteenco{} 2--1 in \autoref{fig:sma_maps} and proceed to discuss the other lines.

\ceighteeno{} 2--1 does not show absorption features and its emission is less extended, as shown in \autoref{fig:sma_maps}. We will discuss its correlation with dust emission in \autoref{subsec:disc_2dxcorr}.

\subsubsection{Dense Gas Tracers}\label{subsubsec:results_smadense}
For the $^{13}$CS line, the CSO data are contaminated by the atmospheric line, while the APEX data do not cover this frequency, so we use the SMA data only. For the other dense gas tracers, including \fmh{}, \cyacet{}, and \mthc{}, we use the combined SMA and CSO/APEX data. As shown in the integrated intensity maps in \autoref{fig:sma_maps}, these lines all present compact emission at 0.1-pc scales, but the emission is not always spatially associated with the dense cores traced by compact dust emission. We will discuss spatial correlations between line emission and dust emission in detail in \autoref{subsec:disc_2dxcorr}.

\subsubsection{Shock Tracers}\label{subsubsec:results_smashock}
The maps of SiO, HNCO, SO, and \methanol{} in \autoref{fig:sma_maps} show the combined SMA and CSO/APEX data. All species present filamentary structures that have been seen in e.g.~G0.253+0.016 as signatures of pc-scale shocks \citep{johnston2014}. We will discuss the enhancement of these tracers in \autoref{subsec:disc_shocks}.

\begin{figure}[!t]
\centering
\includegraphics[width=0.45\textwidth]{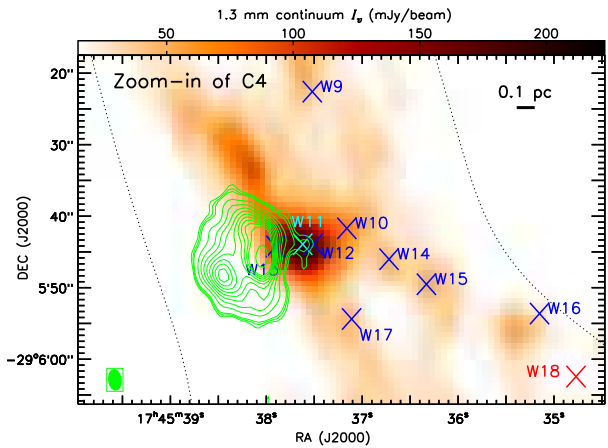}
\caption{Green contours show the 1.3~cm continuum emission in the C4 clump. Levels are in steps of 2$\sigma$ between 6$\sigma$ and 10$\sigma$, steps of 10$\sigma$ between 10$\sigma$ and 100$\sigma$, and finally 150$\sigma$, in which 1$\sigma$=100~$\mu$Jy\,beam$^{-1}$. The synthesized beam of the VLA observation is shown in the lower left corner. The SMA+BGPS 1.3~mm continuum emission is shown in the background, while \water{} masers detected by the VLA are marked by crosses. The maser W18 is consistent with a known AGB star and is marked by a red cross. The \water{} maser W11, which is spatially coincident with the compact 1.3~cm emission as well the C4-P1 dense core, is highlighted with cyan color. Properties of the other \water{} masers are reported in \citet{lu2015b}.}
\label{fig:ff}
\end{figure}

\subsection{Embedded Ionizing Source in A Dense Core}\label{subsec:results_hii}

As reported in \citet{lu2015b}, the VLA 1.3~cm continuum observations confirm the existence of an \hii{} region in the C4 clump \citep[e.g., SgrA-G in][]{ho1985}, with a peak intensity of 17.5 mJy/beam ($\sim$175$\sigma$ levels). In addition, as shown in \autoref{fig:ff}, there is weak, compact 1.3~cm continuum emission on the western side of the \hii{} region at $\sim$8$\sigma$ levels. The image is not dynamic-range limited \citep{perley1999}, and the weak emission is unlikely due to side-lobes after a comparison with the dirty beam. This compact emission is spatially coincident with the C4-P1 dense core as well as the \water{} maser W11 (shown in \autoref{fig:ff}). It thus likely represents free-free emission from an embedded ultra-compact \hii{} (UC\hii{}) or hyper-compact \hii{} (HC\hii{}) region, which can be confirmed by follow-up radio recombination line observations. No other 1.3~cm continuum emission that can be associated with any dense cores is found in the cloud \citep[see Figure~1 of][for a complete 1.3~cm continuum map]{lu2015b}.

We fitted a 2D Gaussian to it and obtained a flux density of 1.5~mJy. The ionising photon rate, assuming optically thin continuum emission and an electron temperature of 10$^4$~K, is $1.1\times10^{46}$~s$^{-1}$ \citep{mezger1974}, which can be produced by an early B-type star \citep[no earlier than B0.5;][]{panagia1973,vacca1996}. This is consistent with our estimate using \water{} maser luminosities in \citet{lu2015b}, in which W11 was suggested to trace an early B-type protostar. In \autoref{subsec:disc_tkin} we will discuss signatures of internal heating associated with this protostar.

\begin{figure*}[!t]
\begin{tabular}{p{3.6cm}p{3.05cm}p{3.05cm}p{3.05cm}p{3.05cm}}
\includegraphics[height=0.27\textwidth]{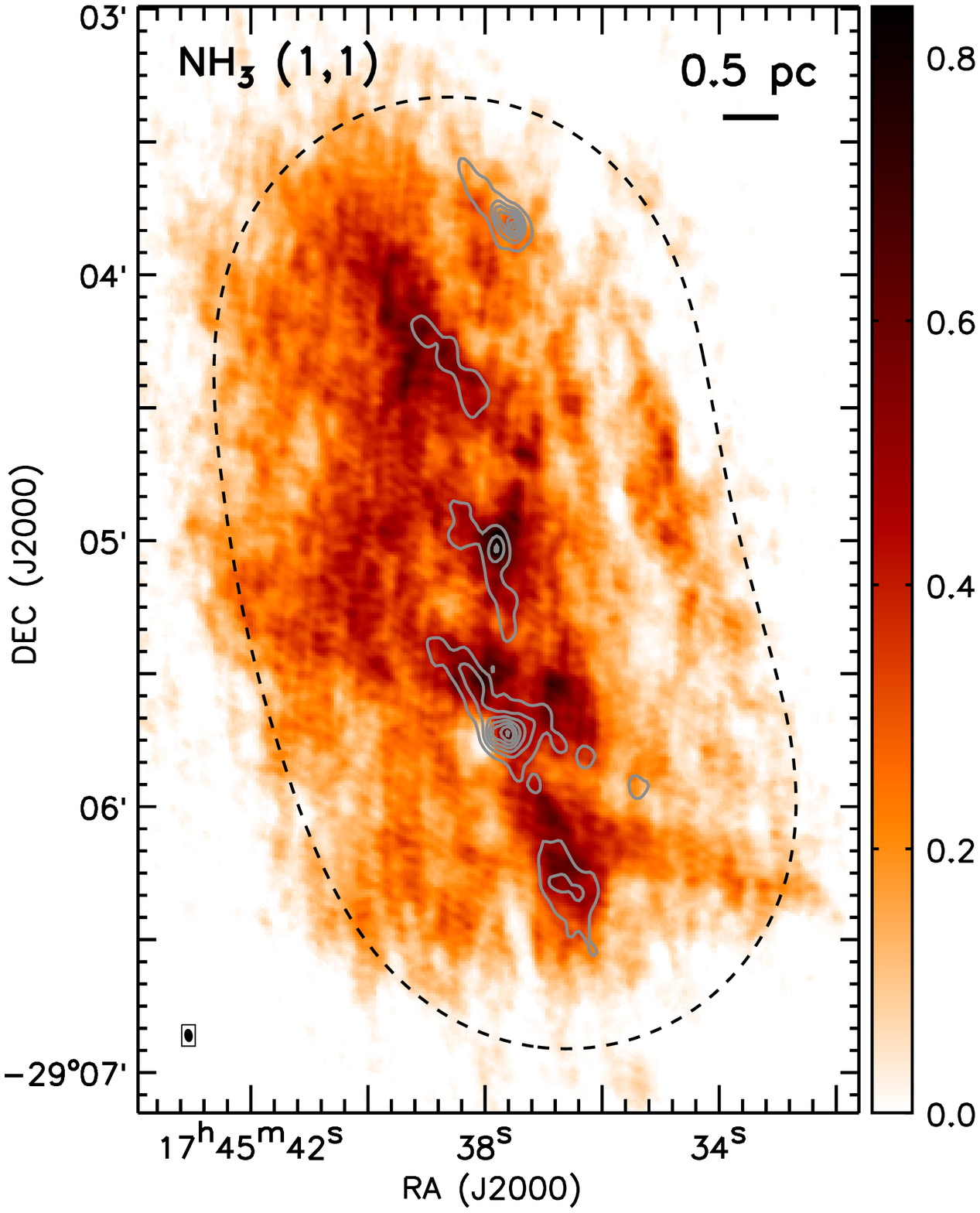} & \includegraphics[height=0.27\textwidth]{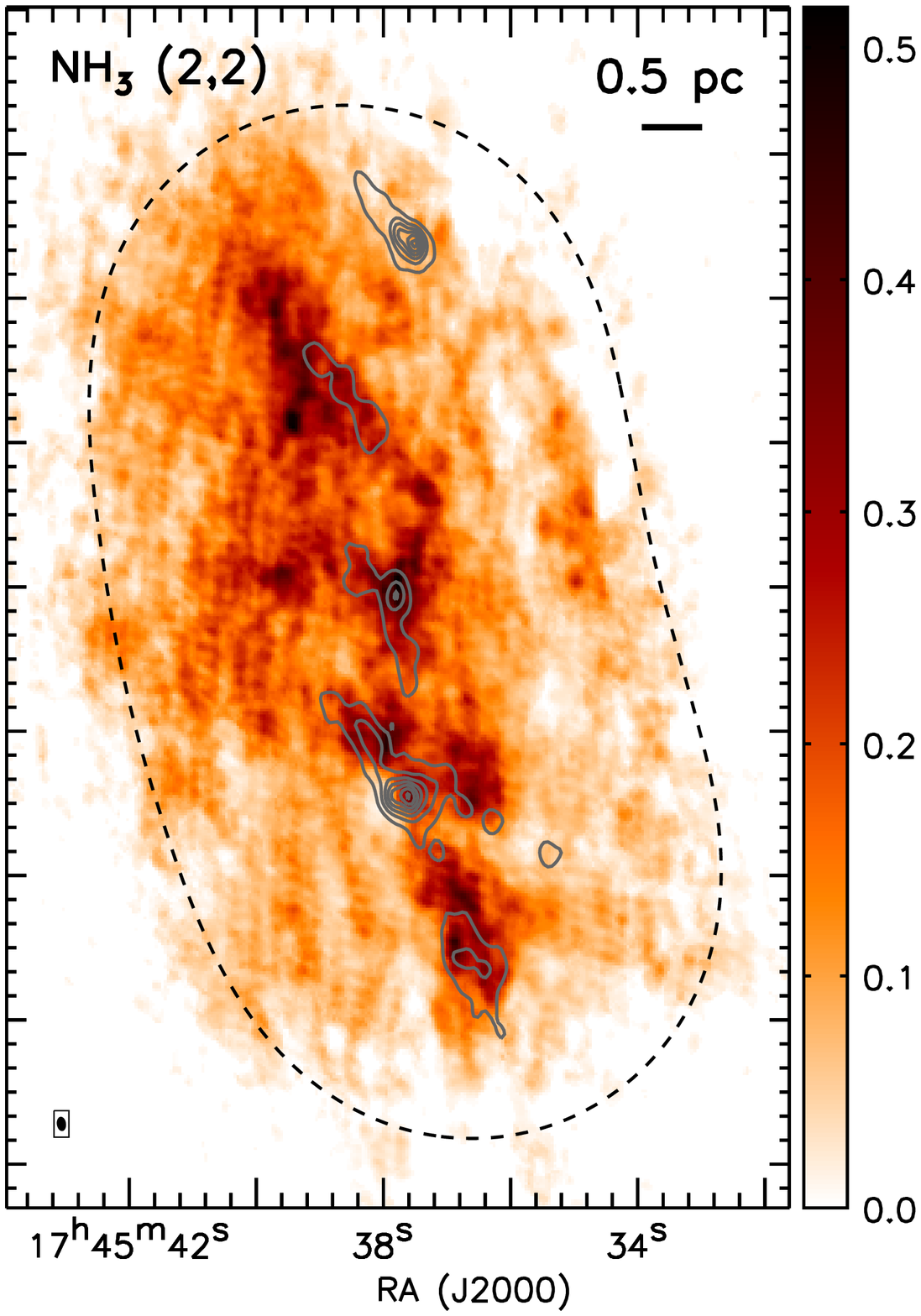} & \includegraphics[height=0.27\textwidth]{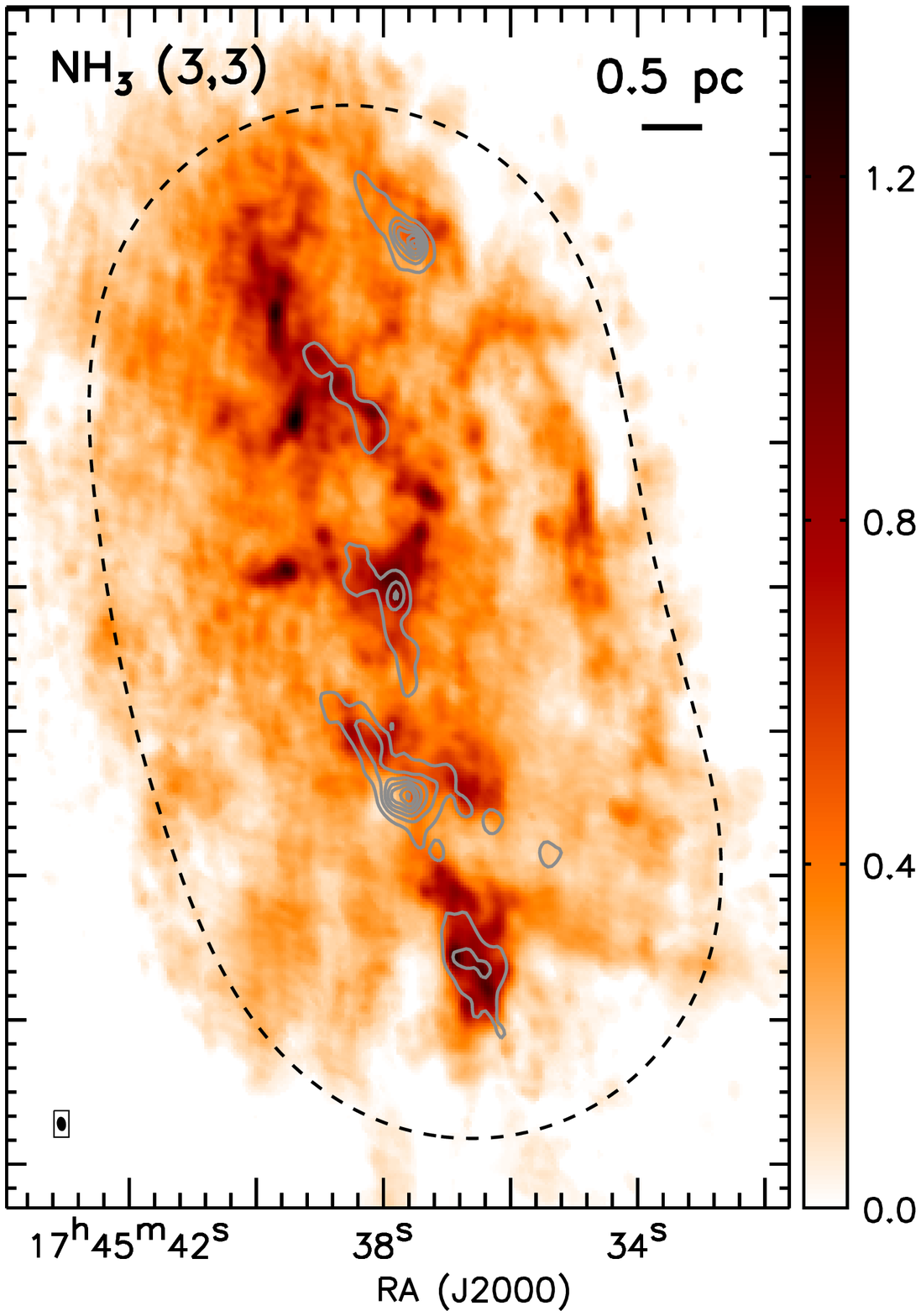} &\includegraphics[height=0.27\textwidth]{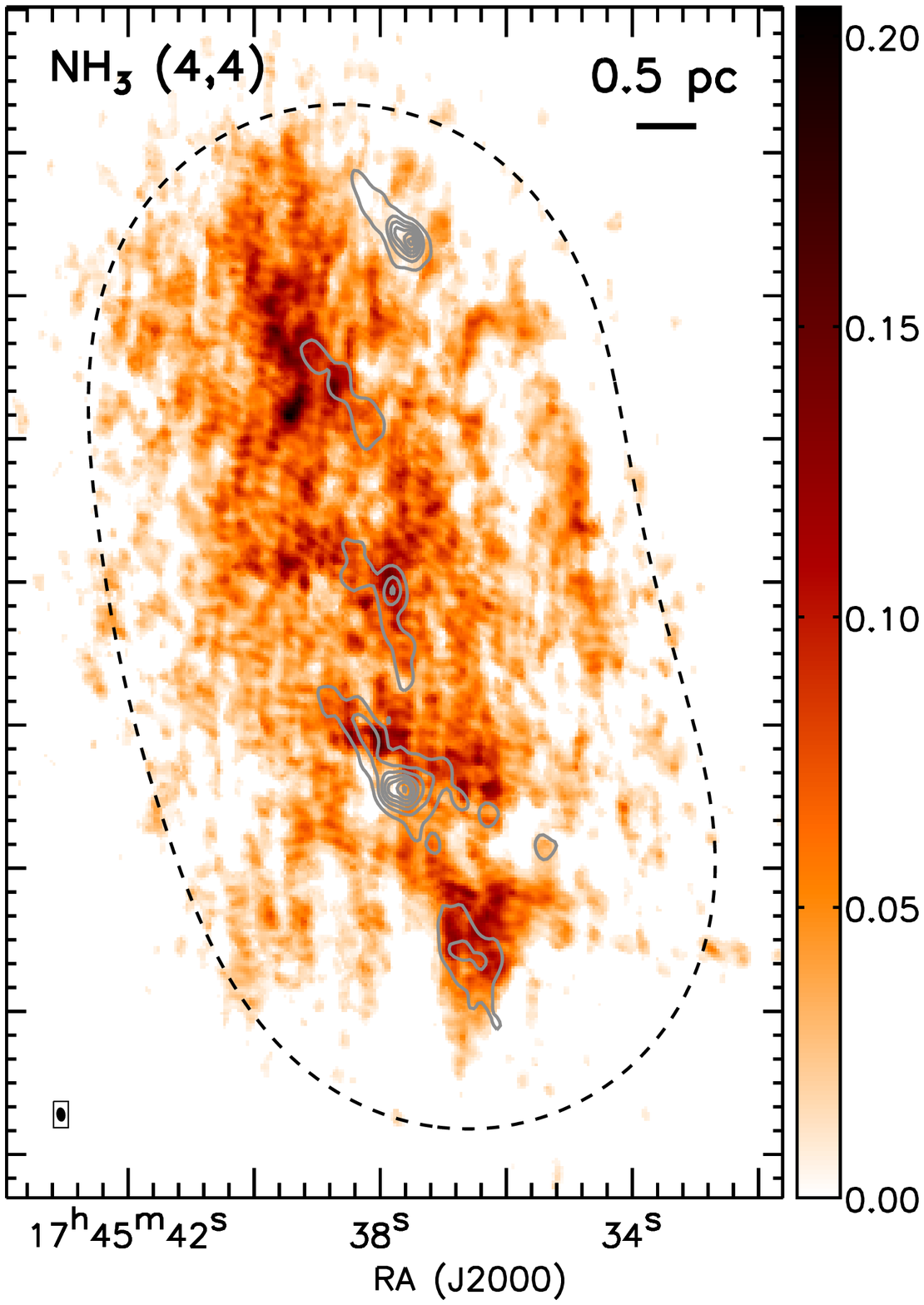} & \includegraphics[height=0.27\textwidth]{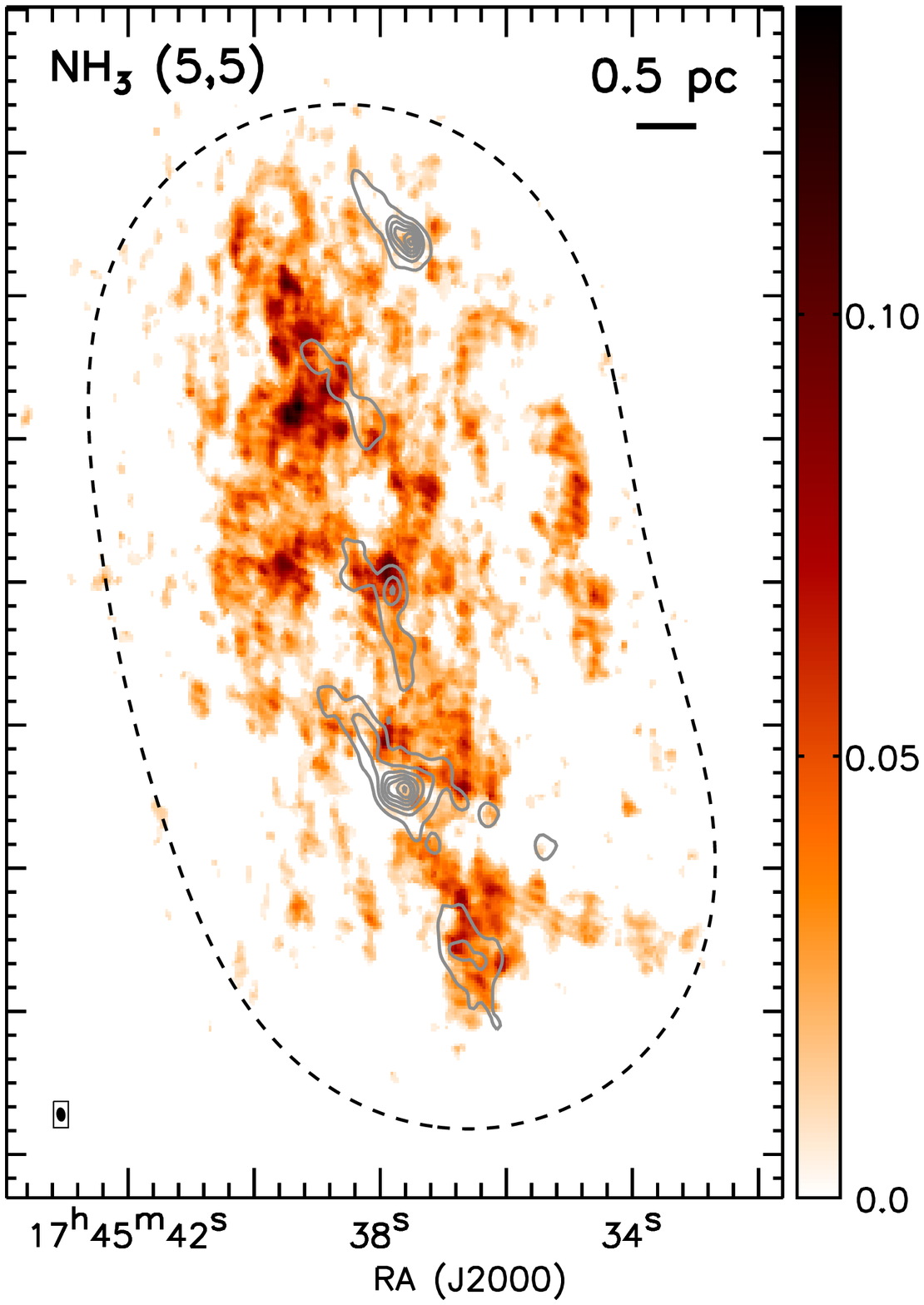}
\end{tabular}
\caption{Integrated intensity maps (not corrected for primary beam responses) of the five \amm{} transitions from VLA+GBT combined data. The unit of the halftone scale is \jypbm{}\,\kms{}. Contours show the SMA+BGPS 1.3~mm continuum emission with identical levels as in the right panel of \autoref{fig:dust}. Dashed loops show the FWHM of the VLA primary beams. The synthesized beam of the VLA is shown in the lower left corner of each panel.}
\label{fig:nh3_maps}
\end{figure*}

\subsection{VLA+GBT \amm{} Lines}\label{subsec:results_nh3}
The integrated intensities of the five inversion transitions of \amm{} represented by the zeroth moment are shown in \autoref{fig:nh3_maps}. For \amm{} (1,~1) and (2,~2), the integrated intensities include all hyperfine components between $V_\text{lsr}$ of $-$30 and 50~\kms{}, because separating channels with only the main hyperfine lines is difficult due to broad linewidths and large velocity gradients across the cloud. For \amm{} (3,~3) to (5,~5), the integrated intensities only include the main hyperfine component between $V_\text{lsr}$ of $-$20 and 40~\kms{}. The signal-to-noise ratio of the \amm{} (3,~3) line is much larger than the other four lines. To suppress noise, pixels below 10$\sigma$ in the (3,~3) datacube were discarded when making the moments maps while for the other four lines pixels below 5$\sigma$ were discarded.

The \amm{} emission, in particular that of \amm{} (3,~3), tends to be diffuse rather than concentrated on SMA dust peaks. Note that \amm{} (3,~3) belongs to the ortho species while the other four \amm{} lines belong to the para species \citep{ho1983}, between which transitions are very slow and can be ignored. Therefore, \amm{} (3,~3) can be treated as a different species from the other four lines, though the ortho/para ratio is usually in the range of 1--3 \citep[e.g.,][]{mills2013}.

The mosaic area of the VLA observations is larger than that of the SMA. In the extra area of the VLA maps that is not covered by the SMA observations, we found a likely filamentary feature on the western side of the cloud in all five transitions. This feature shows an apparently different centroid velocity (at $-5$ to 0~\kms{}) with the central part of the cloud. Its FWHM linewidths, measured with \amm{} (3,~3) lines, are between 6 and 8~\kms{}, similar to those in the \ctt{} but larger than in the Galactic disk clouds \citep[$\sim$1~\kms{},][]{lu2014}. In a more extended mapping obtained in the SMA Legacy Survey of the CMZ (CMZoom), multiple molecular line emission has been detected at the same location and velocity \citetext{C.~Battersby, E.~Keto, et al., in prep.}. The feature may suggest an unassociated gas component in the foreground or background of the \ctt{}, and is likely in the CMZ as well given its large linewidth.

\begin{figure*}[!t]
\centering
\begin{tabular}{p{6.5cm}p{6.5cm}}
\hspace{10em}(a) & \hspace{8em}(b) \\
\includegraphics[height=0.5\textwidth]{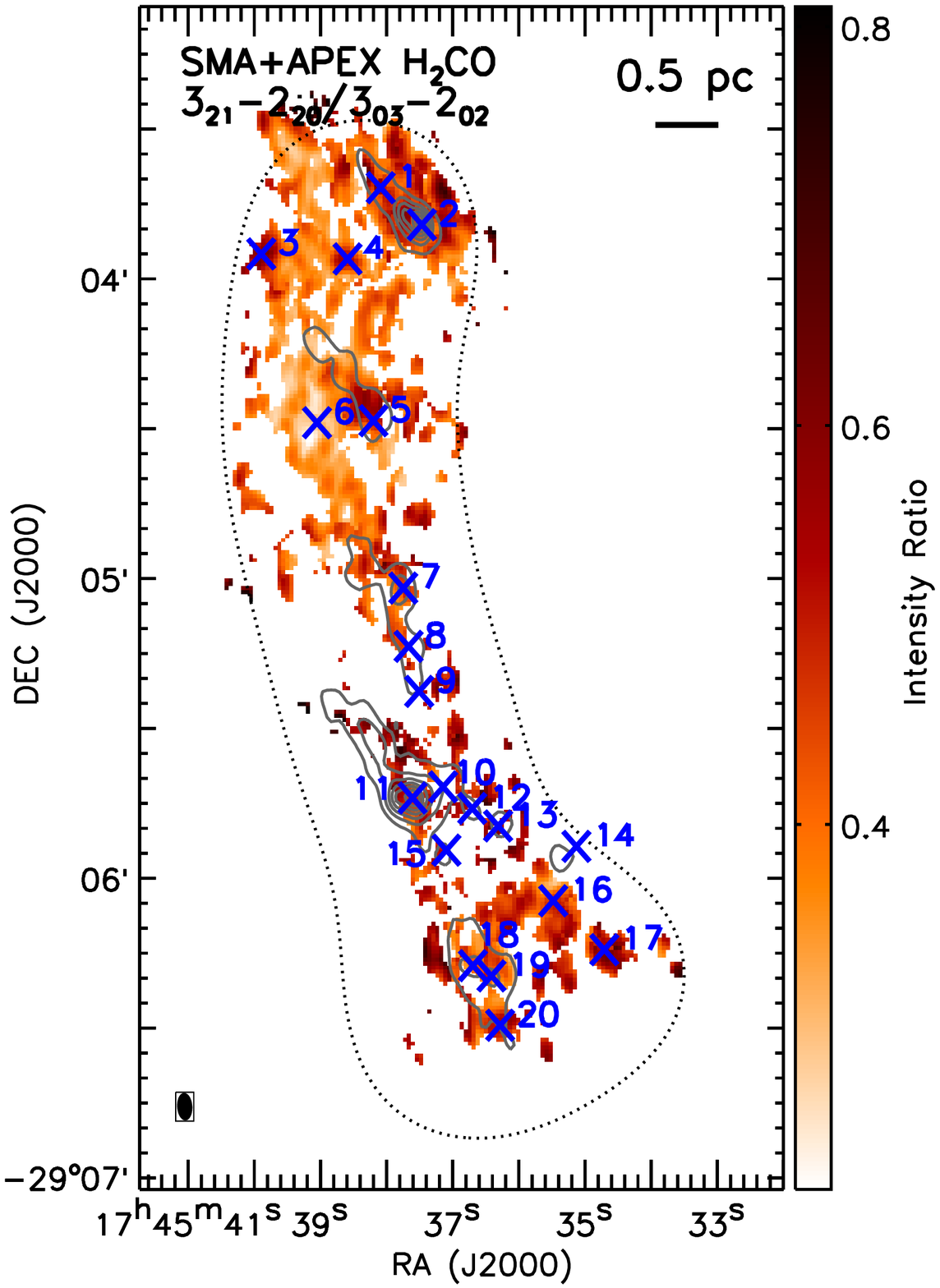} &
\includegraphics[height=0.5\textwidth]{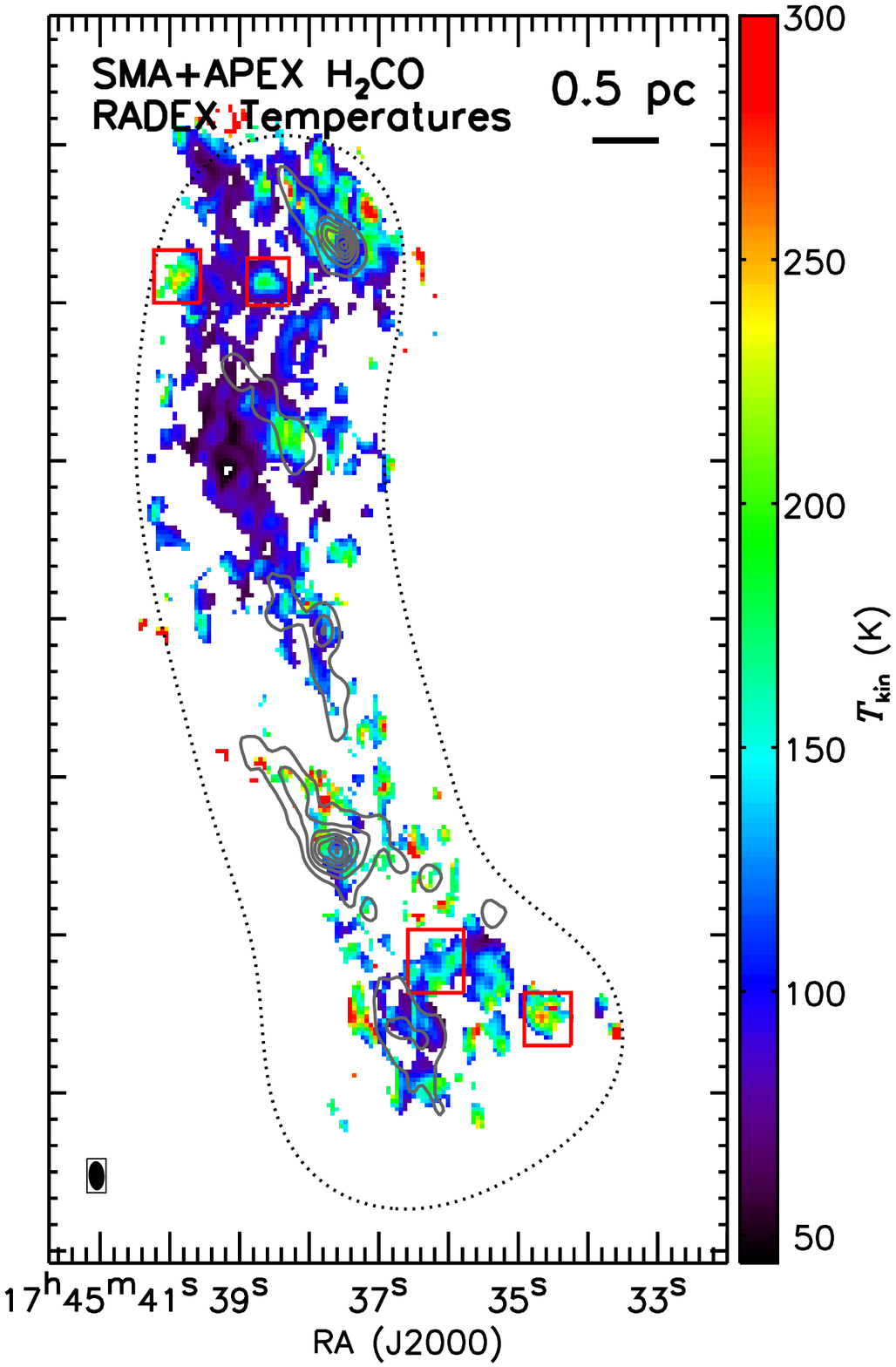}
\end{tabular}
\caption{Contours in both panels represent SMA+BGPS 1.3~mm continuum emission with identical levels as in the right panel of \autoref{fig:dust}. Dashed loops show the FWMH of the SMA primary beams. The enlarged beam after smoothing, 5\farcs{5}$\times$3\farcs{4}, is shown in the lower left corner of each panel. (a) Ratios of peak intensities of SMA+APEX \fmh{} lines. The cyan crosses mark the positions where we extracted the temperatures and linewidths in \autoref{tab:heating} and \autoref{fig:heating}. (b) Kinetic temperatures derived with RADEX modelling in \autoref{subsec:results_tkin}. Four candidates of shock heated regions are marked by boxes (cf.\ the SiO map in \autoref{fig:sma_maps}).}
\label{fig:h2co_tkin}
\end{figure*}

\begin{figure*}[!t]
\centering
\begin{tabular}{p{7.5cm}p{7.5cm}}
\hspace{11em}(a) & \hspace{9em}(b) \\
\includegraphics[height=0.5\textwidth]{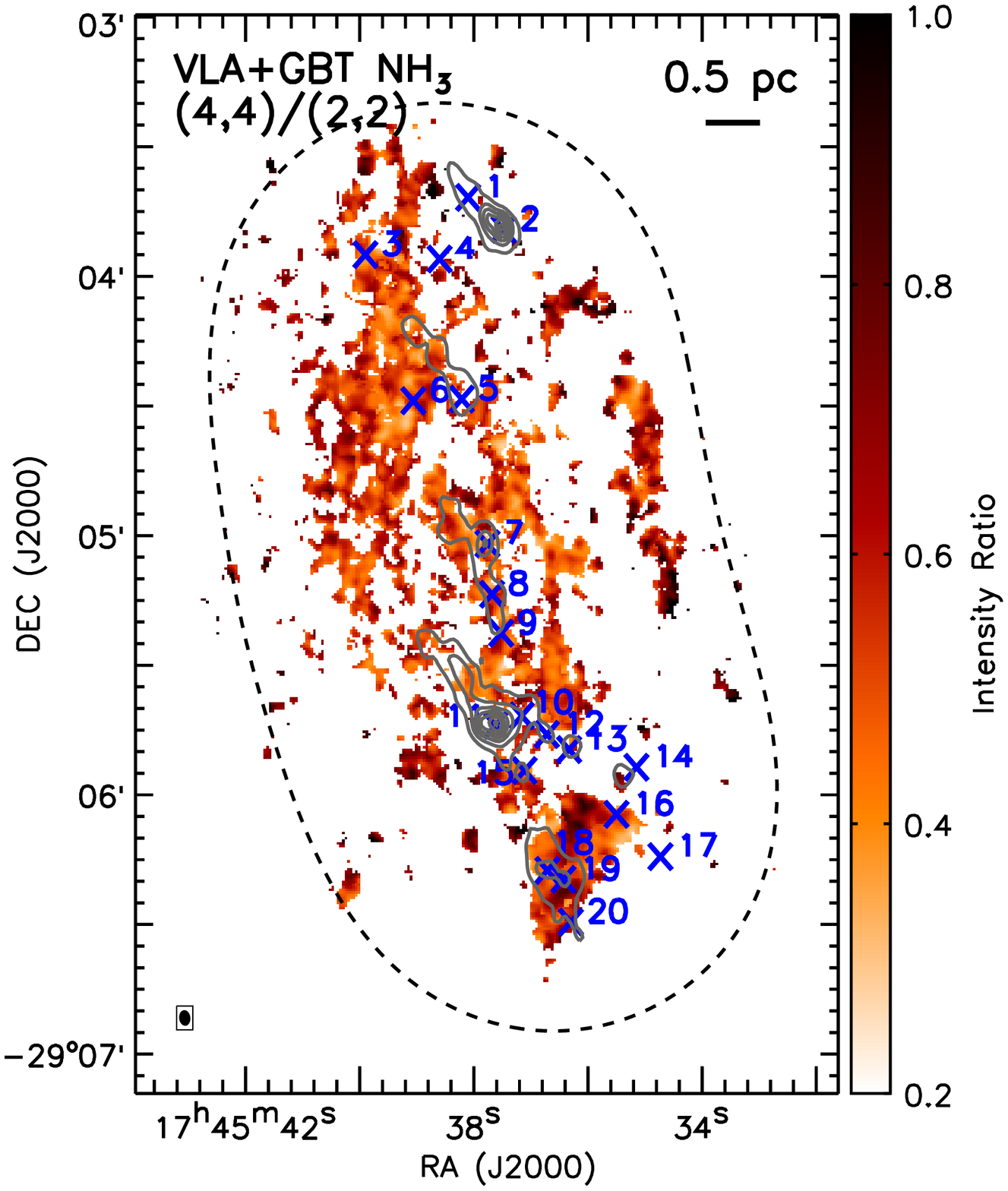} &
\includegraphics[height=0.5\textwidth]{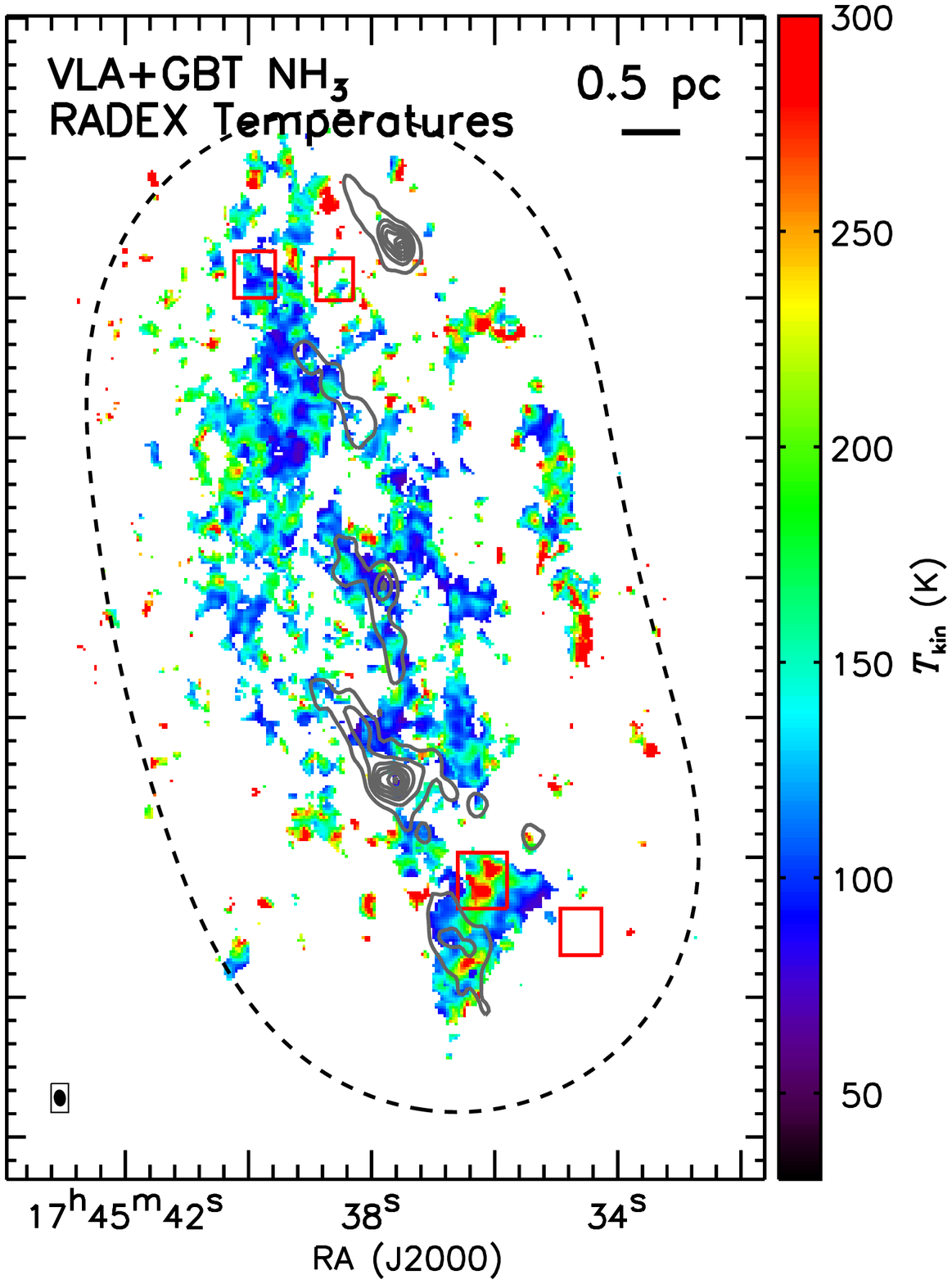}
\end{tabular}
\caption{Contours in both panels represent SMA+BGPS 1.3~mm continuum emission with identical levels as in the right panel of \autoref{fig:dust}. Dashed loops show the FWMH of the VLA primary beams. The enlarged beam after smoothing, 3\farcs{6}$\times$2\farcs{9}, is shown in the lower left corner of each panel. (a) Ratios of peak intensities of VLA+GBT \amm{} lines. The cyan crosses are the same as in \autoref{fig:h2co_tkin}. (b) Kinetic temperatures derived with RADEX modelling in \autoref{subsec:results_tkin}. Four candidates of shock heated regions are also marked by boxes as in \autoref{fig:h2co_tkin}.}
\label{fig:nh3_tkin}
\end{figure*}

\subsection{Kinetic Temperatures of Gas}\label{subsec:results_tkin}
We used the \fmh{} transitions covered in the SMA+APEX observations and the \amm{} transitions covered in the VLA+GBT observations to estimate gas temperatures \citep{ho1983,walmsley1983,mangum1993}. Both transitions can be classified as para or ortho species due to orientations of spins of the multiple H nuclei and usually we only need to consider transitions between the same species.

First, we considered the \fmh{} transitions. The three \fmh{} transitions falling in the SMA+APEX band, 3$_{0,3}$--2$_{0,2}$, 3$_{2,2}$--2$_{2,1}$, and 3$_{2,1}$--2$_{2,0}$, are all para species. The difference between upper level energies of the first transition and the latter two is $\sim$47~K.

We have smoothed the \fmh{} datacubes with a Gaussian kernel of 2\arcsec{} FWHM to increase signal-to-noise ratios, resulting in a larger beam size of 5\farcs{5}$\times$3\farcs{4}. We based the following analysis of temperatures on the smoothed \fmh{} data. In \autoref{fig:h2co_tkin}a we show the peak intensity ratios between the 3$_{2,1}$--2$_{2,0}$ and 3$_{0,3}$--2$_{0,2}$ transitions for the whole cloud. Pixels where \fmh{} 3$_{2,1}$--2$_{2,0}$ or 3$_{0,3}$--2$_{0,2}$ signals are below 5$\sigma$ levels were masked. Since the 3$_{2,1}$--2$_{2,0}$ transition has a higher upper level energy, larger ratios between the two transitions qualitatively represent higher gas temperatures.

We employed the RADEX code \citep{vandertak2007} with a modified solver \texttt{myRadex}\footnote{\url{https://github.com/fjdu/myRadex}} to implement the non-local-thermodynamic-equilibrium (non-LTE) conditions. The collision rates between \fmh{} and ortho/para-H$_2$ were taken from LAMDA \citep{wiesenfeld2013}. The ortho/para ratios of H$_2$ were calculated based on kinetic temperatures. We assumed a uniform spherical geometry, and fixed the velocity gradient to a representative value of 5~\kms{}\,pc$^{-1}$, then generated \fmh{} intensities expected from models when kinetic temperatures range between $10^{0.5}$~K and $10^3$~K in logarithmic steps of 0.1, H$_2$ number densities between $10^2$ and $10^8$~\cc{} in logarithmic steps of 0.1, and column densities between $10^{22.3}$ and $10^{23.8}$~\sqc{} in logarithmic steps of 0.1. The abundance of \fmh{} to H$_2$ is assumed to be $10^{-9}$ \citep{ao2013}, hence the \fmh{} column densities we were using in the models are between $10^{13.3}$ and $10^{14.8}$~\sqc{}. We have assumed a single velocity component and a beam filling factor of 1 throughout the analysis.

We used a customized routine \citep{zzy2014} to determine the most likely gas temperatures as well as H$_2$ densities that can reproduce the observed \fmh{} 3$_{0,3}$--2$_{0,2}$ intensities as well as the observed \fmh{} 3$_{2,2}$--2$_{2,1}$ to 3$_{0,3}$--2$_{0,2}$ line ratios. An example is shown in \autoref{appd:radex} while details of this procedure can be found in \citet{zzy2014}.

One feature we noticed is that the \fmh{} 3$_{2,2}$--2$_{2,1}$ to 3$_{0,3}$--2$_{0,2}$ line ratios, $R$(3$_{2,2}$--2$_{2,1}$/3$_{0,3}$--2$_{0,2}$), are sensitive to kinetic temperatures, but not so to H$_2$ densities or column densities. In light of this, in \autoref{appd:radex} we fitted the line ratios versus the kinetic temperatures and derived an analytical expression:
\begin{equation}\label{equ:h2co_ratio2tkin}
T_\text{kin}=17.5\exp(4.27\,R(3_{2,2}\text{--}2_{2,1}/3_{0,3}\text{--}2_{0,2}))~\text{K}.
\end{equation}
This equation could reproduce the kinetic temperatures from RADEX modelling as accurate as 0.08~dex (20\%) between $T_\text{kin}=30\text{--}300$~K. The temperatures derived from RADEX modelling have uncertainties of 0.15~dex (40\%) themselves, as discussed in \autoref{appd:radex}, which include both systematic errors with assumed density, column density, and abundance, as well as random errors with the observed \fmh{} intensities. The overall uncertainty in the temperatures derived from \autoref{equ:h2co_ratio2tkin} would be 0.17~dex ($\sim$50\%). We applied this relation to the line ratio map in \autoref{fig:h2co_tkin}a and derived a kinetic temperature map as shown in \autoref{fig:h2co_tkin}b.

Second, we considered the VLA \amm{} transitions. Among the five transitions, the \amm{} (1,~1), (2,~2), (4,~4), and (5,~5) are all para species thus can be jointly considered. Unfortunately, the \amm{} (1,~1) lines suffer from strong absorption, while the \amm{} (5,~5) lines are usually weak, therefore we chose the (2,~2) and (4,~4) lines to estimate temperatures. The difference between their upper level energies is $\sim$137~K.

We smoothed the \amm{} datacubes with a Gaussian kernel of 2\arcsec{} FWHM to increase signal-to-noise ratios, resulting in a beam size of 3\farcs{6}$\times$2\farcs{9}. We masked pixels where \amm{} (2,~2) or (4,~4) signals are below 5$\sigma$ and showed the peak intensity ratios between main components of these two transitions for the whole cloud in \autoref{fig:nh3_tkin}a. In addition, similar to the above analysis with the \fmh{} lines, we derived kinetic temperatures under non-LTE conditions with the VLA+GBT \amm{} (2,~2) and (4,~4) transitions. The collision rates between \amm{} and para-H$_2$ were taken from LAMDA \citep{danby1988}. The ortho/para ratios of H$_2$ were calculated based on kinetic temperatures. We assumed a \amm{} to H$_2$ abundance of 3$\times$10$^{-8}$ \citep{harju1993}, and used identical model grids as for \fmh{}. An example is shown in \autoref{appd:radex}. We also derived a relation between \amm{} (4,~4) to (2,~2) line ratios, $R$(44/22) and kinetic temperatures:
\begin{equation}\label{equ:nh3_ratio2tkin}
T_\text{kin}=31.2\exp(2.75\,R(44/22))~\text{K},
\end{equation}
which could reproduce temperatures from RADEX modelling as accurate as 0.08~dex ($\lesssim$20\%) between $T_\text{kin}=30\text{--}300$~K. We assumed an overall uncertainty of 0.18~dex ($\sim$50\%) in the temperatures after adding in the uncertainties in the RADEX modelling as discussed in \autoref{appd:radex}. The kinetic temperatures derived using the observed line ratios with this equation are shown in \autoref{fig:nh3_tkin}b. More discussions on the gas temperatures can be found in \autoref{subsec:disc_tkin}.

\section{DISCUSSION}\label{sec:discussion}

\subsection{Correlation between the Dust and Spectral Line Emission}\label{subsec:disc_2dxcorr}
We study the spatial correlation between dust emission and spectral lines in order to understand the effect of star formation on the molecular gas environment. As demonstrated in \autoref{subsec:results_dust}, the compact dust emission traces gravitationally bound dense cores at 0.1-pc scales in which we found a population of \water{} masers. Therefore, it is a reasonable indicator of star formation in the \ctt{}. Correlations between the compact dust emission from dense cores and spectral lines will help to reveal underlying connections of such lines to star formation.

Here we attempt to quantitatively characterize which lines are spatially correlated with the compact dust emission based on the 2D cross-correlation analysis\footnote{Based on the IDL function `correl\_images.pro': \url{https://idlastro.gsfc.nasa.gov/ftp/pro/image/correl_images.pro}.}.

\begin{table}[!t]
\begin{deluxetable}{lcc}
\tabletypesize{\scriptsize}
\tablecaption{2D cross-correlation coefficients between the dust and the molecular lines.\label{tab:2dxcorr}}
\tablewidth{0pt}
\tablehead{
\multirow{2}{*}{Transitions} & Coefficients with  & Coefficients with  \\
 & SMA-only dust &  SMA+BGPS dust
 }
\startdata
\thirteenco{} 2--1                  & 0.36 & 0.78 \\
\ceighteeno{} 2--1                & 0.40 & 0.80 \\
\fmh{} 3$_{0,3}$--2$_{0,2}$ & 0.37 & 0.78 \\
\fmh{} 3$_{2,2}$--2$_{2,1}$ & 0.41 & 0.58 \\
\fmh{} 3$_{2,1}$--2$_{2,0}$ & 0.43 & 0.61 \\
\mthc{} 12--11, K=0,1           & 0.43 & 0.30 \\
\cyacet{} 24--23                   & 0.19 & 0.19 \\
$^{13}$CS 5--4                    & 0.35 & 0.23 \\
HNCO 10$_{0,10}$--9$_{0,9}$ & 0.48 & 0.79 \\
SO 6$_5$--5$_4$                & 0.54 & 0.74 \\
\methanol{} 4$_{2,2}$--3$_{1,2}$  & 0.56 & 0.79 \\
\methanol{} 8$_{-1,8}$--7$_{0,7}$ & 0.60 & 0.60 \\
SiO 5--4                               & 0.40 & 0.64 \\
\amm{} (1,~1)			   & 0.41 & 0.81 \\
\amm{} (2,~2)			   & 0.39 & 0.78 \\
\amm{} (3,~3)			   & 0.40 & 0.80 \\
\amm{} (4,~4)			   & 0.32 & 0.69 \\
\amm{} (5,~5)			   & 0.27 & 0.58 \\
\enddata
\end{deluxetable}
\end{table}

We take the SMA-only dust emission image and the SMA+BGPS combined dust emission image (the middle and right panels of \autoref{fig:dust}), to represent the compact emission from the dense cores and the diffuse dust emission, respectively. The SMA+BGPS dust emission image is used as a control sample. We reset pixels with negative intensities in these two images to 0 to keep only emission.

In addition to the SMA+CSO/APEX data, we take the VLA+GBT \amm{} integrated intensities maps and convolve them with the SMA primary beams. We are aware that sensitivities and dynamic ranges of the SMA and VLA observations are different, which may make a direct comparison between the 1.3~mm lines and the \amm{} lines problematic. Nevertheless, we include the \amm{} lines here for reference. 

Then we normalize all the images with respect to their maximum intensities. The images under consideration are, or have been regridded, in the same coordinates, so we only need to calculate their cross-correlation coefficients, without spatially shifting one image relative to the other in order to find the maximum cross-correlation. The 2D cross-correlation coefficients derived in such way will be between $-$1 and 1, with $-1$ being anti-correlated, 0 being not correlated at all, and 1 being correlated.

The 2D cross-correlation coefficients between the dust emission images and the spectral line integrated intensity maps are listed in \autoref{tab:2dxcorr}. First, we consider correlations with the compact dust emission represented by the SMA-only dust image. Among the 18 spectral lines, the two \methanol{} lines and the SO line have 2D cross-correlation coefficients with the compact dust emission of $>$0.50. The HNCO line also has a coefficient of 0.48. On the other hand, the three \fmh{} lines and the five \amm{} lines, which are conventional dense gas tracers for Galactic disk clouds, have smaller coefficients of 0.37--0.43 and 0.27--0.41, respectively. Second, we consider correlations with the control sample, the diffuse dust emission represented by the SMA+BGPS dust emission image. Large 2D cross-correlation coefficients of $\sim$0.8 are found for \thirteenco{}, \ceighteeno{}, HNCO, one \fmh{} line, one \methanol{} line, and three \amm{} lines. As seen in Figures~\ref{fig:sma_maps}~\&~\ref{fig:nh3_maps}, these lines do present diffuse emission. The \mthc{}, $^{13}$CS, \cyacet{}, and \methanol{}~8$_{-1,8}$--7$_{0,7}$ lines have the same or smaller 2D cross-correlation coefficients with the diffuse dust emission than with the compact dust emission. This happens when the line emission is concentrated on the compact dust emission.

Based on the 2D cross-correlation analysis, the SO line, the HNCO line, and the two \methanol{} lines are best correlated spatially with the compact dust emission from dense cores among all the detected lines. We also examine the correlations with a visual inspection of the maps in \autoref{fig:sma_maps}. Indeed, several emission peaks of these lines spatially coincident with the dense cores we identified, although emission peaks that are not associated with any dust emission are also found (e.g., in the south-western part of the cloud). The most notable line is \methanol{} 8$_{-1,8}$--7$_{0,7}$, which presents emission peaks toward all five massive clumps and is concentrated on dust cores. These lines have been observed to be enhanced by shocks \citep{bachiller1997,rodriguez2010}, but our result suggests that they also trace the dense cores in the \ctt{} well. A potential explanation is that they are enhanced by star formation embedded in the dense cores, as discussed in \autoref{subsec:disc_shocks}.

The three \fmh{} and the five \amm{} lines appear to be correlated with the compact dust emission in the dense cores, but with smaller 2D cross-correlation coefficients than the four lines above. As shown in Figures~\ref{fig:sma_maps} \& \ref{fig:nh3_maps}, the two species do not present strong emission in the brightest dust peaks in the C4 clump, and \amm{} emission is also absent in the C1 clump. At 0.1-pc scales in this cloud, they are not good tracers of dense gas. The reason could be strong shocks in the cloud that destroy molecules.

At last, although the \mthc{}, \cyacet{} and $^{13}$CS lines have small 2D cross-correlation coefficients, a visual inspection of \autoref{fig:sma_maps} suggests that their emission is detected preferably on the dense cores. They are fainter than the lines discussed above, hence in our sensitivity-limited observations they present a large fraction of area of non-detection thus tend to have small 2D cross-correlation coefficients with the dust emission images. In particular, \mthc{} is usually detected in hot molecular cores \citep[e.g.,][]{araya2005}, and has hyperfine splittings that are suitable to determine gas temperatures of $>$100~K. Therefore, sensitive spectral line observations that are able to detect faint \mthc{} emission will be useful for studying embedded high-mass star formation in these dense cores.

Overall, the 2D cross-correlation coefficients between the dust and spectral line emission we found are larger than the results of \citet{rathborne2015}, which discussed the dust emission and various spectral lines at 3~mm in G0.253+0.016. We stress that a direct comparison between different observations is inappropriate. \citet{rathborne2015} argued that the low 2D cross-correlation coefficients are due to the optically thick molecular line emission. The dust emission traced by 3~mm continuum also shows a lack of compact substructures. These factors may lead to the smaller cross-correlation coefficients between the dust and molecular line emission in G0.253+0.016.

\begin{figure*}[!t]
\centering
\includegraphics[height=0.4\textwidth]{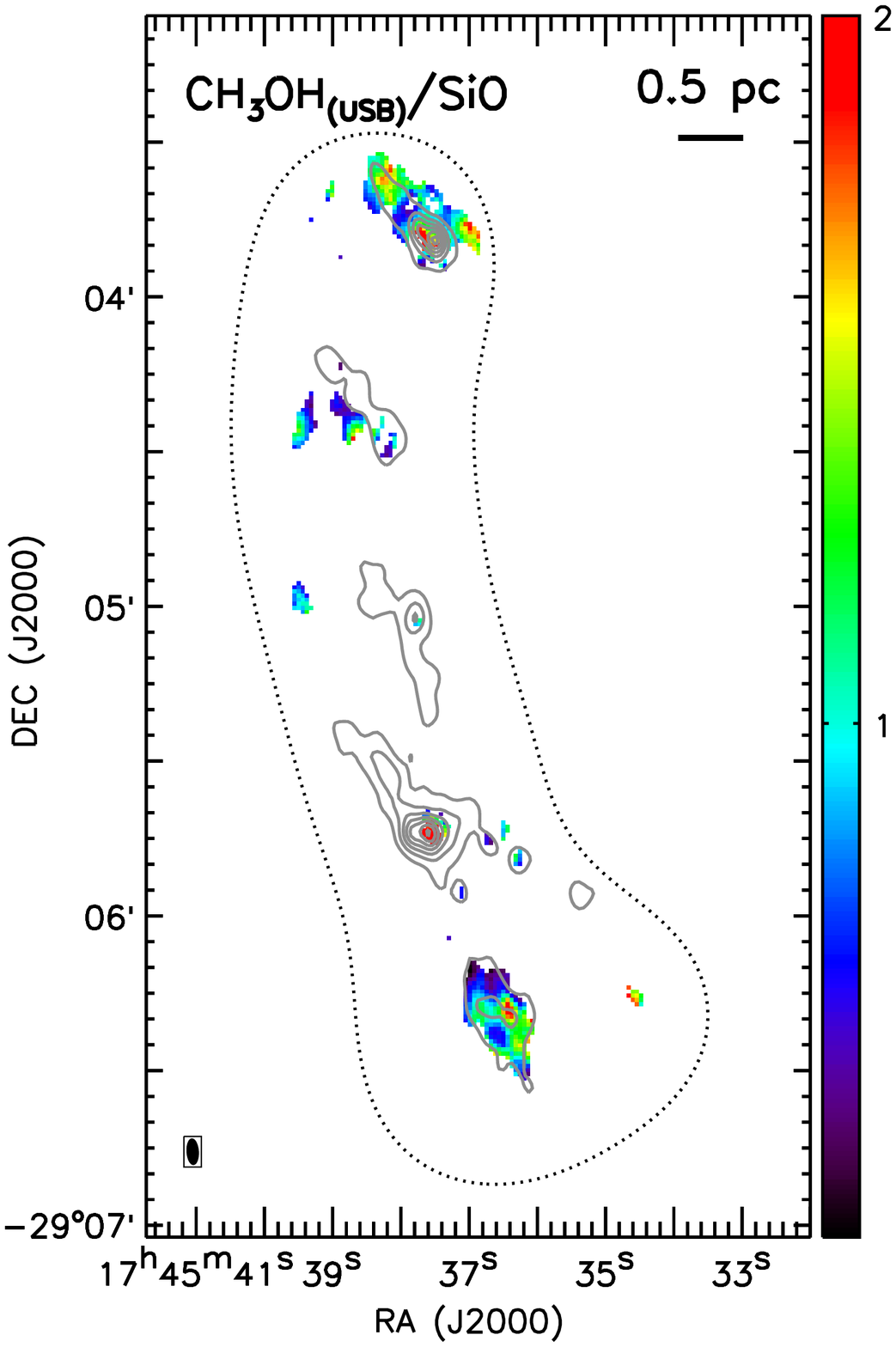}\includegraphics[height=0.4\textwidth]{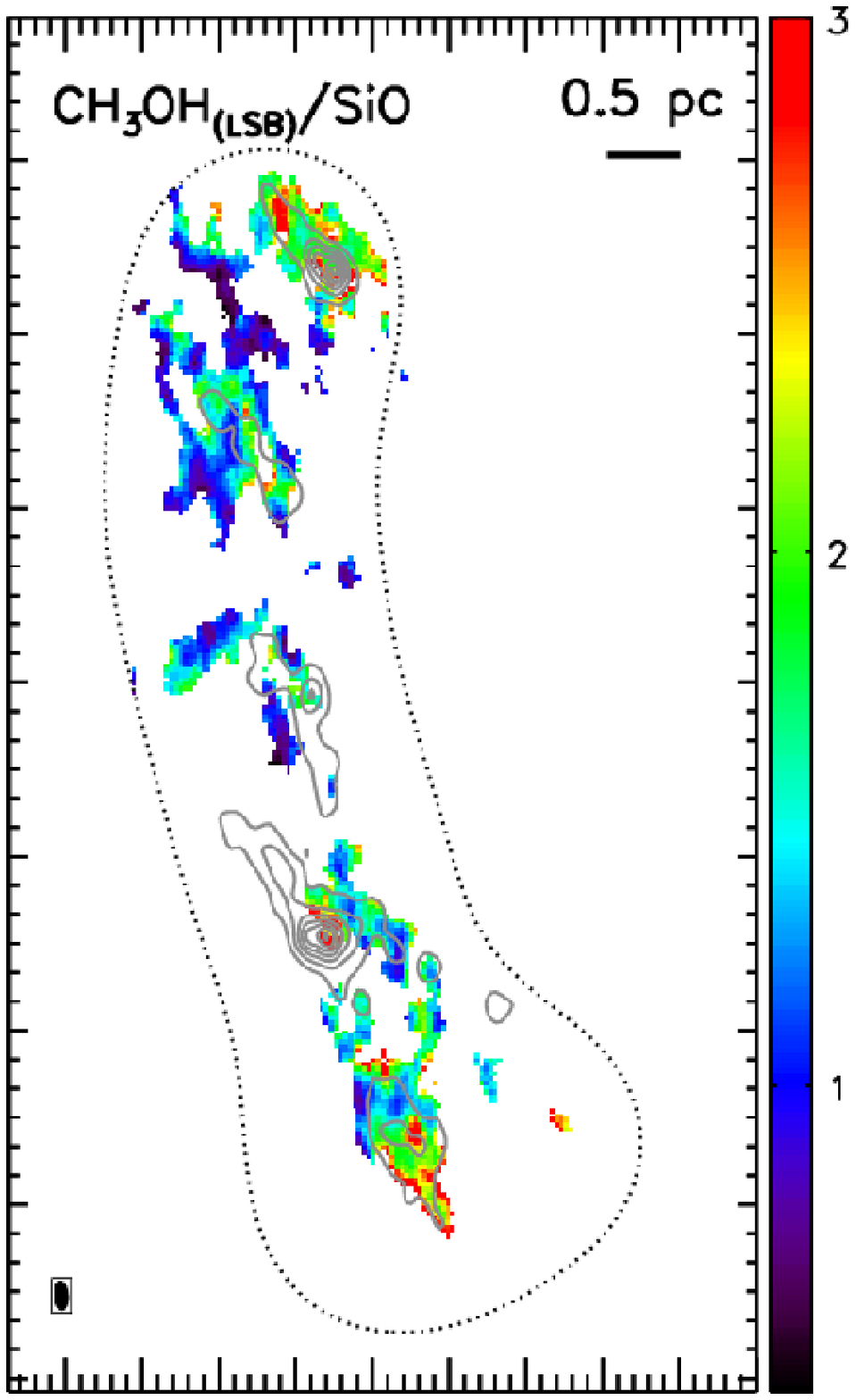}\includegraphics[height=0.4\textwidth]{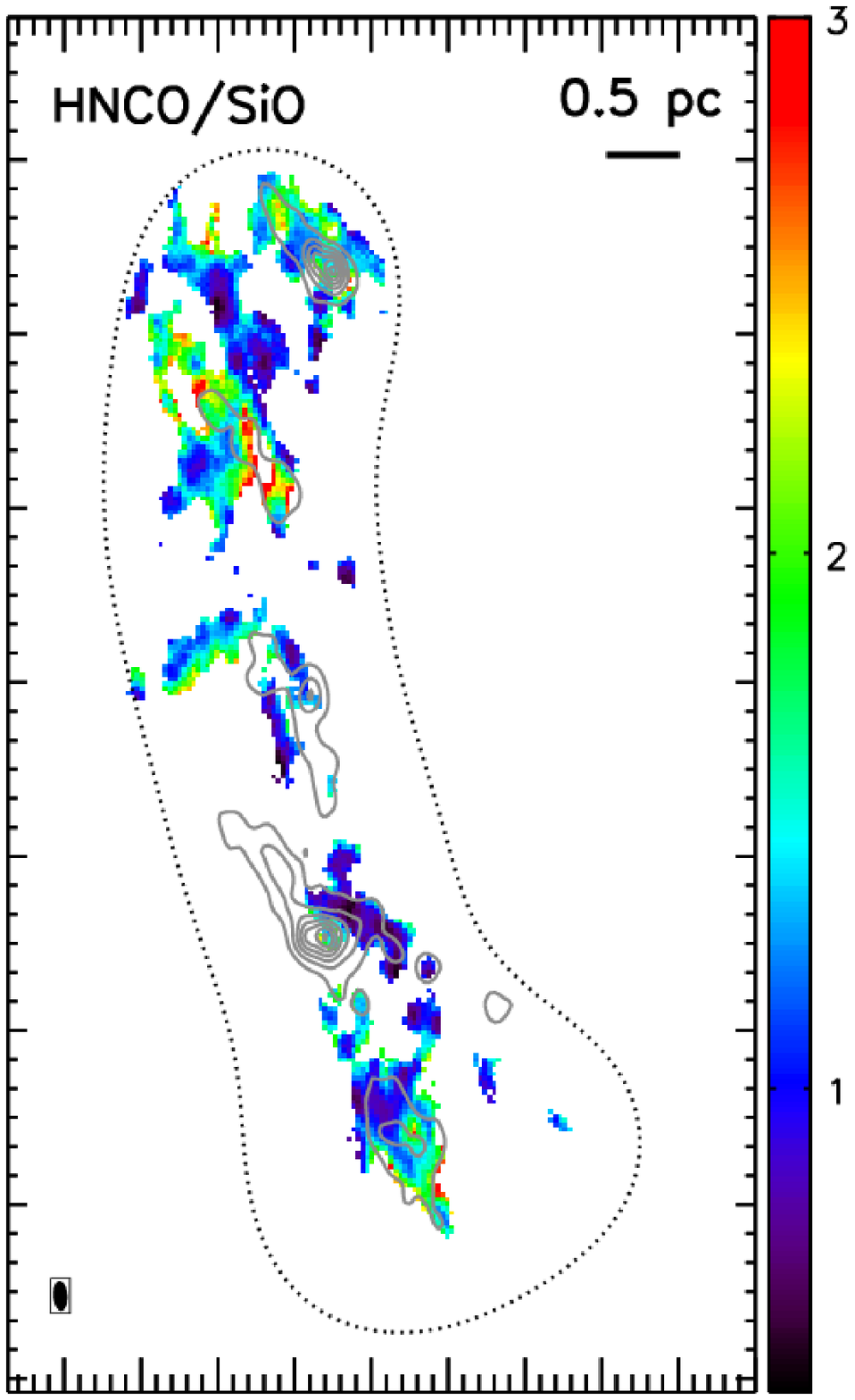}\includegraphics[height=0.4\textwidth]{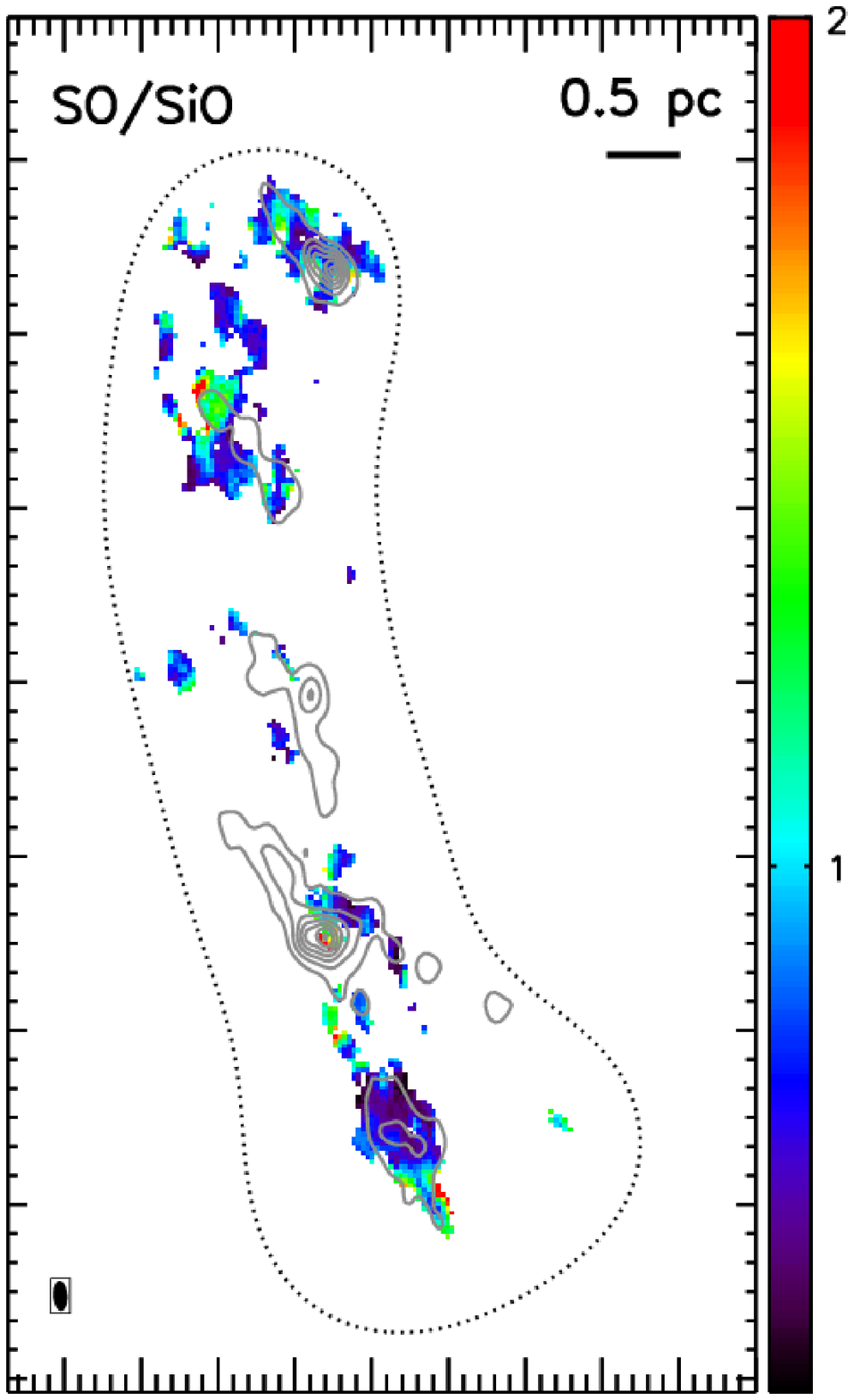}
\caption{Each of the four panels shows integrated intensity ratios between one of the `slow shock tracers' (\methanol{} \mbox{8$_{-1,8}$--7$_{0,7}$} labeled as `USB' \& \mbox{4$_{2,2}$--3$_{1,2}$} labeled as `LSB', HNCO 10$_{0,10}$--9$_{0,9}$, and SO 6$_5$--5$_4$) and SiO 5--4. Contours in all panels represent SMA+BGPS 1.3~mm continuum emission with identical levels as in the right panel of \autoref{fig:dust}. The \methanol{} \mbox{8$_{-1,8}$--7$_{0,7}$}/SiO and SO/SiO ratio larger than 2, and the \methanol{} \mbox{4$_{2,2}$--3$_{1,2}$}/SiO and HNCO/SiO ratios larger than 3 are truncated in the maps, which correspond to about twice of the mean ratios, to better illustrate the enhancement of line ratios.}
\label{fig:shocklineratio}
\end{figure*}

\subsection{The Enhancement in Emission of Shock Tracers towards Massive Clumps}\label{subsec:disc_shocks}
We use the enhancement of line ratios between shock tracers to reveal the relative shock strength in the cloud. The SMA+CSO/APEX observations detected several shock tracers, including SiO, \methanol{}, HNCO, and SO. It has been suggested that relatively fast shocks ($\gtrsim$20~\kms{}) can destroy dust grains and release silicon (Si) into the gas phase \citep{guillet2009}\footnote{Widespread detection of SiO throughout the CMZ \citep{jones2012} raises questions about this interpretation.  Similarly, \citet{jimenezserra2010} detected diffuse SiO emission in a region with only low-velocity shocks.}, while relatively slow shocks ($\lesssim$20~\kms{}) can evaporate ice mantles and release molecules such as \methanol{}, HNCO, and SO \citep{garay2000,rodriguez2010} which in turn could be destroyed by fast shocks \citep{garay2000,kelly2017}. Therefore, the SiO and \methanol{}/HNCO/SO lines can be used as tracers of fast and slow shocks, respectively, and line ratios between them are related to shock strength \citep[e.g.,][]{usero2006,kelly2017}. However it must be noted that ice mantle evaporation in hot molecular cores can also be responsible for the enhancement of \methanol{}, HNCO, or SO emission \citep{garrod2006}.

In \autoref{fig:shocklineratio}, we present the ratios of integrated intensities between \methanol{}/HNCO/SO and SiO. To suppress the low signal-to-noise emission, we have excluded pixels with integrated intensities $<$3.6~\jypbm{}\,\kms{} ($\sim$5$\sigma$) for all the lines when making the ratios.

The line ratios between \methanol{}/HNCO/SO and SiO show clear enhancement toward the compact dust emission as shown in \autoref{fig:shocklineratio}. Based on the spatial distribution of the shock tracers, we speculate that fast shocks of $\gtrsim$20~\kms{} traced by the widespread SiO emission are processing gas in the entire cloud. Meanwhile, the increased line ratios of the `slow shock tracers' toward the compact dust emission suggest two possible scenarios\footnote{A third possibility is that photodissociation, X-ray, or cosmic ray destroy \methanol{}/HNCO/SO molecules in diffuse gas but cannot affect those in dense cores. However the widespread emission of these lines in the cloud makes it unlikely.}. First, slow shocks of $\lesssim$20~\kms{} in these regions can release these molecules. The origin of slow shocks is to be determined. Second, hot molecular cores may evaporate these molecules from dust. In particular, the C1 and C4 clumps are associated with luminous \water{} masers, and C4 is also associated with an UC\hii{} region, hence they likely harbor high-mass protostars. In such scenario, the enhanced \methanol{}, HNCO, and SO emission in these clumps are due to star formation. Future sensitive spectral line observations with e.g.\ the Atacama Large Millimeter/submillimeter Array (ALMA) of \mthc{} lines in these regions will help to tell the existence of hot molecular cores hence verify the two scenarios.

\begin{figure*}[!t]
\centering
\begin{tabular}{cc}
(a) & (b) \\
\includegraphics[height=0.45\textwidth]{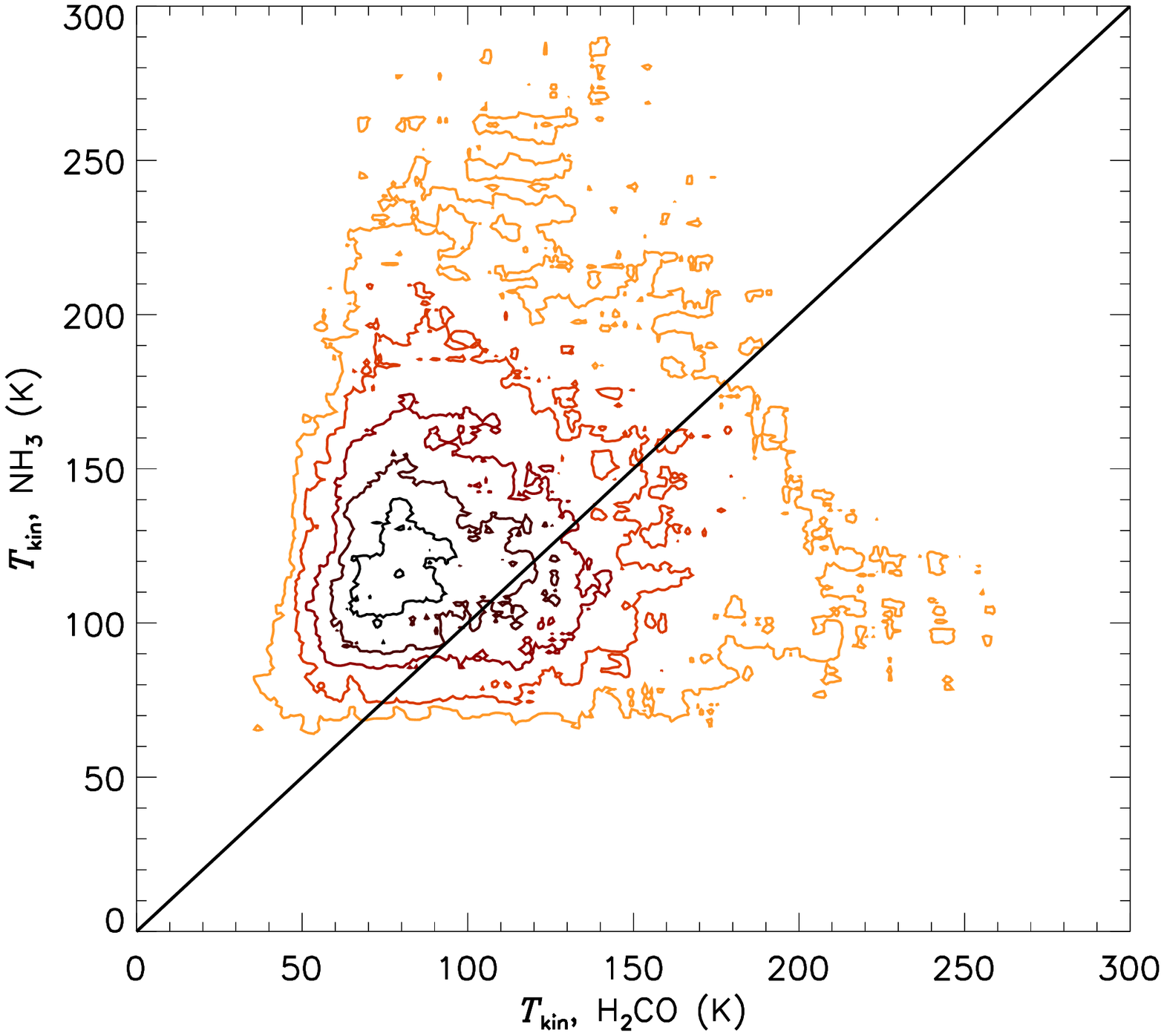} &\includegraphics[height=0.45\textwidth]{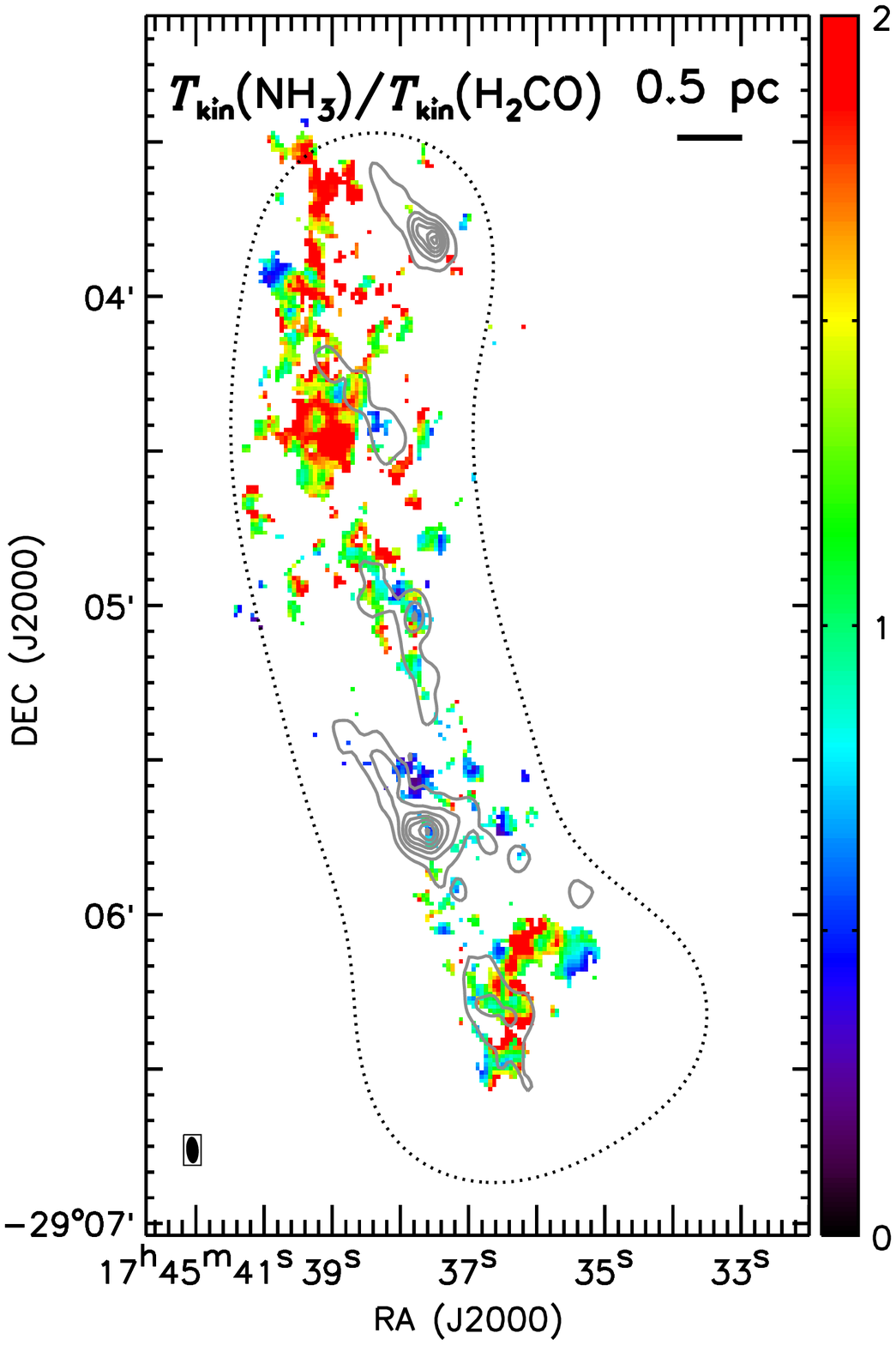}
\end{tabular}
\caption{Comparison between temperatures derived from \amm{} and \fmh{} lines. (a) Density plot of kinetic temperatures derived from \fmh{} and \amm{} lines for pixels where both temperatures are available. Contour levels are between 10\% and 90\%, from lightest to darkest colors, in step of 20\%, meaning the percentage of data points outside of the given level. The diagonal line shows where the two temperatures are equal. (b) Ratio map of kinetic temperatures derived from \amm{} and \fmh{}. Ratios above 2 (up to a maximum of 4.7) are truncated to 2. Only pixels where both temperatures are available are displayed. Contours show SMA+BGPS 1.3~mm continuum emission with identical levels as in the right panel of \autoref{fig:dust}.}
\label{fig:tkin_density}
\end{figure*}

\subsection{Gas Temperatures at 0.1-pc Scales}\label{subsec:disc_tkin}
As shown in \autoref{subsec:results_tkin}, the gas temperatures derived from \fmh{} and \amm{} lines are higher than the dust temperatures in the \ctt{} that are at $\sim$18--30~K \citep{molinari2011}. Such discrepancy between gas and dust temperatures has been observed in the CMZ at $\gtrsim$1~pc spatial scales \citep[e.g.,][]{ao2013,mills2013,ott2014,ginsburg2016,immer2016,krieger2016}. Our observations confirm that in the \ctt{} high gas temperatures of $\gtrsim$50~K continue down to 0.1-pc scales. However, no observations of dust temperatures in this cloud at 0.1-pc scales have been made, making a direct comparison between gas and dust temperatures infeasible.

\subsubsection{Gas Temperatures Derived from Different Tracers}\label{subsubsec:disc_diffTkin}
We compare gas temperatures derived from the \fmh{} and \amm{} lines. We first regrid the \amm{} (2,~2) and (4,~4) datacubes to the \fmh{} maps that have larger pixel size, then derive temperatures based on \amm{} line ratios following the procedures in \autoref{subsec:results_tkin}. The offset between the resulting temperature map and the original \amm{} temperature map in \autoref{fig:nh3_tkin} is $\lesssim$5~K for most of the pixels, but increases to 30~K on boundaries of emission where signal-to-noise ratios are lower. Pixels with large temperature offsets are a small fraction of the map, and the offset is comparable to the uncertainty of the temperatures, therefore we consider the regridded temperature map to be acceptable.

The regridded \amm{} temperature map is compared with the temperature map derived from \fmh{} lines (\autoref{fig:h2co_tkin}b). As shown in \autoref{fig:tkin_density}a, for pixels where both \fmh{} and \amm{} temperatures are available, temperatures derived from the two tracers are poorly correlated. This poor correlation is likely due to the large uncertainties in both temperatures.

One feature in \autoref{fig:tkin_density}a is a slight overpopulation in higher temperatures with \amm{} than with \fmh{}. This can be also seen in the temperature ratio map in \autoref{fig:tkin_density}b. In the northeastern side of the cloud, temperatures with \amm{} are $\sim$200~K, while temperatures with \fmh{} are $\lesssim$100~K, resulting in ratios of $\gtrsim$2. Similar high ratios are found in the southwestern side of the cloud, where temperatures with \amm{} are $>$200~K and those with \fmh{} are $\sim$150~K. Such large temperature ratios are unlikely due to the 50\% uncertainties in temperatures. Given that the \fmh{} lines have higher critical densities than the \amm{} lines ($\sim$10$^5$~\cc{} vs.\ $\sim$10$^{3\text{--}4}$~\cc{}), the two molecules may trace two gas components in these high-temperature-ratio regions: a less dense and warmer component traced by \amm{} lines, and a denser and colder component traced by \fmh{} lines.

On the other hand, temperature ratios of $<$1 are also found. One example is the dense core C4-P1, whose temperature derived from \fmh{} lines is higher than from \amm{} lines ($\sim$120~K vs.\ $\sim$70~K, \autoref{appd:radex}; also see position~11 in \autoref{tab:heating}) despite large uncertainties. The density derived from \fmh{} is also larger than from \amm{}, although the densities highly depend on the assumed column densities and abundances. This may suggest internal heating in C4-P1, if this dense core has a centrally peaked density profile. The heating source could be embedded high-mass protostars, traced by \water{} masers and UC\hii{} regions (see \autoref{subsec:results_hii}). Other candidates of such internally heated dense cores include C1-P1, C2-P1, C4-P4, and C4-P5 (positions~2, 5, 12, and 13, respectively in \autoref{tab:heating}).

In addition, the gas temperatures with \fmh{} at 0.1-pc scales are systematically higher than those at 1-pc scales. \citet{ao2013}, \citet{ginsburg2016}, and \citet{immer2016} derived temperatures of 50--100~K in the \ctt{} with single-dish \fmh{} data. In \autoref{fig:h2co_tkin}c, high tempratures of $>$200~K are found at 0.1-pc scales, especially toward the C1, C2, and C5 clumps. If we smooth our SMA+APEX \fmh{} data to the angular resolution of 30\arcsec{} to match with the single-dish observations, the \fmh{} 3$_{2,2}$--2$_{2,1}$ to 3$_{0,3}$--2$_{0,2}$ line ratios range from 0.27 to 0.45 and the temperatures from RADEX modelling are 50--120~K. Therefore, the discrepancy is likely due to the angular resolution, and our observations suggest that high-temperature spots of $>$200~K exist at 0.1-pc scales.

\begin{table}[!t]
\begin{deluxetable*}{cccDcDcc}
\tabletypesize{\scriptsize}
\tablecaption{Temperatures, linewidths, and \water{} maser luminosities of selected positions.\label{tab:heating}}
\tablewidth{0pt}
\tablehead{
\multirow{2}{*}{Position ID} & R.A.~\& Decl. & $T_\text{kin}$, \fmh{} & \multicolumn2c{FWHM, \fmh{}} & $T_\text{kin}$, \amm{} & \multicolumn2c{FWHM, \amm{}} & $L_\text{H$_2$O}$ & Associated \\
 & (J2000) & (K) & \multicolumn2c{(\kms{})} & (K) & \multicolumn2c{(\kms{})} & (10$^{-7}$~\lsol{}) & \water{} masers
 }
\decimals
\startdata
1  & 17:45:38.10, $-$29:03:41.75 & 250 & 15.7$\pm$1.0  & 177 & 10.7$\pm$1.9 & 7.0   & W1 \\
2  & 17:45:37.48, $-$29:03:49.15 & 105 & 8.5$\pm$1.3    & 87   &   9.5$\pm$1.1 & 61.8  & W3 \\
3  & 17:45:39.90, $-$29:03:55.00 & 215 & 14.6$\pm$1.2  & 114 &  15.7$\pm$0.8 & \nodata & \nodata \\
4  & 17:45:38.60, $-$29:03:56.00 & 162 & 16.6$\pm$1.3  & 153 & 13.7$\pm$1.1  & \nodata & \nodata \\
5  & 17:45:38.21, $-$29:04:28.40 & 151 & 14.9$\pm$0.8  & 138 & 19.4$\pm$1.3 & \nodata & \nodata \\
6  & 17:45:39.06, $-$29:04:28.80 & 64  & 7.7$\pm$0.3     & 138 & 10.0$\pm$0.7 &  \nodata & \nodata \\
7  & 17:45:37.76, $-$29:05:01.91 & 117 & 7.6$\pm$0.5    & 159 &   6.2$\pm$0.7 & 110.0 & W5 \\
8  & 17:45:37.68, $-$29:05:13.57 & 105 & 11.8$\pm$1.0  & 127 &   5.4$\pm$0.6 & 0.3   & W8 \\
9  & 17:45:37.52, $-$29:05:22.59 & 124 & 10.4$\pm$1.4  & 141 &   4.1$\pm$0.6 & 0.7   & W9 \\
10  & 17:45:37.16, $-$29:05:41.70 & 169 & 9.6$\pm$0.8    & 232 &   8.8$\pm$0.9 & 3.2   & W10 \\
11  & 17:45:37.62, $-$29:05:43.92 & 134 & 5.8$\pm$0.8    & 76   &   8.1$\pm$1.0 & 53.0  & W11 \\
12 & 17:45:36.72, $-$29:05:46.02 & 135 & 13.1$\pm$1.3 & 77   &   7.5$\pm$2.3 & 4.0   & W14 \\
13 & 17:45:36.33, $-$29:05:49.52 & 257 & 15.8$\pm$1.1 & 126 &   6.1$\pm$0.8 & 38.7  & W15 \\
14 & 17:45:35.15, $-$29:05:53.62 & 78  & 15.6$\pm$0.9  & 104 &   8.0$\pm$1.5 & 0.8   & W16 \\
15 & 17:45:37.11, $-$29:05:54.38 & 127 & 6.8$\pm$0.6   & 249 &   8.9$\pm$0.6 & 4.9   & W17 \\
16 & 17:45:35.50, $-$29:06:04.50 & 132 & 13.5$\pm$0.5 & 107 &   7.0$\pm$0.6 & \nodata & \nodata \\
17 & 17:45:34.73, $-$29:06:14.30 & 208 & 11.0$\pm$2.0 & 103 &   4.7$\pm$0.3 & \nodata & \nodata \\
18 & 17:45:36.71, $-$29:06:17.50 & 102 & 13.3$\pm$1.1 & 109 &   6.7$\pm$0.4 & \nodata & \nodata \\
19 & 17:45:36.43, $-$29:06:19.55 & 108 & 8.4$\pm$0.9   & 166 &   5.1$\pm$0.3 & \nodata & \nodata \\
20 & 17:45:36.30, $-$29:06:29.43 & 157 & 21.4$\pm$1.2 & 192 &   9.2$\pm$0.7 & \nodata & \nodata \\
\enddata
\end{deluxetable*}
\end{table}

\subsubsection{Turbulent Heating and Protostellar Heating}\label{subsubsec:disc_heating}
We make use of the temperature maps derived from \fmh{} and \amm{} lines to understand heating mechanisms in the cloud. As discussed in \citet{ao2013}, \citet{ginsburg2016}, and \citet{immer2016}, turbulent heating can reproduce the observed gas temperatures at $>$1-pc scales in the CMZ. When it comes to 0.1-pc scales, turbulent heating can be still present. As discussed in \citet{ginsburg2016}, the temperature can be well mixed on milliparsec scales in kiloyear timescales for turbulent heating. Indirect evidence of turbulent heating in the northern end of the \ctt{} has been suggested with arguments from enhanced \amm{} emission and SiO/C$^{34}$S line ratios \citep{wright2001,baobab2013}.

In addition, in the \ctt{} where we have found signatures of star formation, we need to take protostellar heating into account. Gas temperatures of $>$100~K have been observed at $<$0.1~pc spatial scales towards hot molecular cores \citep{araya2005}. If the luminous \water{} masers we detected indeed trace high-mass star formation, then it is possible to observe temperatures of $>$100 K at 0.1-pc scales once the protostars evolve to a phase of hot molecular cores. If such protostars exist, they should warm up the dust and produce significant mid-infrared emission, which is in contrast with the non-detection of point sources other than the known \hii{} region at mid-infrared wavelengths in the \ctt{} \citep{yusefzadeh2009}. The reason could be strong extinction: with a characteristic hydrogen column density of 4$\times$10$^{23}$~\sqc{} in the \ctt{} \citetext{C.~Battersby et~al.\ in~prep.}, the extinction at mid-infrared ($\sim$5--30~\micron{}) is $\sim$10--20~mag \citep{draine2003}. High-mass star-forming regions with typical bolometric luminosities of $\sim$10$^5$~\lsol{} \citep{sridharan2002} have an absolute bolometric magnitude of $-7.8$, and an apparent magnitude of $\sim$6.8 at the distance of the CMZ. Adding the extinction will bring their apparent magnitude down to 16.8--26.8, hence make them below the detection limit at mid-infrared in \citet{yusefzadeh2009}.

We take two approaches to explore the effect of turbulent and protostellar heating. First, we search for correlations between temperatures and linewidths, assuming the latter to be an indicator of turbulent strength \citep[cf.][]{ginsburg2016,immer2016}, and between temperatures and \water{} maser luminosities, assuming the latter to be correlated with protostellar luminosities \citep{palla1993}. Second, we attempt to look for direct signatures of turbulent or protostellar heating from our temperature maps. Evidence of strong turbulence is based on SiO emission, while evidence of star formation comes from masers and UC\hii{} regions in dense cores.

In the first approach, we select 20 representative positions in Figures~\ref{fig:h2co_tkin} \& \ref{fig:nh3_tkin}, and derive their temperatures from the mean \fmh{} or \amm{} spectra within a 5\arcsec{}-diameter circle ($\sim$0.2 pc at the distance of the CMZ). The size of 0.2 pc is chosen because protostellar heating usually affect local gas in a radius of $\lesssim$0.1 pc \citep{longmore2011} out of which the temperatures drop to $<$50~K. These positions locate both on dense cores and offset from any dense cores.

\begin{figure}[!t]
\includegraphics[width=0.45\textwidth]{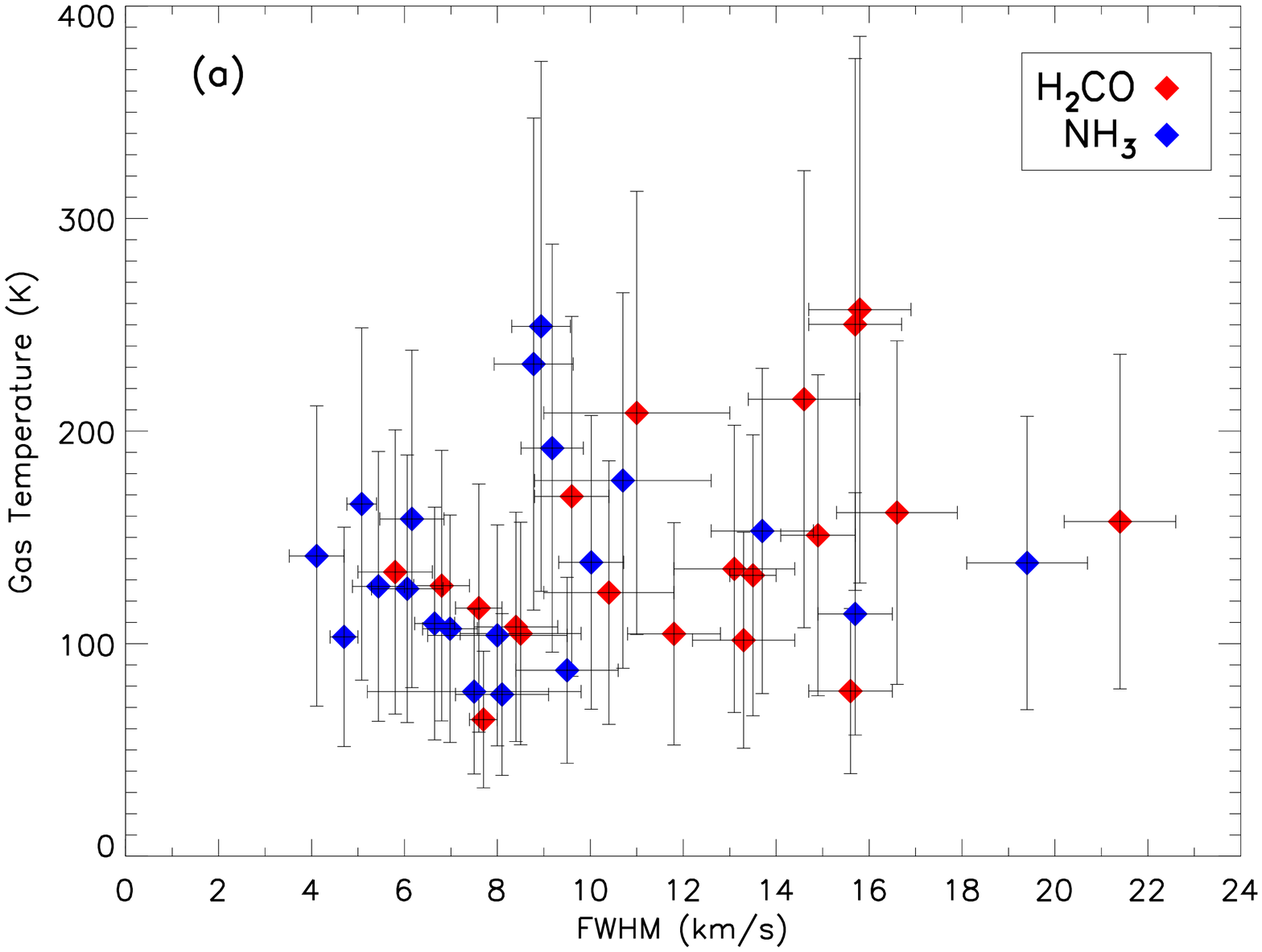} \\
\includegraphics[width=0.45\textwidth]{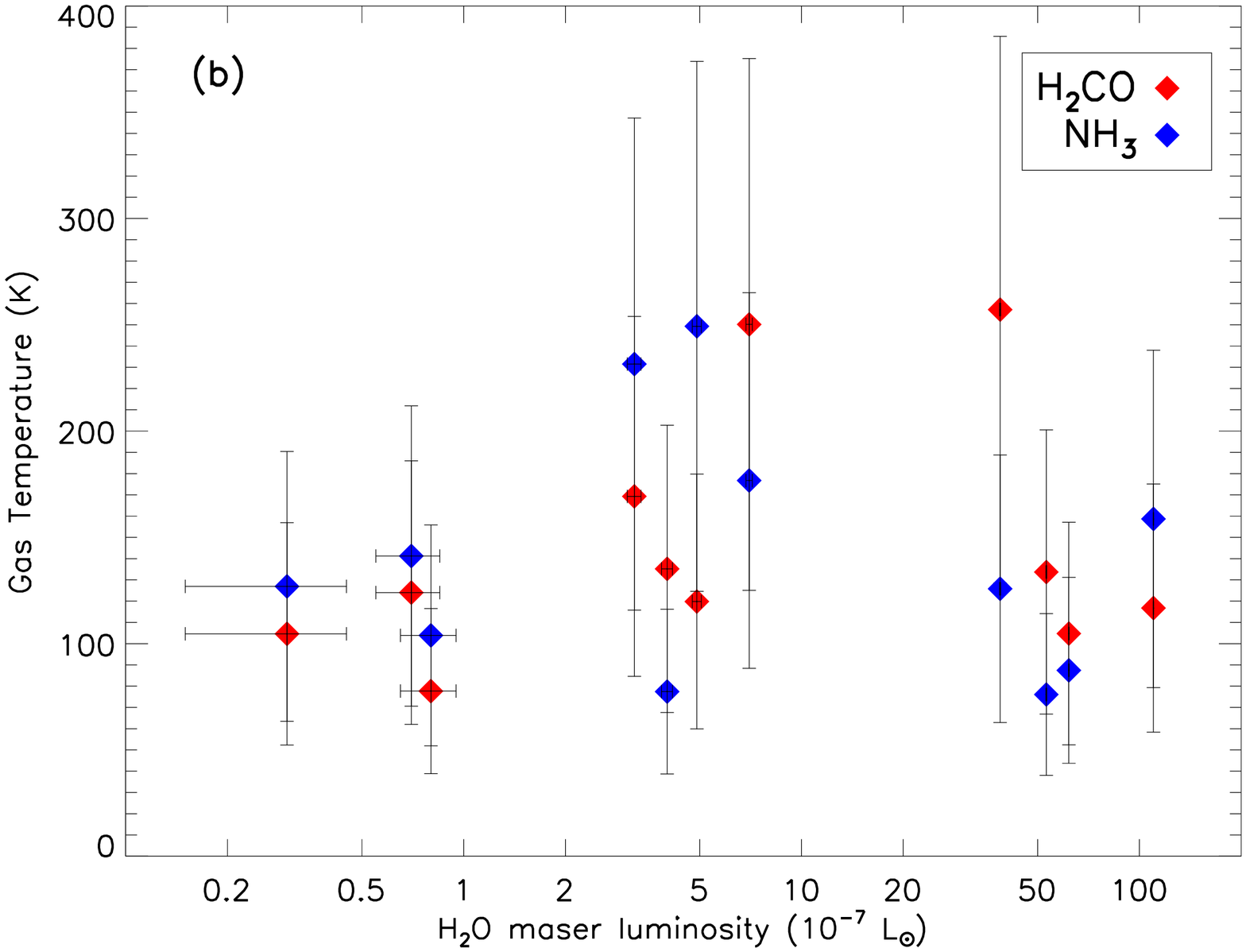}
\caption{(a) The mean temperatures and FWHM linewidths from 18 representative positions in Figures~\ref{fig:h2co_tkin}a \& \ref{fig:nh3_tkin}a. The errors in the temperatures are assumed to be 50\%, while the errors in the linewidths are taken from \autoref{tab:heating}. (b) The mean temperatures and \water{} maser luminosities from 11 positions in \autoref{fig:h2co_tkin}. The errors in the temperatures are the same as in panel (a) and the errors in luminosities are based on the rms of 6~mJy per 0.2~\kms{} of the VLA \water{} maser observations \citep{lu2015b}.}
\label{fig:heating}
\end{figure}

The results are listed in \autoref{tab:heating}. In \autoref{fig:heating}a, we plot the temperatures versus the linewidths. The temperatures are derived directly from the observed line ratios using Equations~\ref{equ:h2co_ratio2tkin} \& \ref{equ:nh3_ratio2tkin} to simplify the process. The linewidths are measured with the mean \fmh{} 3$_{2,1}$--2$_{2,0}$ or \amm{} (4,~4) spectra by fitting a Gaussian profile, with 1-$\sigma$ errors from the fittings listed in \autoref{tab:heating}. For spectra with obvious multiple velocity components, we try to fit two Gaussian profiles simultaneously and take the stronger one.

We derive the Pearson correlation coefficients between the temperatures and the linewidths, using inverse squares of the uncertainties in temperatures (50\%) as weighting factors. We do not include the uncertainties of linewidths in the weighting factors because they are less significant. The weighted correlation coefficients, with \fmh{} data only, \amm{} data only, and all data together, are 0.32, 0.13, and 0.22, respectively. They suggest weak to no correlations between the temperatures and the linewidths. This is in contrast to the statistically significant correlation found at 1-pc scales in several other CMZ clouds in \citet{immer2016}.

In \autoref{fig:heating}b, we plot the temperatures versus the luminosities of the associated \water{} masers, if there are any, in \autoref{fig:heating}b. A \water{} maser must be within 3\arcsec{} projected of the selected position ($\sim$0.1 pc projected distance) to be considered as associated. At last, we find 11 positions with associated \water{} masers. The weighted Pearson correlation coefficient with all data together is 0.08. Therefore, no correlation is found between the temperatures and maser luminosities. 

The temperature-linewidth and temperature-maser luminosity data suffers from several uncertainties. Although we have excluded most of the systematic motions in the cloud by resolving the dense cores at 0.1-pc scales, the linewidth may still be contaminated by unresolved multiple velocity components thus may not be a good indicator of turbulence. Likewise, using \water{} masers as a star formation tracer has large uncertainties (e.g., the occurrence rate and variability of \water{} masers). Therefore, such non-correlations are not straightforward for understanding heating of gas.

In the second approach, we directly search for turbulent or protostellar heated regions in the high-angular-resolution maps from the SMA and VLA observations. This avoids the ambiguities, e.g., using linewidths to trace turbulence, hence is more straightforward.

We notice four candidates of shock heated regions that are marked by boxes in Figures~\ref{fig:sma_maps}, \ref{fig:h2co_tkin}, \& \ref{fig:nh3_tkin}. On the one hand, in the SiO map in \autoref{fig:sma_maps}, these regions show SiO emission that usually traces fast shocks, but are offset from compact dust emission hence are unlikely associated with protostars. On the other hand, as shown in Figures~\ref{fig:h2co_tkin} \& \ref{fig:nh3_tkin}, temperatures in these regions ($\gtrsim$200~K) tend to be higher than their environment. Therefore, these regions are likely heated by shocks.

We also find a candidate of internally heated dense cores, C4-P1, as discussed in \autoref{subsubsec:disc_diffTkin}. Combining with evidence of high-mass star formation in this dense core (\autoref{subsec:results_hii}), it is likely internally heated by embedded high-mass protostars. Several dense cores, such as C1-P1, C2-P1, C4-P4, and C4-P5 also show higher temperatures with \fmh{} than with \amm{} thus are likely internally heated, although their protostellar nature is to be confirmed. Future sensitive spectral line (e.g., multiple \mthc{}) observations with ALMA in these candidates will help to derive their temperature profiles and confirm the embedded heating sources.

Therefore, with the second approach we find evidence of both turbulent and protostellar heating at 0.1-pc scales in the \ctt{}. Turbulent heating seems to be widespread in the cloud, while protostellar heating is localized in dense cores around protostars.

\section{CONCLUSIONS}\label{sec:conclusions}
We have used the SMA 1.3~mm spectral line observations, the VLA \amm{} line observations, as well as complementary single-dish observations to study properties of dense gas in the \ctt{}, one of the massive molecular clouds in the CMZ. 

The main results are:
\begin{itemize}
\item Various molecular lines are detected with the SMA and VLA observations, most of which are widespread and not always spatially coincident with the dense cores traced by dust emission. Analysis based on 2D cross-correlations between the dust and spectral line emission suggests that the \methanol{}, SO, and HNCO lines are best spatially correlated with compact dust emission from dense cores, which may suggest connections between the excitation of these lines and star formation in the dense cores.
\item The line ratios between the `slow shock tracers' (\methanol{}, SO, and HNCO) and the `fast shock tracer' SiO show clear enhancement toward the compact dust emission, indicating the presence of slow shocks or hot molecular cores in these regions.
\item Using multiple transitions of \fmh{} and \amm{}, we estimate gas kinetic temperatures at $\sim$0.1-pc scales under non-LTE conditions. Temperatures derived from \amm{} lines are $\gtrsim$2 times higher than those from \fmh{} lines in several regions, which may suggest two gas components, while toward a dense core C4-P1 temperatures from \fmh{} is higher, which may suggest internal heating. The high angular resolution data reveal high temperatures of $>$50~K, and at some positions $>$200~K at 0.1-pc scales, which are smeared to 50--100~K in lower angular resolution observations.
\item Comparisons between kinetic temperatures and linewidths, as well as between kinetic temperatures and maser luminosities suggest no strong correlations. However, direct evidence of shock heating is found based on the high-angular-resolution temperature maps, with evidence of shocks from the SiO emission. Several likely protostellar heated dense cores are also discussed.
\end{itemize}

In summary, our observations reveal two potential impacts of turbulence and star formation on the molecular gas environment at 0.1-pc scales in the \ctt{}. First, the two factors may affect the chemical composition of the molecular gas. This is supported by widespread SiO emission in the cloud, which is likely related to fast shocks, as well as spatial correlation and enhancement of \methanol{}/HNCO/SO emission toward compact dust emission from dense cores, which is likely related to slow shocks or embedded high-mass star formation. Second, the two factors may heat the molecular gas. Candidate shock heated regions of 0.1-pc scales and $>$200~K temperatures are found throughout the cloud, while signatures of localized heating by embedded star formation are found in several likely internally heated dense cores. Such high-angular-resolution spectral line observations have been proved to be powerful in understanding gas properties in the complex environment in the CMZ. Future observations using ALMA that provide better spectral line sensitivity will be important for the study of star formation and dense gas in the CMZ clouds.

\acknowledgments

We thank the anonymous referee for a constructive review. We thank Simon Radford and William Tan for help with the CSO observation and data reduction, and Junhao Liu for helpful discussions on RADEX modelling. XL acknowledges the support of a Smithsonian Predoctoral Fellowship and the program A for outstanding PhD candidate of Nanjing University. JMDK gratefully acknowledges support in the form of an Emmy Noether Research Group from the Deutsche Forschungsgemeinschaft (DFG), grant number KR4801/1-1. CB is supported by the National Science Foundation under Award No.~1602583. ZYZ acknowledges support from the European Research Council in the form of the Advanced Investigator Programme, 321302, {\sc cosmicism}. This research made use of Astropy, a community-developed core Python package for Astronomy \citep{astropy2013}, and NASA's Astrophysics Data System. This material is based upon work at the Caltech Submillimeter Observatory, which was operated by the California Institute of Technology. This publication is also based on data acquired with the Atacama Pathfinder Experiment (APEX). APEX is a collaboration between the Max-Planck-Institut fur Radioastronomie, the European Southern Observatory, and the Onsala Space Observatory.

\vspace{5mm}
\facilities{SMA, VLA, CSO, APEX, GBT}

\software{CASA, MIRIAD, MIR, GILDAS/CLASS, Astropy \citep{astropy2013}, RADEX \citep{vandertak2007}, myRadex, correl\_images.pro}

\bibliographystyle{aasjournal}

\appendix

\section{Estimate of Kinetic Temperatures Using RADEX}\label{appd:radex}

\begin{figure}[!h]
\centering
\includegraphics[width=0.45\textwidth]{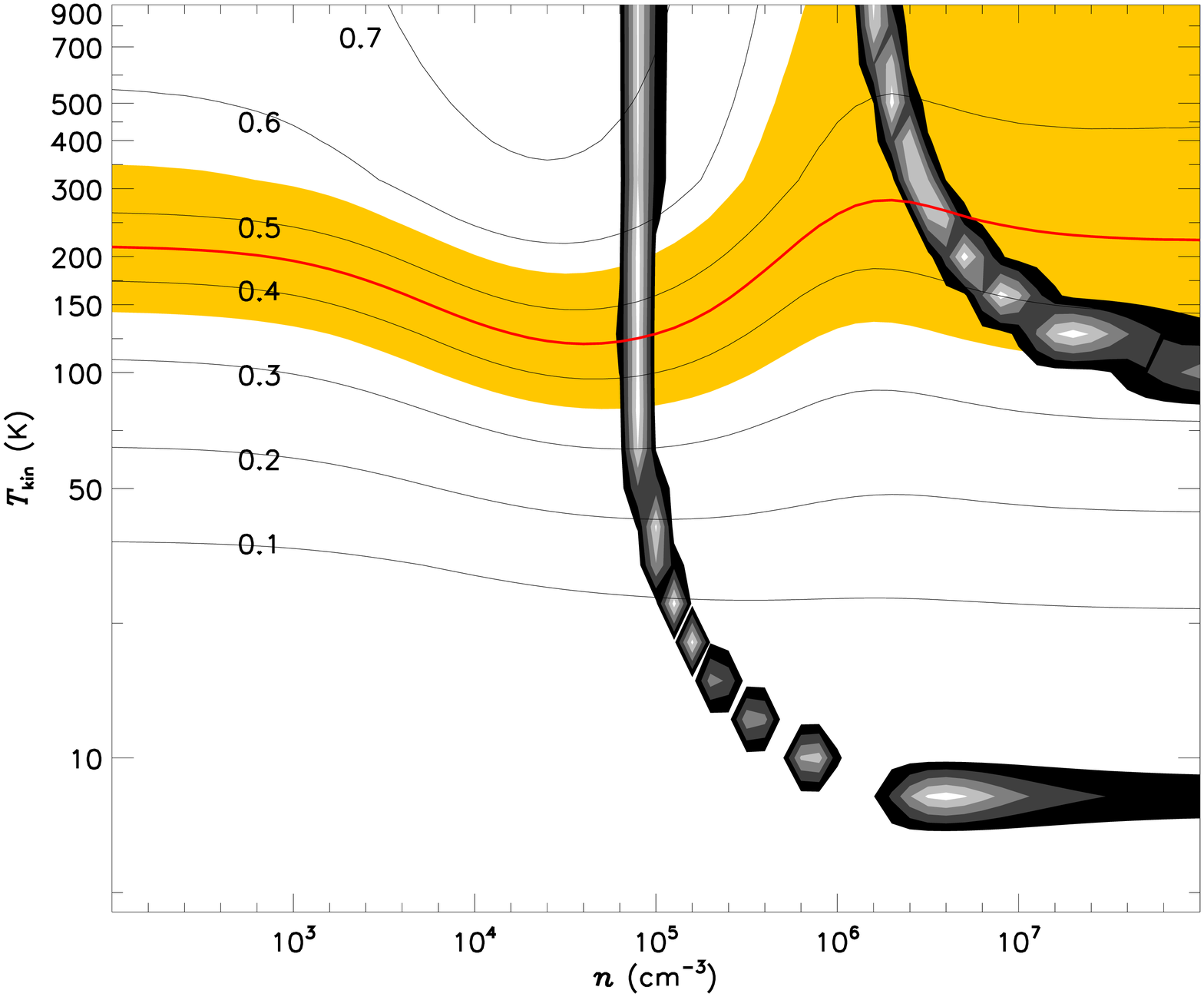}
\caption{RADEX modelling of SMA+APEX \fmh{} lines toward the C4-P1 dense core. The grey shades show the normalized likelihood of the observed \fmh{} 3$_{0,3}$--2$_{0,2}$ peak brightness temperatures being reproduced in RADEX, with values from 0.1 to 0.9 in steps of 0.2 from the darkest to the brightest. The observed \fmh{} 3$_{2,1}$--2$_{2,0}$ and 3$_{0,3}$--2$_{0,2}$ peak intensity ratio is highlighted with a red contour, while the yellow shade around it represents the 1$\sigma$ error of the line ratio due to the observational uncertainty of 0.1~\jypbm{} in the line intensities.}
\label{fig:h2co_radex}
\end{figure}

After running \texttt{myRadex} with the grids discribed in \autoref{subsec:results_tkin}, we derived the likelihood of the observed \fmh{} 3$_{0,3}$--2$_{0,2}$ peak brightness temperatures being reproduced at given gas temperatures, H$_2$ densities and several typical column densities. The grey shades in \autoref{fig:h2co_radex} show the normalized likelihood for the observed \fmh{} transitions towards the C4-P1 dense core in the SMA+APEX data at a fixed column density of 10$^{23}$~\sqc{}. Details can be found in \citet{zzy2014}. As pointed out by \citet{ao2013}, the line ratios between \fmh{} 3$_{2,2}$--2$_{2,1}$ and 3$_{0,3}$--2$_{0,2}$ are more sensitive to gas densities therefore are not as good thermometers as the line ratios between \fmh{} 3$_{2,1}$--2$_{2,0}$ and 3$_{0,3}$--2$_{0,2}$. In \autoref{fig:h2co_radex} we plot the latter line ratios from the models as black contours, which indeed are almost independent of gas densities while sensitive to kinetic temperatures.

\begin{figure}[!t]
\centering
\includegraphics[width=0.6\textwidth]{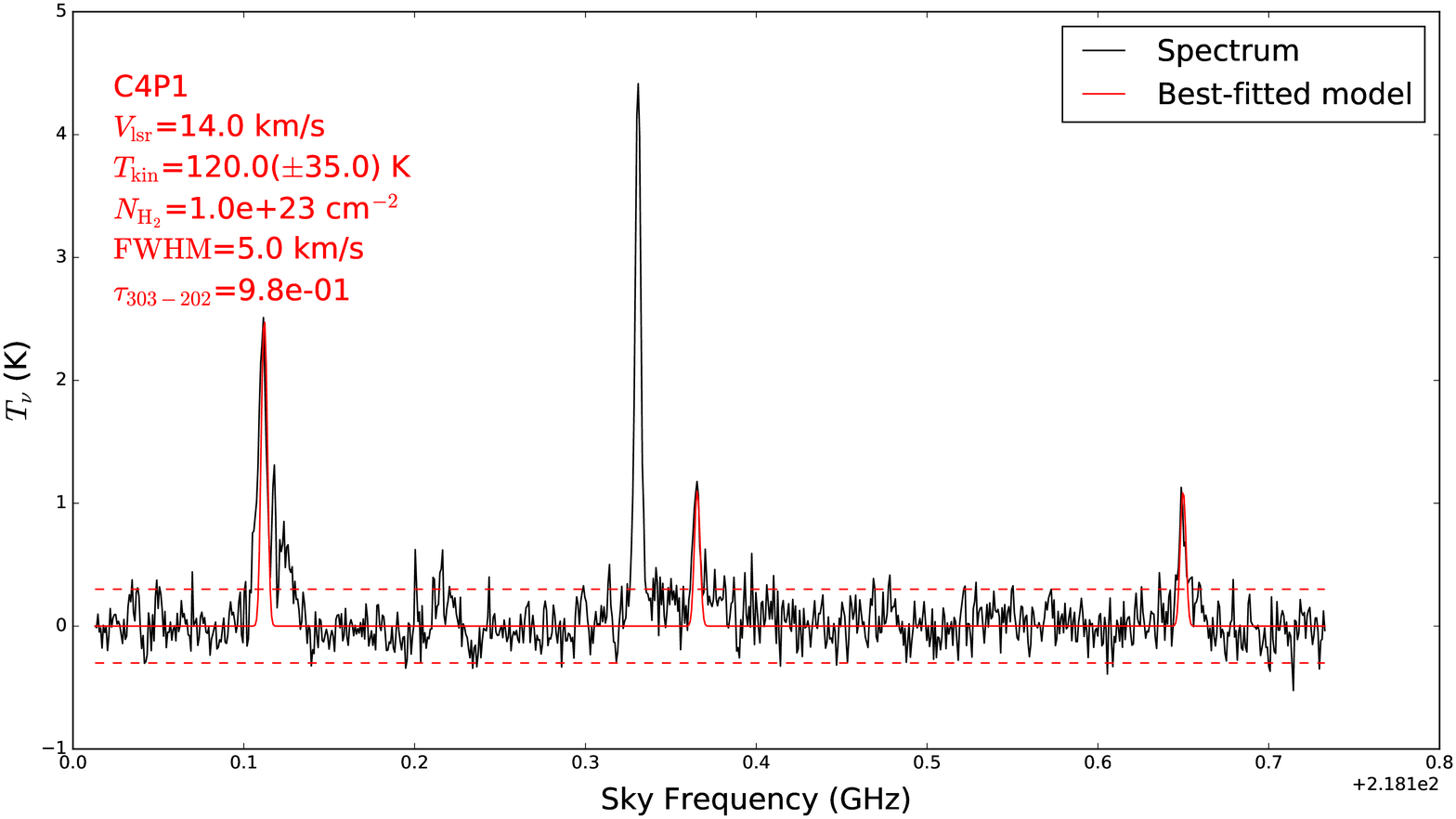}
\caption{The observed SMA+APEX \fmh{} spectrum toward C4-P1, shown in black curve, and the expected spectrum from the RADEX modelling, shown in red curve. The input parameters of the model spectrum are shown in the figure. The horizontal dashed lines represent $\pm$3$\sigma$ levels.}
\label{fig:h2co_fitting}
\end{figure}

\begin{figure}[!b]
\centering
\includegraphics[width=0.4\textwidth]{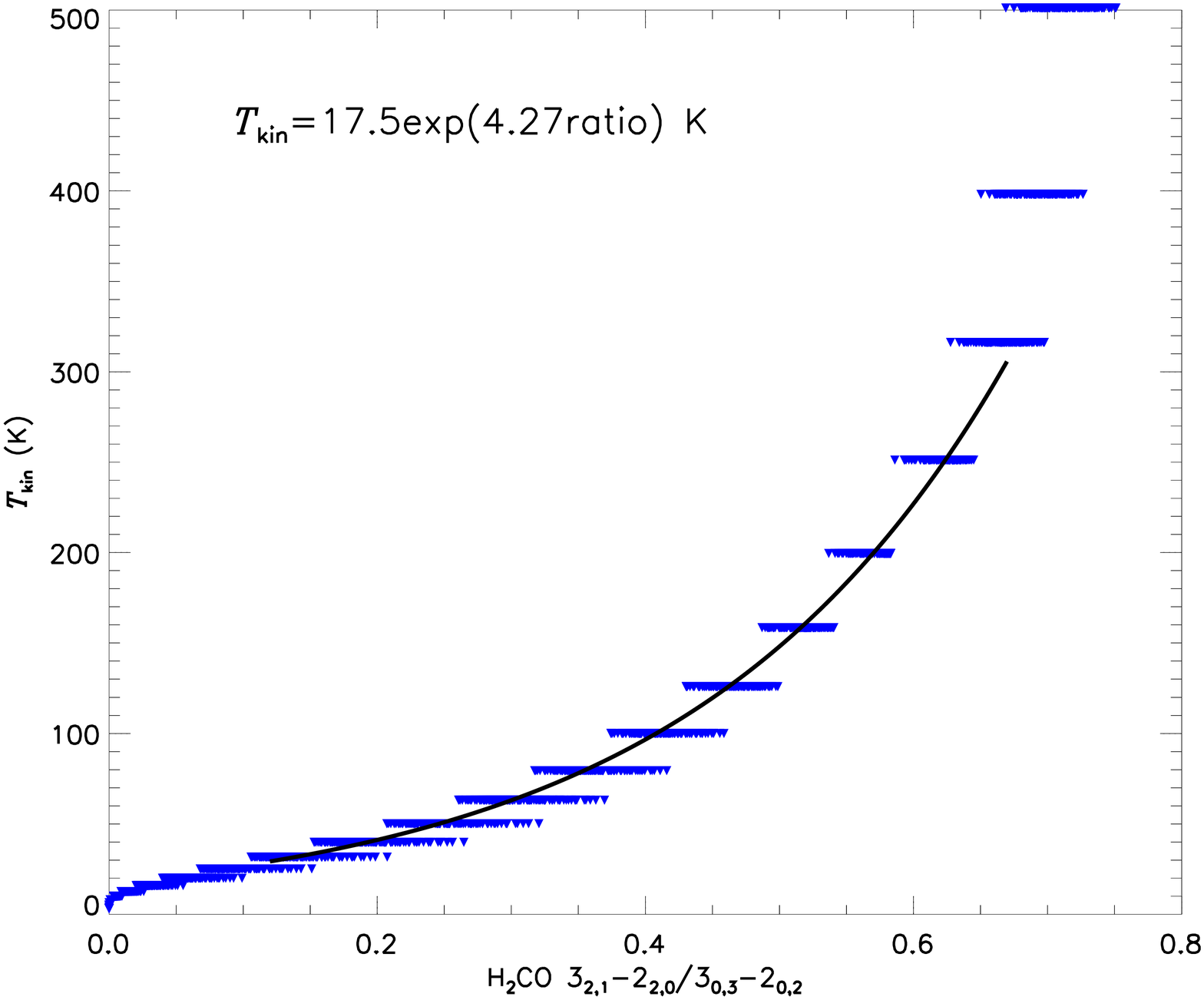}
\caption{The \fmh{} 3$_{2,1}$--2$_{2,0}$/3$_{0,3}$--2$_{0,2}$ line ratios and the kinetic temperatures from RADEX models. At each temperature, we took the line ratios from models with several H$_2$ densities between $10^4$ and $10^5$~\cc{}, and column densities between $5\times10^{22}$ and $5\times10^{23}$~\sqc{}. The solid line is a least-square fit to the ratio--log($T_\text{kin}$) relation between $T_\text{kin}=30\text{--}300$~K.}
\label{fig:h2co_tkin_vs_ratio}
\end{figure}

For the \fmh{} lines toward the C4-P1 dense core in the SMA+APEX data, the peak brightness temperatures of the \fmh{} 3$_{2,1}$--2$_{2,0}$ and 3$_{0,3}$--2$_{0,2}$ transitions are 1.13~K and 2.51~K, which leads to a ratio of 0.45. Therefore, the intersection between the ratio=0.45 contour (highlighted in red in \autoref{fig:h2co_radex}) and the highest likelihood for \fmh{} brightness temperatures (brightest grey shades in \autoref{fig:h2co_radex}) represent the most likely gas temperatures and H$_2$ densities that can simultaneously reproduce the observed \fmh{} 3$_{0,3}$--2$_{0,2}$ intensities and the observed line ratios. Hence the estimated gas temperature is $\sim$120~K and the estimated gas density is $\sim$10$^5$~\cc{}. The estimated temperature and density depend on the assumed column density and \fmh{} abundance. For example, varying the abundance from 10$^{-9}$ to 2$\times$10$^{-9}$ while keeping the other conditions unchanged will bring the estimated temperature to 110~K and the estimated density to 4$\times$10$^4$~\cc{}. In addition, the optical depth of the \fmh{} 3$_{0,3}$--2$_{0,2}$ transition derived at the same time is 0.98, which suggests the \fmh{} line is getting optically thick therefore sensitivity to temperatures is disappearing \citep{ginsburg2016}. The rms of observed \fmh{} line intensities also bring in an uncertainty of 30\%, as shown in \autoref{fig:h2co_radex}. Taking all these factors into account, we assume an uncertainty of 0.15~dex (40\%) for the kinetic temperatures. Note that the collision rates we were using were extrapolated above 300~K (\citealt{wiesenfeld2013}; discussions in \citealt{ginsburg2016}), therefore temperatures $>$300~K are not reliable (they are derived when large \fmh{} line ratios are found due to optically thick emission therefore have large uncertainties themselves anyway).

To examine the quality of the modelling, in \autoref{fig:h2co_fitting} we plot the \fmh{} spectrum extracted towards C4-P1 in the SMA+APEX data and the model constructed with the parameters estimated above. The model reproduces the observed \fmh{} spectrum well. The blue-shifted `line wings' seen in \fmh{} 3$_{0,3}$--2$_{0,2}$ and 3$_{2,2}$--2$_{2,1}$ transitions could be a different velocity component.

In \autoref{subsec:results_tkin} we have noticed that the \fmh{} 3$_{2,1}$--2$_{2,0}$ to 3$_{0,3}$--2$_{0,2}$ line ratios are sensitive to kinetic temperatures, but not so to H$_2$ densities or column densities. We limited H$_2$ densities within $10^4$--$10^5$~\cc{}, which are the range we obtained in RADEX models, and column densities within 5$\times$$10^{22}$--5$\times$$10^{23}$~\sqc{}, which are from \textit{Herschel} observations. Then we plot the estimated temperatures against the \fmh{} line ratios in \autoref{fig:h2co_tkin_vs_ratio}. A least-square fit to the line ratio--log($T_\text{kin}$) relation between $T_\text{kin}=30\text{--}300$~K led to the relation shown in \autoref{fig:h2co_tkin_vs_ratio}. A similar relation has been derived in \citet{ginsburg2016} (see their Figure~6). In \autoref{subsec:results_tkin} we used this relation to directly convert observed \fmh{} line ratios to kinetic temperatures.

\begin{figure}[!t]
\centering
\includegraphics[width=0.45\textwidth]{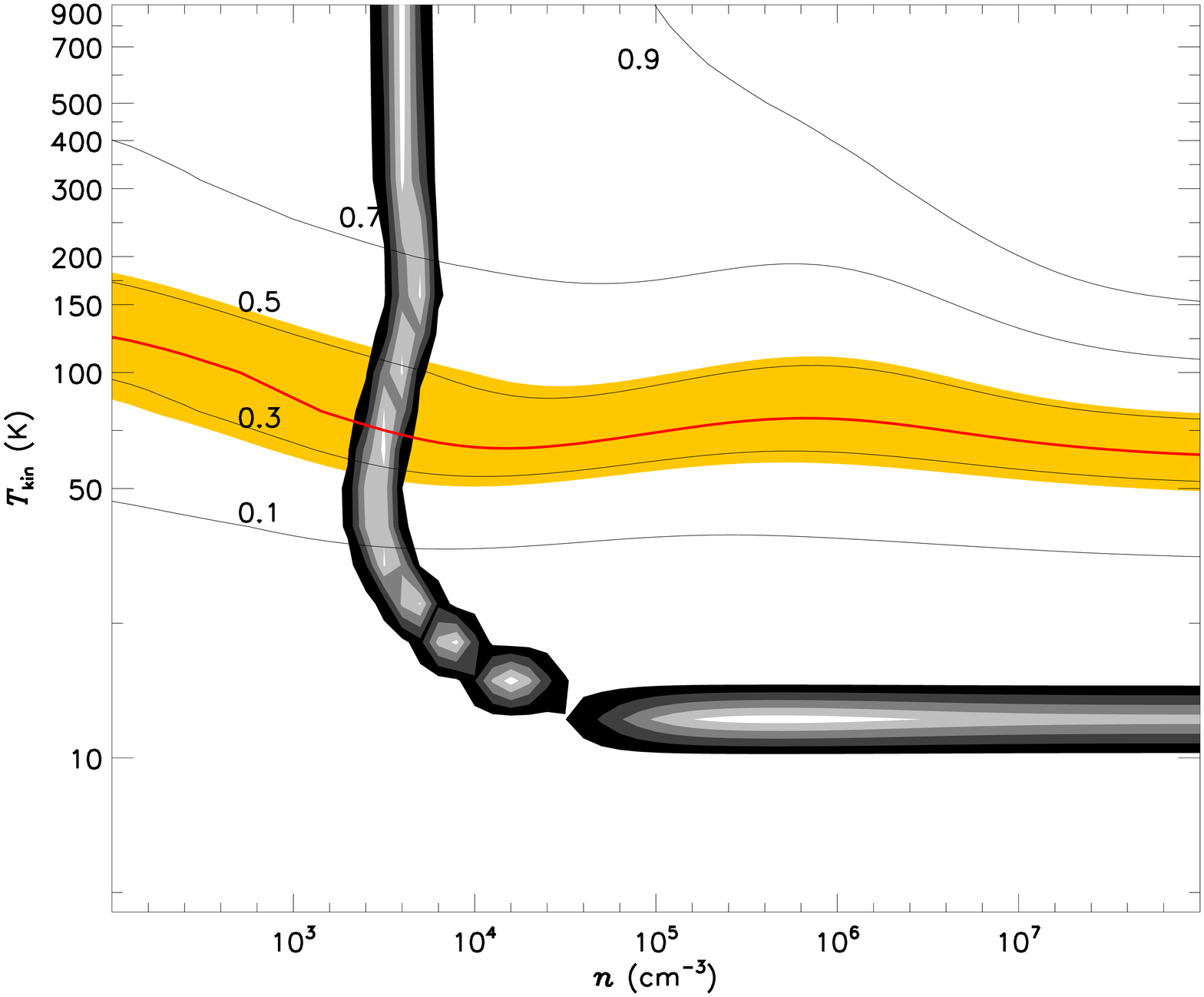}
\caption{RADEX modelling of VLA \amm{} lines toward the C4-P1 dense core. The shaded regions show the normalized likelihood of the observed \amm{} (2,~2) peak brightness temperatures being reproduced in RADEX, with values from 0.1 to 0.9 in steps of 0.2 from the darkest to the brightest. The observed \amm{} (4,~4) and (2,~2) line ratio is highlighted with a red contour, while the yellow shaded region represents the 1$\sigma$ error of the line ratio due to the observational uncertainty of 2~\mjypbm{} in the line intensities.}
\label{fig:nh3_radex}
\end{figure}

Then we applied the above analyses to the VLA \amm{} (2,~2) and (4,~4) spectra. In \autoref{fig:nh3_radex}, we plot the modelling result for the \amm{} lines toward the C4-P1 dense cores at a fixed column density of 10$^{23}$~\sqc{}, but using an \amm{} abundance of 1.8$\times$10$^{-7}$ instead of 3$\times$10$^{-8}$ in order to fit the hyperfine structures of \amm{}~(2,~2). The black contours show the ratios between \amm{} (4,~4) and (2,~2) peak brightness temperatures from the models and the grey shades show the normalized likelihood of the observed \amm{} (2,~2) peak brightness temperature (6.42~K) being reproduced in the models. The observed \amm{} (4,~4)/(2,~2) line ratio in C4-P1 is 0.34. Therefore the estimated kinetic temperature is $\sim$70~K and the estimated density is $\sim$3$\times$10$^3$~\cc{}. The optical depth of the \amm{} (2,~2) main hyperfine component derived at the same time is 12. For the same reason of extrapolated collision rates \citep{danby1988}, temperatures above 300~K are not reliable either. Note that if we use an \amm{} abundance of 3$\times$10$^{-8}$ as mentioned in \autoref{subsec:results_tkin}, the estimated temperature would be $\sim$100~K but the model could not fit the \amm{} (2,~2) satellite components.

\begin{figure}[!t]
\centering
\includegraphics[width=0.6\textwidth]{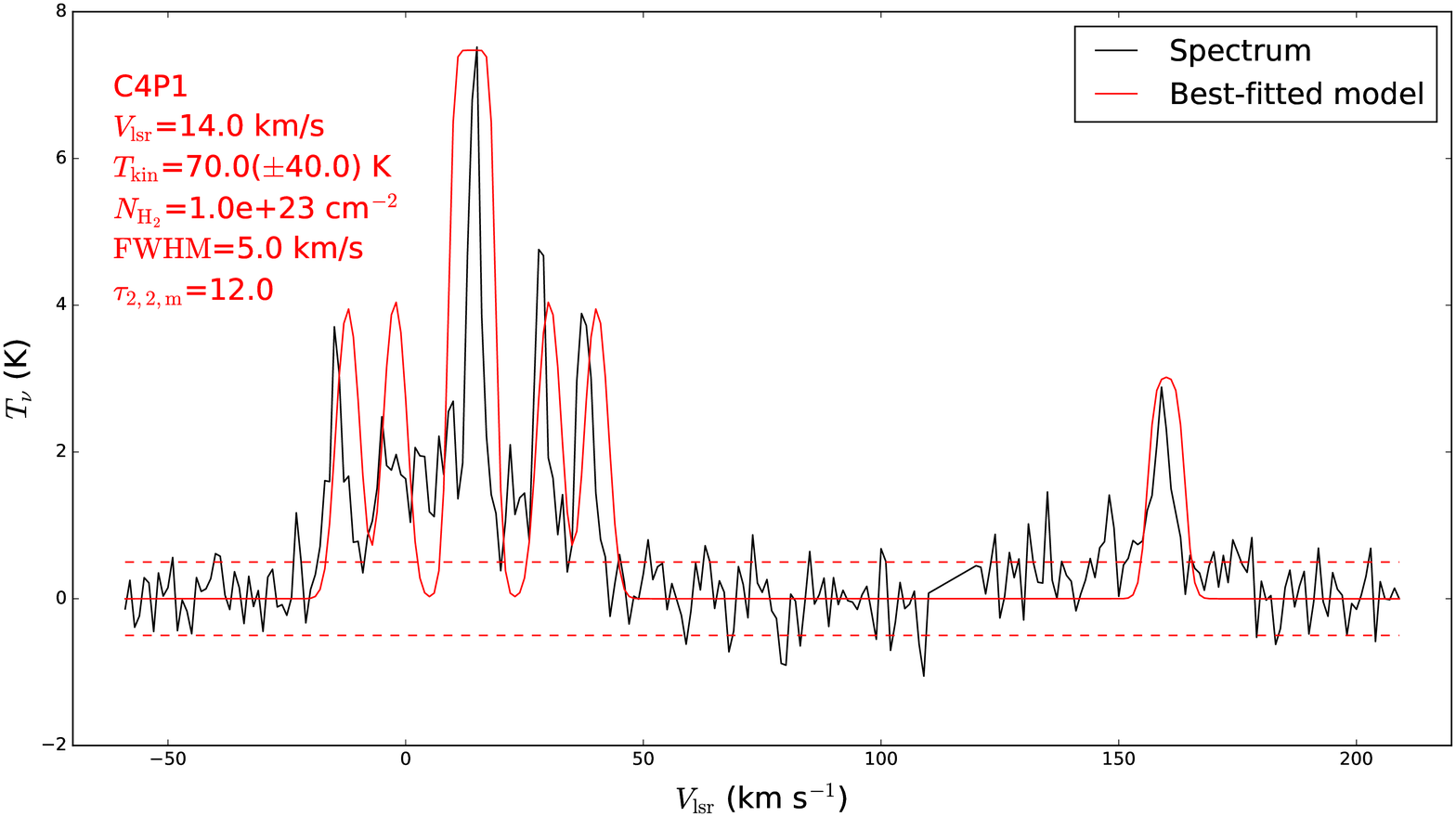}
\caption{The observed VLA+GBT \amm{} (2,~2) and (4,~4) spectra toward C4-P1, shown in black curve, and the expected spectra from the RADEX modelling, shown in red curve. Note that the \amm{} (4,~4) spectrum has been horizontally shifted by $+$150~\kms{}. The input parameters of the model spectra are shown in the figure. The horizontal dashed lines represent $\pm$3$\sigma$ levels.}
\label{fig:nh3_fitting}
\end{figure}

In \autoref{fig:nh3_fitting}, we also compare the \amm{} spectra extracted towards C4-P1 in the VLA+GBT data with the model constructed with the parameters estimated above. The model can reproduce intensities of the observed \amm{} (2,~2) and (4,~4) main hyperfine components, as well as the satellite hyperfine components of \amm{} (2,~2). However, due to the large optical depth, the model lines are much broader than observed. It is possible that multiple optically thick components with small filling factors exist, which lead to the large satellite-to-main line ratio in \amm{} (2,~2), but do not produce broad lines because of their smaller intrinsic linewidths.

The strong \amm{} (2,~2) satellite components shown in \autoref{fig:nh3_fitting} are not universal in the \ctt{}. We only find such spectra toward several massive dense cores, including C4-P1 and C5-P1. For most of the area in the cloud, the \amm{} (2,~2) satellite components are weak, suggesting optically thin emission, and the assumed \amm{} abundance of 3$\times$10$^{-8}$ can fit the observed spectra well. Therefore, we continued to use the abundance of 3$\times$10$^{-8}$, but it must be kept in mind that in several dense cores a higher abundance is likely and resulting temperatures can be lower by 30\%. Again the rms of observed \amm{} line intensities bring in an uncertainty of 30\% as shown in \autoref{fig:nh3_radex}. Overall, we assume an uncertainty of 0.16~dex (45\%) for the kinetic temperatures.

\begin{figure}[!t]
\centering
\includegraphics[width=0.4\textwidth]{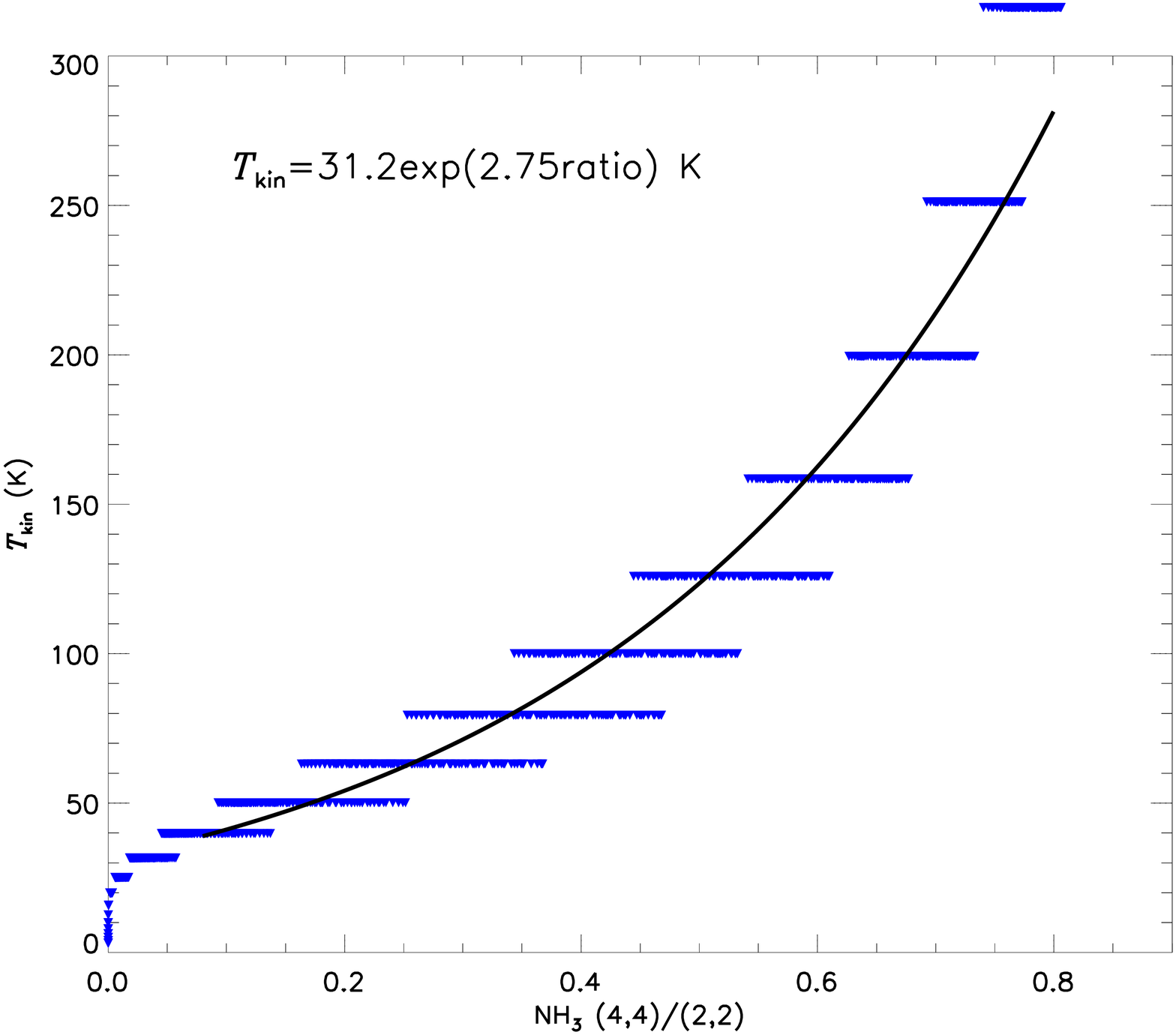}
\caption{The \amm{} (4, 4)/(2, 2) line ratios and the kinetic temperatures from RADEX models. At each temperature, we took the line ratios from models with H$_2$ densities between $10^4$ and $10^5$~\cc{}, and column densities between $5\times10^{22}$ and $5\times10^{23}$~\sqc{}. The solid line is a least-square fit to the ratio--log($T_\text{kin}$) relation between $T_\text{kin}=30\text{--}300$~K.}
\label{fig:nh3_tkin_vs_ratio}
\end{figure}

In \autoref{fig:nh3_tkin_vs_ratio}, we plot the estimated kinetic temperatures against the \amm{} line ratios from the models, within the same ranges of H$_2$ densities and column densities as for the \fmh{} and an \amm{} abundance of 3$\times$10$^{-8}$. Then a least-square fit between $T_\text{kin}=30\text{--}300$~K led to the relation shown in the figure, with which we obtained the kinetic temperature map in \autoref{subsec:results_tkin}.

\end{document}